\documentclass[preprint,12pt]{elsarticle}

\usepackage[english]{babel}
\usepackage{graphicx}
\usepackage{amsmath}
\usepackage{amssymb}
\usepackage{dsfont}
\usepackage{braket}
\usepackage{siunitx}
\usepackage[table]{xcolor}
\usepackage{appendix}
\usepackage{listings}
\usepackage{float}
\usepackage{hyperref}

\graphicspath{{./png/} {./eps/} {./eps/jap/}}

\newcommand{\tsc}[1]{$_{\text{#1}}$}

\journal{arXiv}

\begin{document}

\begin{frontmatter}

\title{Winterface: a postprocessing tool for ab initio quantum transport simulations}


\author[a]{Christian Stieger}
\author[a]{Mathieu Luisier}

\cortext[author] {Corresponding author.\\\textit{E-mail address:} stiegerc@iis.ee.ethz.ch}
\address[a]{Integrated Systems Laboratory\\Gloriastrasse 35, 8092 Z{\"u}rich, Switzerland}

\frontmatter



\selectlanguage{english}
\begin{abstract}
\label{chap:abstract}

In this work a framework for quantum transport simulation from first principles is introduced, focusing on the coherent case. The model is based on the non-equilibrium Green's function (NEGF) formalism and maximally localized Wannier functions (MLWFs). Any device simulation, here based on two-dimensional (2-D) materials, starts by identifying a representative unit cell, computing its electronic structure with density functional theory (DFT), and converting the plane-wave results into a set of MLWFs. From this localized representation of the original unit cell, the device Hamiltonian can be constructed with the help of properly designed upscaling techniques. Here, a powerful tool called Winterface is presented to automatize the whole process and interface the initial MLWF representation with a quantum transport solver. Its concepts, algorithms, and general functionality are discussed on the basis of a molybdenum disulfide (2-D) monolayer structure, as well as its combination with tungsten disulfide. The developed approach can be considered as completely general, restricted only by the capability of the user to perform the required DFT calculations and to "wannierize" its plane-wave results.

\end{abstract}

\begin{keyword}
\texttt{C++} \sep Wannier functions \sep quantum-transport \sep DFT \sep 2-D materials
\end{keyword}

\end{frontmatter}



\let\clearpage\relax
\section{Introduction}
\label{chap:intro}

The inevitable end of Moore's scaling law \cite{moore} calls for novel transistor concepts that can deliver reliable logic performance in future ultra-scaled technology nodes. Although the semiconductor industry has already moved to three-dimensional FinFETs \cite{intel22nm}, innovations at the architecture and material levels will be required for next-generation devices. Besides the widely studied gate-all-around (GAA) nanowire (NW) \cite{gaa_nw_paper1,gaa_nw_paper2,gaa_nw_paper3} and ultra-thin-body \cite{utb_paper} field-effect transistors, recent years have seen the emergence of a new class of two-dimensional (2-D) materials consisting of atomically thin layers connected by van der Waals forces. Its first and probably most famous member is graphene, which was discovered by Novoselov et. al. in 2005 \cite{graphene_omgomg}. Despite impressive carrier mobility values ($>$ 100,000 V/ms), graphene does not lend itself to logic applications due to the lack of a band gap. However, the available design space for 2-D materials is huge, according to recent theoretical investigations \cite{Mounet2018}. Currently, a strong accent is set on transition metal dichalcogenides (TMD), such as MoS\tsc{2}, which appear more promising as future channel materials than graphene. A transistor made of a single layer of MoS\tsc{2} was experimentally realized in 2011 \cite{mos2_exp_paper}. Applications involving few-layer of heterostructures of 2-D materials have already been demonstrated, e.g. light-emitting diodes \cite{Withers2015}, photodetectors \cite{Hong2014,Lopez-Sanchez2013,Esmaeili-Rad2013}, memory cells \cite{Bertolazzi2013} or memristors \cite{Ge2018}. Among the 2-D materials that have received wide attention, black phosphorous (BP) stands out \cite{bp_paper1,Buscema2014}. As compared to TMDs, BPs exhibit highly anisotropic electrical and thermal properties \cite{Qiao2014}, which could pave the way for other original applications. \\

At this point, it is not clear whether 2-D materials can compete with other technologies and if so, which component or heterostructure is the most suitable at performing a given task. Whilst such a large number of possible configurations offers exciting opportunities in terms of novel device concepts, it also requires improved solutions, both experimental and theoretical, to explore the available design space. Technology computer aided design (TCAD) represents a powerful and well-established approach to address this challenge. Thanks to its cost- and time-effectiveness, TCAD can help experimentalists to rapidly converge towards the most promising contenders. However, at the current nanometer scale of the transistor dimensions, classical and semi-classical simulation methods such as the drift-diffusion or Boltzmann transport equations should be replaced by a quantum mechanical treatment of the device properties. To describe the electronic structure of 2-D materials, different methods exist, from the most fundamental ones, such as density functional theory (DFT) \cite{DFT_paper}, up to empirical ones, for example tight-binding (TB) \cite{TB_paper}. When coupled with a quantum transport solver, both approaches suffer from their own limitations,  such as the size of the system that can be handled for DFT, or the need to create physically meaningful parameters sets for TB. As a compromise between them, a maximally localized Wannier function (MLWF) representation of the DFT results can be used \cite{mlwf1,mlwf2}. This can be seen as a first step towards ab initio device investigations. \\

The idea consists of identifying a primitive unit cell that is representative of the system of interest, perform a DFT calculation of it, convert the plane-wave results into a set of MLWF, and finally scale up the obtained TB-like Hamiltonian matrix to the size of the considered device. Once the general concept is established, what remains to be developed is a technique capable of providing the necessary throughput to screen the large design space of 2-D materials. This paper aims at providing a toolbox, called \emph{Winterface}, to automatically upscale the MLWF representation of comparatively small unit cells to desired, sometimes complex, geometries, e.g. heterostructures. Since the problem to be solved depends on numerical outputs and cannot be directly formulated in terms of equations, practical examples will be provided to demonstrate the principle of the proposed approach, identify its shortcomings, and give evidence of its utility to simulate nano-devices based on 2-D materials. The atomic structures selected as testbeds will ideally illustrate the key concepts that have been implemented. Although the physics of these examples might be interesting, this is not the selection criterion that was applied. Here, what matters is the robustness and versatility of the created Winterface code. So far, hundreds of material configurations have been constructed and the transport properties of all of them could be successfully investigated, thus demonstrating the quasi-universality of the methodology. However, at this point, a complete automation was not possible, because critical components are missing, in particular an initial guess projection for the MLWF \cite{LinLin}. Additionally, acceptable approximations depend on a compromise between computational burden and physical accuracy, which have to be prioritized by the user of the upscaling algorithms available in Winterface. \\

While the focus of this work is on coherent transport, dissipative scattering mechanism can be considered as well. They are expected to play a significant role in ultra-scaled structures due to, for example, the close proximity of electrons and phonons and their increased coupling \cite{pop2010}. Few studies mainly concerned with graphene \cite{Balandin2008,Seol213,pop_varshney_roy_2012}, but also molybdenum disulfide \cite{Yan2014,Yalon2017} and black phosphorous \cite{bp_therm} have discussed these issues. Still, it can be generally said that the thermal behavior of 2-D materials is not completely understood yet. The Winterface code can be extended to produce all required inputs for such simulations \cite{MoS2_paper}. \\

The outline of the paper is as follows: in Section \ref{chap:transport}, the toolchain mentioned above and its components are introduced. Then, in Section \ref{chap:ideal} the notion of upscaling is explained on the basis of an ideal example. Subsequently in Section \ref{sec:wbh}, the first step of a general approach to upscaling, the generation of interactions along atomic bonds, is presented. This preliminary work greatly simplifies the actual process of interfacing plane-wave DFT results to the quantum transport (QT) code, and allows for additional approximate upscaling techniques dedicated to heterostructures, as discussed in Section \ref{chap:hctor}. Section \ref{sec:interface} is devoted to the coupling of Winterface with an actual QT Solver, OMEN \cite{OMEN_paper}. The notion of approximate upscaling techniques is discussed in Section \ref{sec:approx}, before results are presented in Section \ref{chap:results} in the form of transmission functions calculated by OMEN from Winterface inputs. Finally, the paper is concluded and an outlook is provided in Section \ref{chap:conclusion}.


\section[Quantum Transport]{Quantum Transport From First-Principles}\label{chap:transport}

Semiconductor and other solid-state devices have long been modeled using classical theories such as the drift-diffusion equations. However, as the channel length of modern transistors is reaching 10nm and below, a full quantum mechanical treatment has become unavoidable. To this end, empirical models such as tight-binding \cite{tb_method} have been employed, but when it comes to the investigation of heterostructures \cite{dft_hetero1,dft_hetero2}, metal-insulator-metal junctions \cite{met_ins_dft}, or novel 2-D materials \cite{Mounet2018}, an approach from first-principles is better indicated.

\subsection{Density Functional Theory and Wannier Functions}
\label{sec:mlwf}

A popular approach for ab initio simulations is \emph{density functional theory} (DFT) \cite{DFT_paper,pbe_paper,hse06_paper,hse06_mega}, where the physics are often expressed in terms of a plane-wave (PW) basis. Since quantum transport simulations require a representation in terms of a basis set localized in real space, as well as system sizes far exceeding the DFT limit (typically 1000 atoms), further processing of the PW data is needed. As a solution to these problems, the DFT results can first be converted to \emph{Wannier functions} through a process called \emph{wannierization} \cite{mlwf1}. Secondly, only a relatively small unit cell is simulated in DFT and the PW outputs transformed into maximally localized Wannier functions. The produced Hamiltonian matrix can then be upscaled to the device dimensions and used in quantum transport simulations. This technique is the main focus of this paper and will be explained in detail in Sections \ref{chap:ideal}, \ref{sec:wbh}, and \ref{chap:hctor}. \\

The Kohn-Sham orbitals from DFT obey Bloch's Theorem, i.e.

\begin{equation}\label{sec:mlwf:eq1}
 \psi_{n,\mathbf{k}}(\mathbf{r}) = u_{n,\mathbf{k}}(\mathbf{r}) \cdot e^{i\mathbf{k}\mathbf{r}},
\end{equation}

where $u_{n,\mathbf{k}}(\mathbf{r})$ is a periodic function in real-space with band index $n$ and wave vector \textbf{k}. A Bloch wave is periodic in real-space up to a phase factor $e^{i\mathbf{k}\mathbf{R}}$ where \textbf{R} is a vector pointing to a neighboring unit cell. This representation in terms of energy eigenfunctions is very convenient for many applications because the Hamiltonian matrix elements obey $\mathcal{H}_{mn}(\mathbf{k}) = \braket{\psi_{n,\mathbf{k}}^* | \mathcal{H} | \psi_{m,\mathbf{k}}} = \epsilon_{mn} \delta_{mn}$. However, if the objective is to find a Hamiltonian operator localized in a super cell, a plane-wave representation is no more convenient as Bloch waves extend over all space. The process of \emph{wannierizing} is essentially a Fourier transform where the fact that Bloch waves are defined only up to a phase factor $e^{i\phi(\mathbf{k})}$ offers a large degree of freedom, which can be cast into a unitary matrix $U_{mn}^{(\mathbf{k})}$. Wannier functions are defined as:

\begin{equation}\label{sec:mlwf:eq2}
 w_{n\mathbf{R}}(\mathbf{r}) = \frac{V}{(2\pi)^3} \int_{BZ} d^3\mathbf{k} \bigg[ \sum_m U_{mn}^{(\mathbf{k})}\psi_{m\mathbf{k}}(\mathbf{r}) \bigg] e^{-i\mathbf{k}\mathbf{R}}.
\end{equation}

\begin{figure}
 \centering
 \includegraphics[width=9cm]{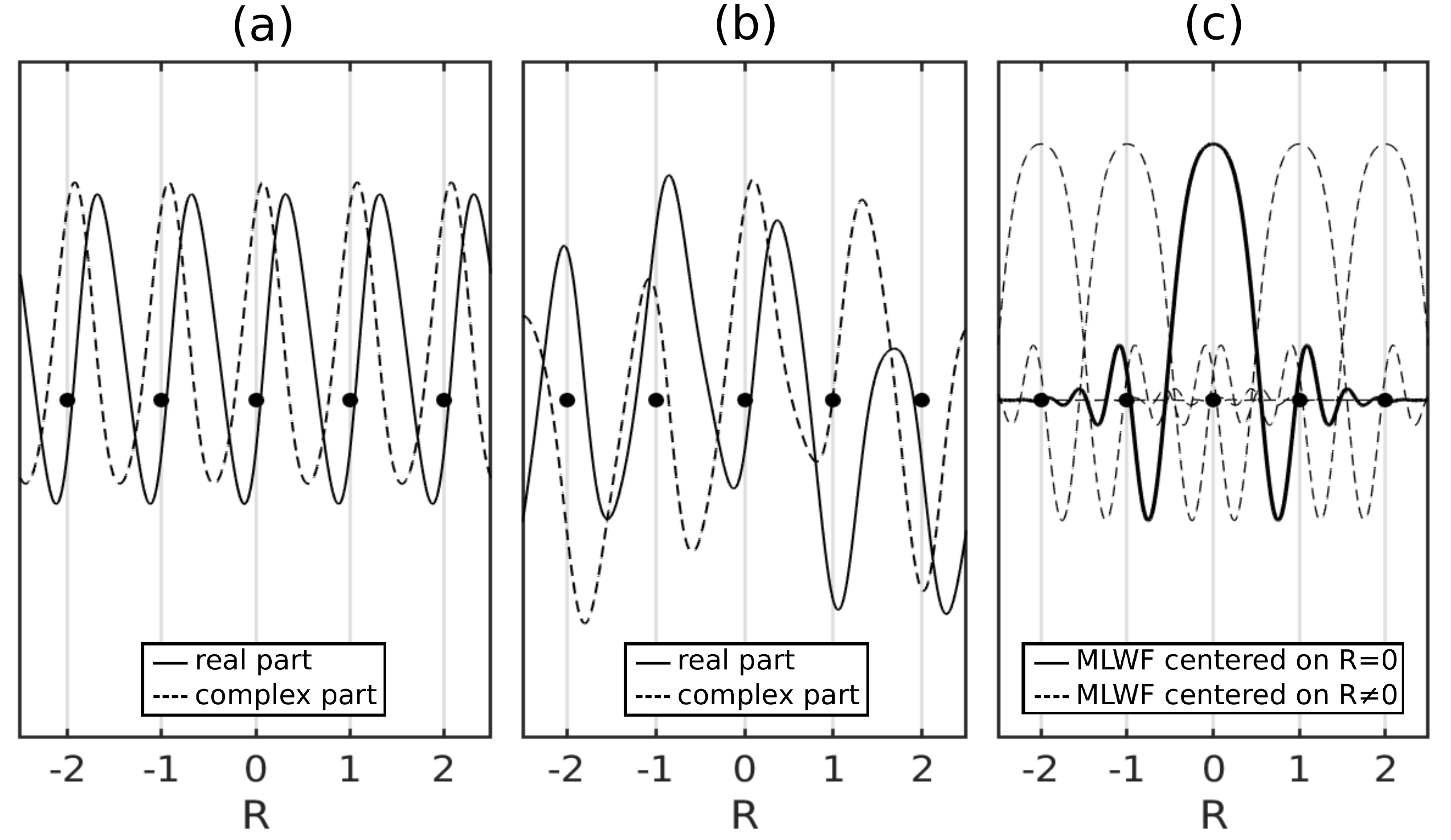}
 \caption{Illustration of plane-wave and Wannier representations of a quantum mechanical ground state on a one-dimensional grid of five unit cells. The parameter R gives the position of each cell relative to the center and the black dots represent atomic positions. (a) Cell-periodic part of a Bloch wave $u_{n,\mathbf{k}}(\mathbf{r})$. (b) Corresponding Bloch wave $\psi_{n,\mathbf{k}}(\mathbf{r}) = u_{n,\mathbf{k}}(\mathbf{r}) \cdot e^{i\mathbf{k}\mathbf{r}}$. (c) Example of a maximally localized Wannier function (MLWF). This is a real function localized in real space and centered on a specific site. Here, the solid line represents the MLWF centered at R = 0, whereas the dashed lines refer to equivalent images centered at R $\neq$ 0, thus illustrating the periodicity of the crystal in the Wannier picture.}\label{sec:mlwf:fig1}
\end{figure}

In Eq. \eqref{sec:mlwf:eq2}, the set of momentum(\textbf{k})-dependent wave functions $\psi_{m\mathbf{k}}(\mathbf{r})$ of band index m are replaced through a unitary transform by a set of Wannier functions $w_{n\mathbf{R}}(\mathbf{r})$ with Wannier index n that are assigned to the unit cell situated at vector \textbf{R} with respect to the origin. The integration is performed over the entire Brillouin zone and V is the volume of the atomic unit cell in real space. For the case where the unitary matrix $U_{mn}^{(\mathbf{k})}$ is chosen such that the spread functional

\begin{equation}\label{sec:mlwf:eq3}
\Omega = \sum_n \big[ \langle w_{n \mathbf{0}} (\mathbf{r}) \rvert r^2 \rvert w_{n \mathbf{0}} (\mathbf{r}) \rangle - \rvert \langle w_{n \mathbf{0}} (\mathbf{r}) \rvert \mathbf{r} \rvert w_{n \mathbf{0}} (\mathbf{r}) \rangle \rvert^2 \big]
\end{equation}

is minimal, we speak of \emph{Maximally Localized Wannier Functions} (MLWF). In this configuration the Wannier functions themselves as well as the Hamiltonian matrix elements can be proved to be real. A comparison of Bloch waves and MLWF is presented in Fig. \ref{sec:mlwf:fig1}. \\

The translation symmetry of the lattice is expressed through \textbf{R} vectors, each of them corresponds to a translated (with respect to a reference unit cell), but otherwise identical Wannier function placed at position \textbf{R}. Due to the spacial localization, pairwise interactions among Wannier functions extend only over a finite subset of \textbf{R} vectors, where the origin may be set arbitrarily. This localization in space is achieved at the cost of localization in energy, i.e. Wannier functions form an orthonormal basis set, but they are not eigenfunctions of the Hamiltonian operator. Therefore, the Wannier index n is not a band index and in general the whole set of Wannier functions contributes to each band. The Hamiltonian operator can be expressed in terms of pairwise interactions among Wannier functions among a finite range of unit cells described by \textbf{R} vectors, i.e.

\begin{equation}\label{sec:mlwf:eq4}
\mathcal{H}_{nm}(\mathbf{R}) = \braket{w_{n\mathbf{0}}|\mathcal{H}|w_{m\mathbf{R}}}.
\end{equation}

Each $\mathcal{H}(\mathbf{R})$ describes the interactions of the Wannier functions shifted according to $\mathbf{R}$ with those assigned to the \emph{home cell} at $\mathbf{R} = 0$. This spatially localized representation has "tight-binding-like" characteristics, as needed for upscaling a set of Hamiltonian matrices $\mathcal{H}(\mathbf{R})$ defined on a cluster of unit cells to a set of Hamiltonian matrices $\tilde{\mathcal{H}}(\tilde{\mathbf{R}})$ describing interactions among a cluster of super cells.

\subsection[NEGF Formalism]{Transport in the NEGF Formalism}
\label{sec:negf}

Density functional theory, when expressed in a plane-wave basis, lends itself perfectly to electronic structure calculations of periodic structures or small molecules. The purpose of this work is, however, to evaluate the "current vs. voltage" characteristics of nano-devices. To do that, the atomic system of interest must be driven out-of-equilibrium by an external voltage source. Such situations can be realized by attaching reservoirs to the simulation domain, from which electrons can be injected and collected. Furthermore, to allow for the definition of the required open boundary conditions (OBCs) \cite{obc1,obc2}, the Hamiltonian that describes the electronic properties of the device must be expressed in a localized basis. Here, a set of MLWFs, as explained in the previous Section, is chosen. For a two-dimensional system (2-D) with transport along the x-axis, confinement along the y-axis, and the z-axis assumed periodic, the following system of equations must be solved for the electron population:

\begin{eqnarray}\label{sec:negf:eq8}
\left\{
\begin{array}{l}
\sum_l\left(\mathbf{E}\delta_{li}-\mathcal{H}_{il}(k_z)-\mathbf{\Sigma}^{RB}_{il}(E,k_z)\right)\cdot \mathbf{G}^{R}_{lj}(E,k_z)=\delta_{ij}, \\
\mathbf{G}^{\gtrless}_{ij}(E,k_z)=\sum_{lm}\mathbf{G}^{R}_{il}(E,k_z)\cdot\mathbf{\Sigma}^{\gtrless B}_{lm}(E,k_z) \cdot \mathbf{G}^{A}_{mj}(E,k_z).
\end{array}
\right.
\end{eqnarray}

The $\mathbf{G}_{ij}(E,k_z)$'s represent the electron Green's Functions at energy E and momentum $k_z$ between atoms i and j situated at position $\mathbf{r}_i$ and $\mathbf{r}_j$, respectively. They are of size $N_{orb,i}\times N_{orb,j}$, where $N_{orb,i}$ is the number of orbitals (basis components) describing atom $i$. The $\mathbf{G}_{ij}(E,k_z)$'s can be either retarded ($R$), advanced ($A$), lesser ($<$), or greater ($>$). The same conventions apply to the self-energies $\mathbf{\Sigma}_{ij}(E,k_z)$. Note that $\mathbf{\Sigma}_{ij}(E,k_z)$ can be computed iteratively \cite{Sancho_1985}, from (generalized) eigenvalue problems \cite{gevp1,gevp2,Sascha_paper} or from contour integral techniques \cite{Mauro_Sascha}. The Hamiltonian entries $\mathcal{H}_{il}(k_z)$ are expressed in the selected MLWF basis. Their contruction is the subject of this paper and will be discussed in the next Sections. The electron concentration $n(\mathbf{r}_i)$ for each atomic position $\mathbf{r}_i$ is given by

\begin{equation}\label{sec:negf:eq9}
 n(\mathbf{r}_i) = -i \sum_{k_z} \int \frac{dE}{2\pi} tr \Big \{ \mathbf{G}_{ii}^{<}(E,k_z) \Big \},
\end{equation}

which can then be plugged into Poisson's equation. The Schr{\"o}dinger equation in Eq. \eqref{sec:negf:eq8} implicitly contains an unknown Hartree component $V_H(\mathbf{r})$ which must be self-consistently calculated with Poisson's equation since the charge in the device gives rise to an electrostatic potential through

\begin{equation}\label{sec:wf:eq4}
 \nabla^2 V_H(\mathbf{r}) = -\frac{\rho(\mathbf{r})}{\varepsilon(\mathbf{r})}.
\end{equation}

The position-dependent charge density $\rho(\mathbf{r})$ may include several components such as the acceptor and donor concentrations as well as the electrons $n(\mathbf{r})$ and hole $p(\mathbf{r})$ densities. Since the charge density $\rho(\mathbf{r})$ depends on $V_H(\mathbf{r})$ through the Schr{\"o}dinger equation and $V_H(\mathbf{r})$ on $\rho(\mathbf{r})$ through the Poisson equation, resulting dependencies must be resolved self-consistently until convergence is reached. Once the out-of-equilibrium state of the system has been determined in this way, the electrical current flowing between two adjacent unit cells labeled $s$ and $s+1$ of a 2-D device structure can also be extracted from the Green's functions

\begin{equation}\label{sec:negf:eq10}
 \begin{split}
 I_{d,s \rightarrow s+1} = \frac{e}{\hbar} \sum_{k_z} \sum_{i \in s} \sum_{j \in s+1} \int \frac{dE}{2\pi} tr \Big \{ \mathcal{H}_{ij}(k_z) \cdot \mathbf{G}_{ji}^{<}(E,k_z) \\
 - \mathbf{G}_{ij}^{<}(E,k_z) \cdot \mathcal{H}_{ji}(k_z) \Big \}.
 \end{split}
\end{equation}

\begin{figure}
 \centering
 \includegraphics[width=11cm]{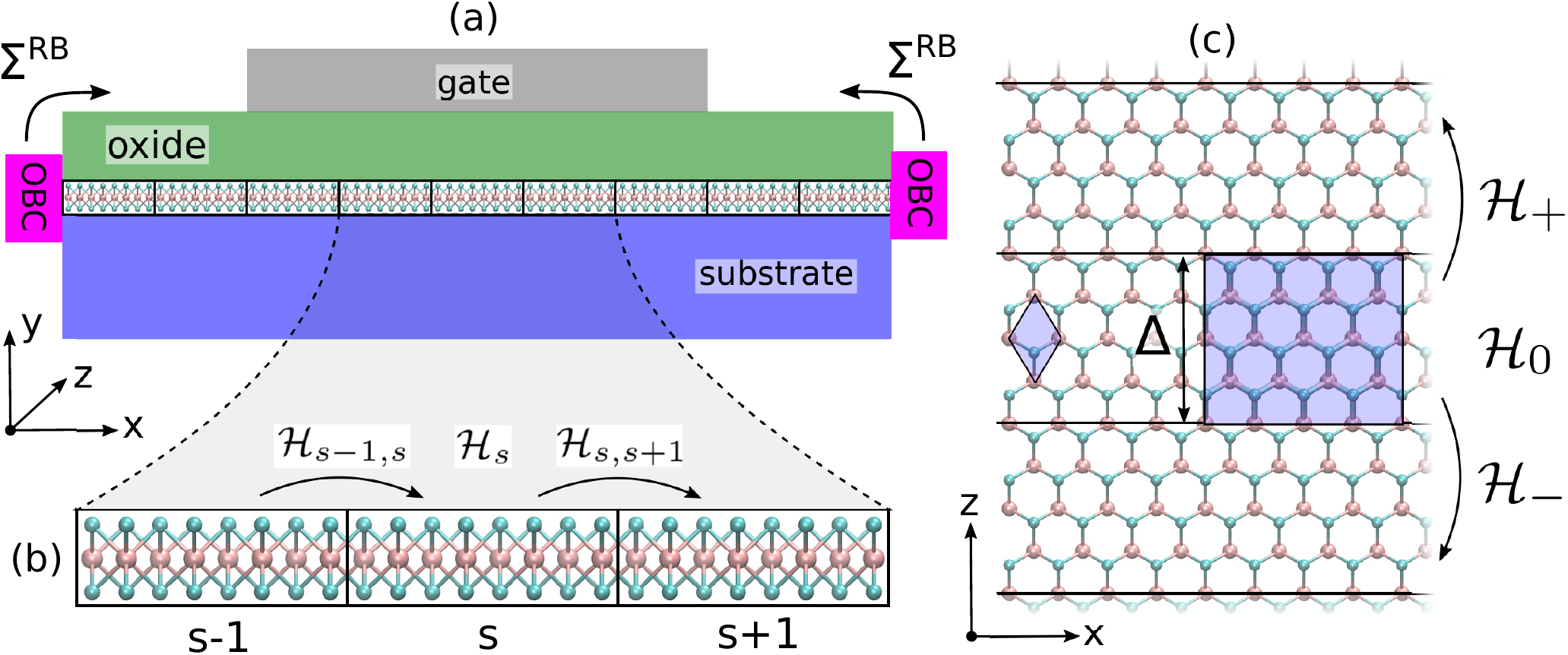}
 \caption{Schematic of a 2-D single-gate transistor structure for quantum transport made of a MoS\tsc{2} monolayer in the channel region. (a) The considered device is composed of 9 orthorhombic unit cells in sequence along the transport direction. The open boundary conditions on either side manifest themselves in terms of boundary self-energies $\mathbf{\Sigma}^{RB}$. The transport direction is along the x-axis, the y-axis is the direction of confinement, and the z-axis is assumed periodic. (b) Zoom into the center 3 unit cells from (a), designated $s-1$, $s$ and $s+1$. Interactions are such that they exist only between next-neighbor cells, i.e. $\mathcal{H}_{s,s+i} \neq 0$ only for $-1 \leq i \leq 1$. (c) Top view of the channel region. The periodicity in z-direction is given by $\Delta$. The small hexagonal unit cell on the left is the primitive unit cell of monolayer MoS\tsc{2}, whereas the larger orthorhombic unit cell on the right, is the one used as the building block for the device in (a). The Hamiltonian  matrix $\mathcal{H}_{0}$ describes the self-interactions of the central slab, $\mathcal{H}_{+}$($\mathcal{H}_{-}$) the interactions of the central slab with the one above(below) it.}\label{sec:negf:fig1}
\end{figure}

Here $\hbar$ is Planck's reduced constant and $e$ the elementary charge. The calculation can be simplified by grouping atoms together into orthorhombic unit cells arranged sequentially along the transport direction such that interactions exist between next neighbor cells only. An example for a 2-D device made of a MoS\tsc{2} monolayer structure is depicted in Fig. \ref{sec:negf:fig1}.

In case of ballistic transport, Eq. \eqref{sec:negf:eq10} can be written in the well-known Landauer-B{\"u}ttiker formalism \cite{Landauer_1989,Buttiker_1986}

\begin{equation}\label{sec:negf:current}
 I_d = - \frac{e}{\hbar} \sum_{k_z} \int \frac{dE}{2\pi} T(E,k_z) \Big ( f_L(E,E_{F,L} - f_R(E,E_{F,R} \Big ).
\end{equation}

In Eq. \eqref{sec:negf:current},

\begin{equation}\label{sec:negf:transm}
 T(E,k_z) = tr( \mathbf{G}_{1N}^R \Gamma_{NN} \mathbf{G}_{N1}^A \Gamma_{11} )
\end{equation}

is the energy- and momentum dependent transmission function through the considered device, $\Gamma_{11}$ ($\Gamma_{NN}$) the broadening function of the left (right) contact and $f_L(E,E_{F,L})$ ($f_R(E,E_{F,R})$) the corresponding Fermi distribution function at energy $E$ and with the Fermi level $E_{F,L}$ ($E_{F,R}$). \\

The momentum-dependent Hamiltonian matrix for the full device includes three components (blocks):

\begin{equation}\label{sec:negf:eq11}
 \mathcal{H}(k_z) = \mathcal{H}_0 + \mathcal{H}_{+} e^{ik_z\mathbf{\Delta}} + \mathcal{H}_{-} e^{-ik_z\mathbf{\Delta}},
\end{equation}

all of them being defined in Fig. \ref{sec:negf:fig1}. For quantum transport from first-principles, the Hamiltonian matrices $\mathcal{H}_{0}$, $\mathcal{H}_{+}$, and $\mathcal{H}_{-}$ must be generated using DFT. Because this is computationally very demanding, only small unit cells can be simulated. They must then be upscaled to match the dimensions of a full device.

\subsection[Toolchain]{Description of the Toolchain}
\label{sec:toolchain}

\begin{figure}
 \centering
 \includegraphics[width=10cm]{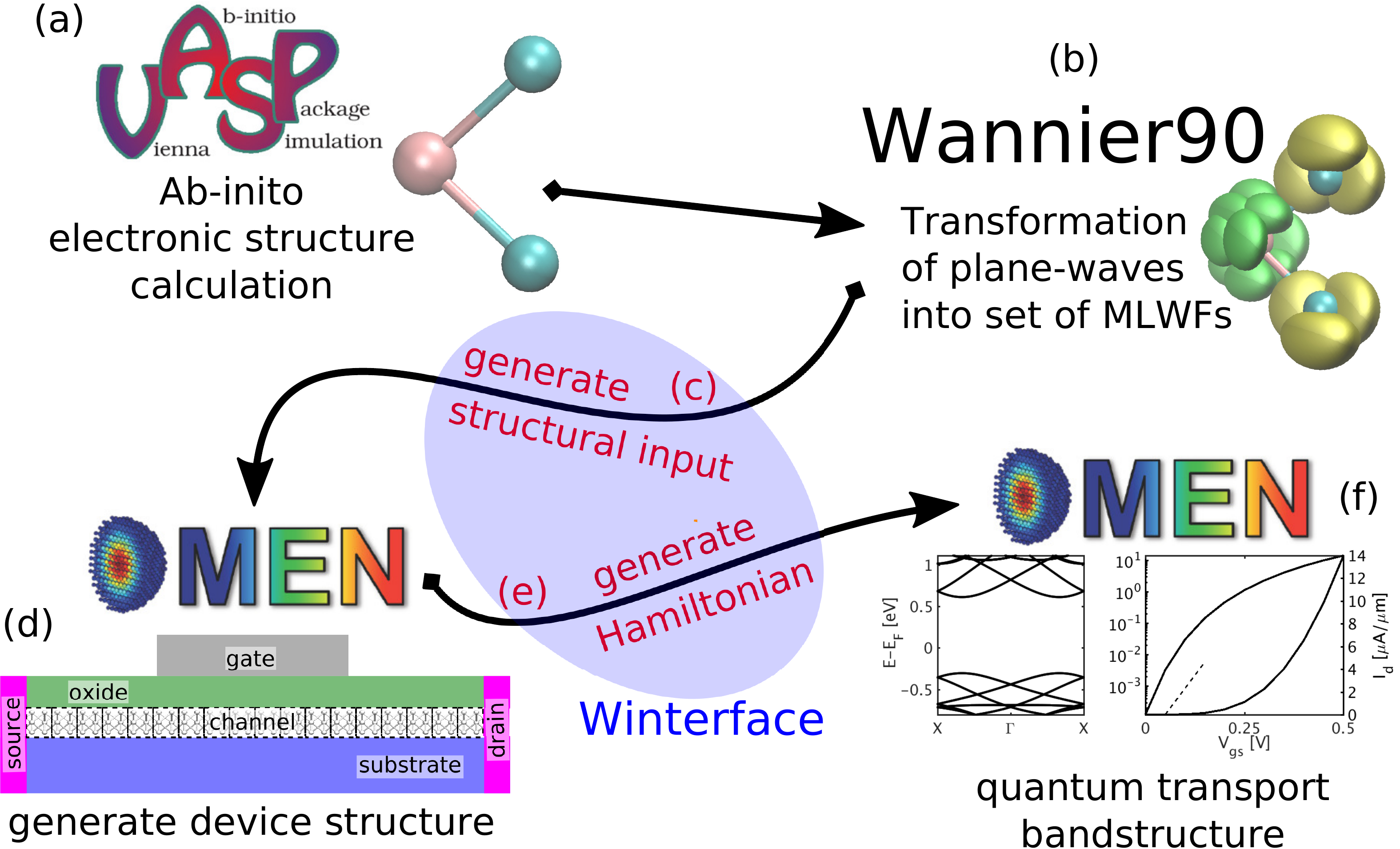}
 \caption{Developed toolchain to perform quantum transport simulations from first-principles. The interfacing parts (c,e) are in red. They represent the core of an open source package called ''Winterface'' (a) The ground state energy of the considered system is first calculated with DFT using a plane-wave representation of the wavefunctions. (b) The resulting Hamiltonian is then transformed into a tight-binding like basis with the help of MLWF and the Wannier90 code. (c) The MLWF data is analyzed and reorganized to generate structural inputs that are passed to OMEN. (d) This input is used by OMEN to create a device structure according to user specifications. (e) The Hamiltonian matrices corresponding to the channel region of this device are produced and transferred to OMEN. (f) Device simulations from first-principles are finally possible within OMEN.}\label{sec:toolchain:fig1}
\end{figure}

In this work, an infrastructure was implemented to generate tight-binding like Hamiltonian matrices for use in device simulations. Its main task is to interface a quantum transport solver, here OMEN \cite{OMEN_paper}, with the combination of a density functional theory (DFT) package, here VASP \cite{VASP_package}, and the Wannier90 code \cite{wannier90_paper}. The created infrastructure called Winterface is therefore part of a toolchain that is visualized in Fig. \ref{sec:toolchain:fig1} and described in the following paragraphs. \\

\emph{OMEN} is a nanodevice simulator capable of modeling the "current vs. voltage" characteristics of up to tens of thousands of atoms in a full-band framework and at the atomic scale. It was originally designed to rely on the semi-empirical $sp^3d^5s^*$ tight-binding model \cite{TB_paper}, but has since been modified to accept structural and Hamiltonian data supplied externally \cite{Sascha_paper}. This update will allow us to use the quantum transport algorithms implemented in OMEN with Hamiltonian matrices constructed from first-principles. Any other QT package, e.g. NanoTCAD ViDES \cite{vides} or TB\_Sim \cite{tbsim} would work as well. \\

\emph{VASP} is a code that provides the quantum mechanical ground state of atomic systems from first-principles, within the DFT approximation. Such calculations can be computationally very expensive and are typically limited to a small number of atoms (up to about a thousand). In VASP the Hamiltonian is represented in a plane-wave basis. Using its results as the basis for quantum transport simulations from first-principles requires a switch from plane-waves to MLWF and an additional upscaling technique. Note that instead of VASP any other plane-wave DFT package providing an interface to Wannier90 can be used, e.g. Quantum Espresso \cite{QE_package} or Abinit \cite{abinit_package}, as the work presented here depends only on Wannier90's outputs. \\

\emph{Wannier90} is a tool to efficiently transform the plane-wave representation used in DFT packages into a set of maximally localized Wannier functions based on a unitary transform. It has been designed in such a way that it only requires a couple of input files produced by DFT codes, but is otherwise completely independent of the specificities employed to derive them. The first file, called the \emph{MMN} file, contains the overlaps of the periodic parts $u_{m\mathbf{k}}(\mathbf{r})$ of the Bloch states sampled on a grid in the Brillouin zone

\begin{equation}\label{sec:toolchain:eq1}
 M_{mn}^{(\mathbf{k},\mathbf{b})} = \langle u_{m\mathbf{k}} \rvert u_{n\mathbf{k+b}} \rangle.
\end{equation}

The \textbf{k} points correspond to the k-grid used in the DFT calculation and the \textbf{b} vectors point to next neighbors. One of the great strengths of Wannier90 is that it can operate on a subset of the total number of bands used in DFT. The band indices m and n are thus elements of an index set specified by the user. In practice, we typically exclude low-lying core states and unoccupied states far above the Fermi energy. For quantum transport, we ideally include only states around the band gap. An example of this process is provided in Section \ref{sec:interface}. \\
The second file, called the \emph{AMN} file, contains overlaps of the wave functions $\psi_{m\mathbf{k}}(\mathbf{r})$ with localized trial orbitals $g_n(\mathbf{r})$

\begin{equation}\label{sec:toolchain:eq2}
 A_{mn}^{(\mathbf{k})} = \langle \psi_{m\mathbf{k}} \rvert g_n \rangle.
\end{equation}

The $g_n(\mathbf{r})$ in Eq. \eqref{sec:toolchain:eq2} must be specified by the user and can be placed anywhere in the unit cell. It has been observed that placing them on atomic positions is typically a successful strategy. These overlaps serve as the initial guesses for the steepest ascent method implemented in Wannier90 to find the MLWF. The quality of the results depends greatly on sensible inputs. Wannierization can be a very time-consuming process as finding a proper selection of bands in Eq. \eqref{sec:toolchain:eq1} and a suitable configuration of trial orbitals in Eq. \eqref{sec:toolchain:eq2} is a difficult task. It should be noted that an automated initial guess generator was proposed in Ref. \cite{autoguess_paper}, but it has not yet been tested for specific application of quantum transport. \\

\emph{Winterface}, whose code is the main focus of this work, interfaces OMEN to outputs from Wannier90 in a two-stage process. First, the data produced by Wannier90 is analyzed to produce inputs specifically designed for quantum transport, as implemented in OMEN (step (c) in Fig. \ref{sec:toolchain:fig1}). Second, after OMEN generated a device structure on the basis of these inputs, the corresponding Hamiltonian matrices $\mathcal{H}_0$, $\mathcal{H}_{+}$, and $\mathcal{H}_{-}$, as introduced in Eq. \eqref{sec:negf:eq11}, are constructed (step (e) in Fig. \ref{sec:toolchain:fig1}). All steps involved in this procdess will be explained in detail in Sections \ref{sec:wbh} and \ref{chap:hctor}. The code is open source and available online \cite{winterface_git}. \\

\section{Lattices, Unit Cells, and Ideal Upscaling}
\label{chap:ideal}

\subsection{Atomic Lattices and Unit Cells}
\label{sec:lattice}

A periodic structure such as an atomic lattice can be described in terms of a unit cell, whose repetition according to the corresponding translational symmetry allows to recover all positions within the lattice. The main purpose of this Section is the manipulation of unit cells, such as the transition from one translational symmetry to another, as well as the adaptation of the representation of the underlying physics. It is therefore prudent to first introduce the notion of a unit cell and the conventions used therein. \\

\subsubsection{Unit Cell}
Any arrangement of atomic positions that, in combination with a translational symmetry, results in unique coordinates for all positions is a valid representation of an atomic lattice. A unit cell is defined by the following components:

\begin{itemize}
 \item A matrix $\mathbf{B} = [\mathbf{b}_1,...,\mathbf{b}_N] \in \mathbb{R}^{N \times N}$, where N is the dimension of space and the columns $\mathbf{b}_i$ describe the translational symmetry of the lattice.
 \item A matrix $\mathbf{A} = [\mathbf{p}_1^{(1)},\mathbf{p}_2^{(1)},...,\mathbf{p}_{N_a}^{(N_t)}] \in [0,1)^{3 \times N_a} \subset \mathbb{R}^{N \times N_a}$ containing the atomic positions as columns expressed in basis \textbf{B}. N\tsc{a} is the total number of atomic positions and N\tsc{t} the total number of atomic types, i.e. $\mathbf{p}_i^{(j)}$ is the i-th position in the unit cell and is occupied by an atom of type j.
 \item A list of strings \textbf{id} of size N\tsc{t} containing information about the atomic type for each position in \textbf{A}.
\end{itemize}

Neither the choice of \textbf{B} nor of \textbf{A} is unique. To ensure a clear representation of all possible unit cells, the matrices \textbf{B} and \textbf{A} are normalized so that all entries of \textbf{A} lie within the interval $[0,1)$. Therefore, all atomic positions are found within the parallelepiped spanned by the columns of \textbf{B}. Setting the origin and sorting the atomic positions along the atomic types and coordinates completely determine the contents of a unit cell given in the basis \textbf{B}. For our purposes, the most important attribute of a unit cell is its volume defined as $vol(\mathbf{B}) = |det(\mathbf{B})|$. \\

The conventions introduced above allow for a unified treatment of atomic lattices since any representation of a given lattice takes the form of a unit cube $[0,1)^N$ when viewed in its own basis. The description of an infinite lattice is completed by a grid of vectors $\mathbf{R} \in \mathds{Z}^N$ (in basis \textbf{B}), where $\mathbf{R} = \mathbf{0}$ is the home unit cell and all $\mathbf{R} \neq \mathbf{0}$ correspond to image cells. The explicit example of a monolayer structure of MoS\tsc{2} can be found in Appendix \ref{app:mos2_ex}.

\subsubsection{Basis Expansions}
The basis \textbf{B} can always be replaced by another basis $\tilde{\mathbf{B}}$. To ensure that the new basis describes the same lattice, it is best expressed as a linear expansion of the old one

\begin{equation}\label{sec:lattice:eq4}
 \tilde{\mathbf{b}}_i = \sum_{j} c_{ji} \cdot \mathbf{b}_j \quad \Leftrightarrow \quad \tilde{\mathbf{B}} = \mathbf{B} \cdot \mathbf{C}
\end{equation}

where $det(\mathbf{C}) \neq 0$, $c_{j,i} \in \mathds{Q}$ in general, but $c_{j,i} \in \mathds{Z}$ if the initial unit cell is primitive. The transformation of coordinates into the new basis $\tilde{\mathbf{B}}$ can be written as

\begin{equation}\label{sec:lattice:eq5}
 \mathbf{A}_{\tilde{B}} = \tilde{\mathbf{B}}^{-1} \cdot \mathbf{B} \cdot \mathbf{A}_{B} = (\mathbf{B} \cdot \mathbf{C})^{-1} \cdot \mathbf{B} \cdot \mathbf{A}_{B} = \mathbf{C}^{-1} \cdot \mathbf{A}_{B}.
\end{equation}

The new coordinates will not necessarily lie within the unit cube described by the new basis, but might be situated in an image cell. For expansions \textbf{C} with $|det(\mathbf{C})| > 1$, leading to an expansion of the unit cell, additional atomic positions will have to be found to match the increase in volume. It should be noted that without the convention of coordinates in the interval $[0,1)$, the choice of the positions to include in a unit cell described by a basis $\tilde{\mathbf{B}}$ with $vol(\tilde{\mathbf{B}}) > vol(\mathbf{B})$, as compared to the initial unit cell is arbitrary.

\subsection{Upscaling Technique for Ideal Cases}
\label{sec:upscaling}

\begin{figure}
 \centering
 \includegraphics[width=10cm]{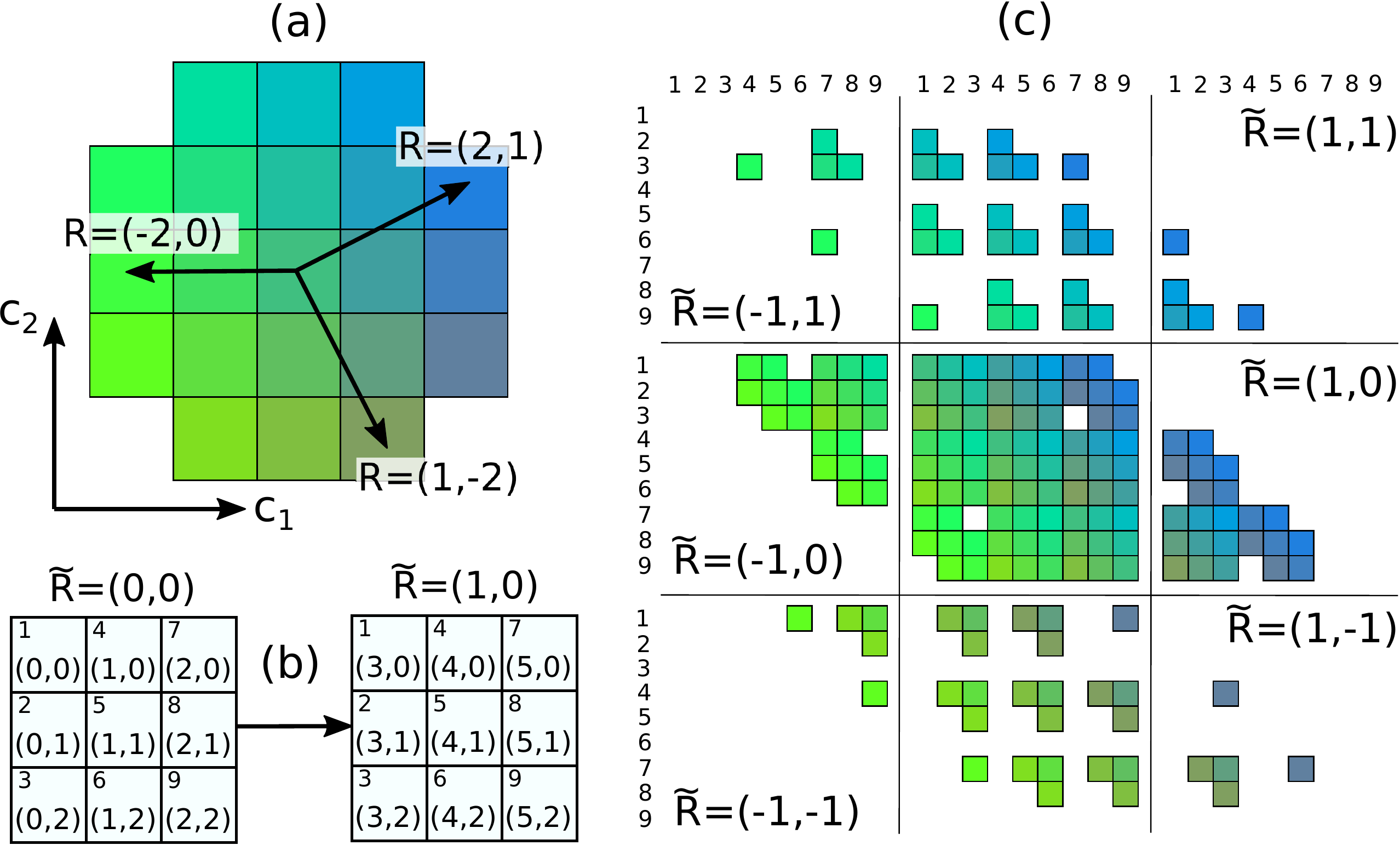}
 \caption{Upscaling technique of a set of $\mathcal{H}(\mathbf{R})$ describing interactions among unit cells in basis \textbf{B}, to a set of $\tilde{\mathcal{H}}(\tilde{\mathbf{R}})$ describing interactions among super cells in basis $\tilde{\mathbf{B}} = 3 \cdot \mathbf{B}$. (a) Schematic of interactions among a set of 21 primitive cells. Each tile represents a unit cell shifted according to \textbf{R} and the corresponding Hamiltonian matrix $\mathcal{H}(\mathbf{R})$. (b) Schematic of the home super cell and the first neighbor super cell at $\tilde{\mathbf{R}} = (1,0)$. Each super cell holds 9 primitive unit cells from (a) arranged in a specific order, as indicated by the index $1 \leq i \leq 9$. Inside each subcell, the mapping $\mathbf{R} = 3 \cdot \tilde{\mathbf{R}} + \mathbf{r}$ to the grid of primitive unit cells is given. (c) Schematic of the set of $\tilde{\mathcal{H}}(\tilde{\mathbf{R}})$ describing interactions in the super cell. The interaction range for the super cell is next neighbor only leading to 9 $\tilde{\mathcal{H}}(\tilde{\mathbf{R}})$ as compared to 21 $\mathcal{H}(\mathbf{R})$ for the primitive cell. Each tile represents a whole matrix $\mathcal{H}(\mathbf{R})$ from (a) determined by the relative positioning of primitive cells given in Eq. \eqref{sec:upscaling:eq1}. The blocks of $\tilde{\mathcal{H}}(\tilde{\mathbf{R}})$ are found by scanning over all of the primitive cells in each super cell, where a high degree of sparsity is achieved for $\tilde{\mathbf{R}} \neq 0$.}\label{sec:upscaling:fig1}
\end{figure}

To illustrate the process of upscaling, consider an expansion $\mathbf{C} = 3$ on the primitive unit cell of a two-dimensional lattice. The task at hand is now to construct the set of Hamiltonian matrices $\tilde{\mathcal{H}}(\tilde{\mathbf{R}})$ describing the physics in terms of the super cell, from the initial set of $\mathcal{H}(\mathbf{R})$ corresponding to the primitive cell.  To this end, the coordinate $\mathbf{r}_i^{(m)} \in \{0,1,2\}^2$ with $1 \leq i \leq 9$ identifying each primitive cell inside the super cell at $\tilde{\mathbf{R}}_m$ is introduced. Since $\mathbf{R} =  \mathbf{C} \cdot \tilde{\mathbf{R}} = 3 \cdot \tilde{\mathbf{R}}$, the i-th primitive cell inside the m-th super cell can be mapped onto the grid of primitive unit cells according to $\mathbf{R}_i^{(m)} = 3 \cdot \tilde{\mathbf{R}}_m + \mathbf{r}_i^{(m)}$. Thus, the relative positioning of two primitive cells is

\begin{equation}\label{sec:upscaling:eq1}
 \delta \mathbf{R}_{ij}^{(m,n)} = 3 \cdot (\tilde{\mathbf{R}}_m - \tilde{\mathbf{R}}_n) + (\mathbf{r}_i^{(m)} - \mathbf{r}_j^{(n)}),
\end{equation}

which determines the Hamiltonian matrix $\mathcal{H}(\delta \mathbf{R}_{ij}^{(m,n)})$ describing the interactions between two primitive cells. Each $\tilde{\mathcal{H}}(\tilde{\mathbf{R}}) = \tilde{\mathcal{H}}(\tilde{\mathbf{R}}_m - \tilde{\mathbf{R}}_n)$ can be constructed by scanning across all primitive cells forming each super cell. Therefore, the Hamiltonian matrices describing interactions among super cells consist of 81 blocks of Hamiltonian matrices describing interactions among the primitive cells contained in them. Some of the data is repeated since $\delta \mathbf{R}_{ij}^{(m,n)}$ in Eq. \eqref{sec:upscaling:eq1} can have the same value for different combinations of m, n, i, and j. A graphical illustration is presented in Fig. \ref{sec:upscaling:fig1}.

The shape of the matrices in Fig. \ref{sec:upscaling:fig1} is determined by the order in which the primitive cells are arranged in the super cell. For quantum transport, a block-tridiagonal shape is needed, but any arrangement is valid since the corresponding $\tilde{\mathcal{H}}(\tilde{\mathbf{R}})$'s are equivalent up to a unitary transform. \\

It should be noted that the case presented here represents the simplest possible scenario and serves only to introduce the notion of upscaling. In general the process is more involved, for example, the concept of subcells as coordinates inside a super cell may not work when scaling to super cells belonging to a different symmetry group, since some of the primitive cells may be only partially located inside the super cell. A general formalism will be introduced in the next two Sections.
\section{Hamiltonian Data in Terms of Bonds}
\label{sec:wbh}

The main purpose of this Section is the reformulation of the raw Wannier Hamiltonian data as produced by Wannier90 into a representation better suited for the task of upscaling. The reasons are mostly twofold. Firstly, Hamiltonians in Wannier representation have a number of degrees of freedom, which must be handled carefully. Secondly, Wannier centers often follow a distinct distribution in space, allowing for a matching of multiple Wannier centers to a representative position. Therefore, a reformulation of the Hamiltonian data in terms of interactions between such positions, i.e. along bonds, is better suited for the task of upscaling than the initial Wannier Hamiltonian data. Two output files from Wannier90 are of special interest for this purpose. The main output file, called \emph{wout}, contains the Wannier centers and spreads of each Wannier function. The second file, called \emph{hrdat}, contains the Hamiltonian in a Wannier representation in the form of a list of matrix elements $\mathcal{H}_{ij}(\mathbf{R}) = \braket{w_{i\mathbf{0}}|\mathcal{H}|w_{j\mathbf{R}}}$. This is all the information needed to generate a Hamiltonian in terms of bonds. In a first step, suitable representative positions must be identified, as explained the next Section. \\

\subsection{Matching Wannier Centers to Atomic Positions}
\begin{figure}
 \centering
 \includegraphics[width=10cm]{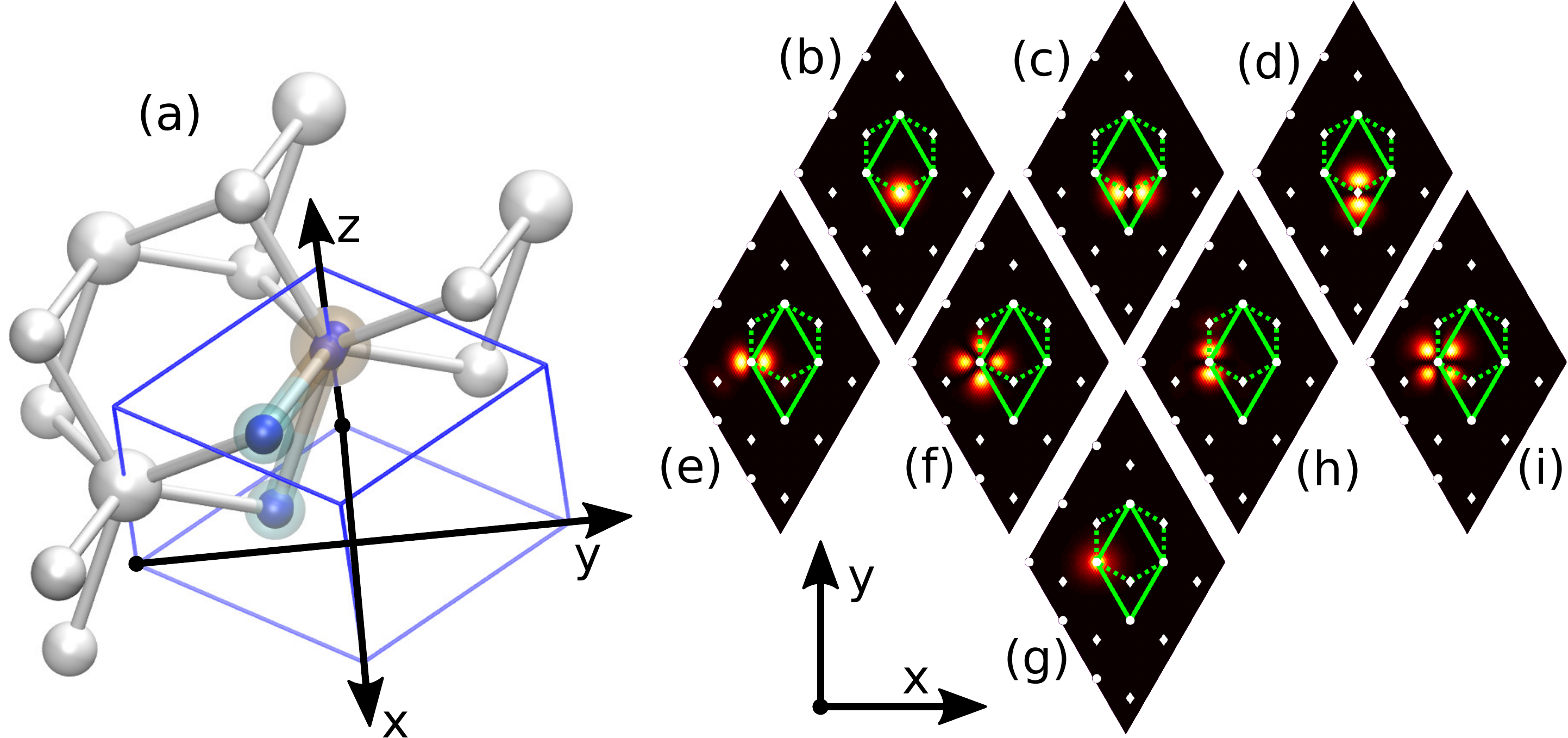}
 \caption{Wannierization of a MoS\tsc{2} monolayer structure. (a) Primitive unit cell holding one molybdenum and two sulfur atoms. The Wannier centers, represented by the small blue marbles are tightly clustered on the atomic positions. 5 d-like orbitals are found on molybdenum and 3 p-like orbitals on each sulfur atom. (b-d) Wannier centers on sulfur. The green parallelepiped represents the primitive unit cell of MoS\tsc{2} seen from above, the dashed line shows one honeycomb of the same volume. (e-g) Wannier centers on molybdenum.}\label{sec:wbh:fig2}
\end{figure}

\begin{figure}
 \centering
 \includegraphics[width=10cm]{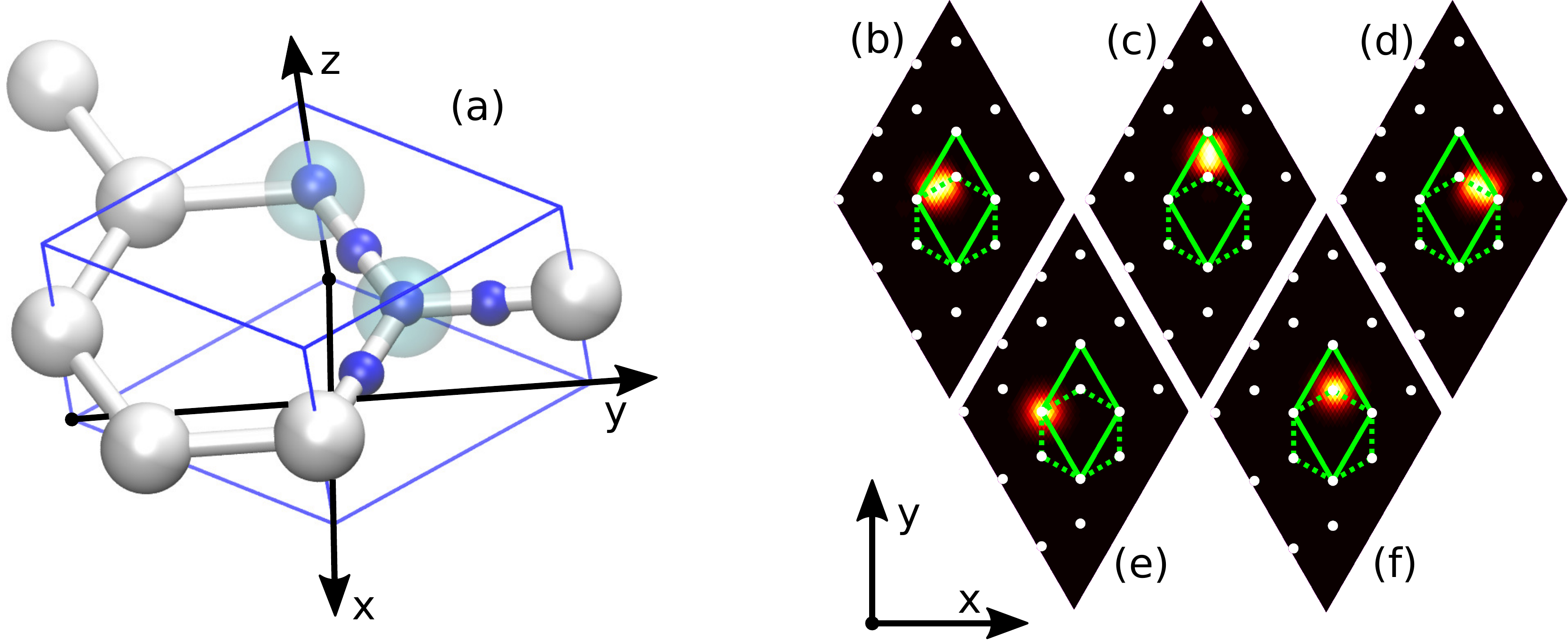}
 \caption{Wannierization of graphene. (a) Primitive unit cell holding two carbon atoms. The Wannier centers are located on both atoms, but also on the bond centers. (b-d) Wannier centers on bond centers. The green parallelepiped represents the primitive unit cell of graphene seen from above, the dashed line shows one honeycomb of the same volume. (e,f) Wannier centers on the carbon atoms.}\label{sec:wbh:fig3}
\end{figure}

DFT simulations are carried out with a description of the material in terms of atomic positions. If possible it would be convenient to keep the same positions when generating interactions along bonds. Since Wannier90 returns a set of Wannier centers, the first step consists of establishing whether these centers can be assigned to the existing atomic positions. In many cases the Wannier centers are clustered around the atoms whose orbitals they represent, as shown for a MoS\tsc{2} monolayer structure in Fig. \ref{sec:wbh:fig2}. In other cases, such as the graphene layer presented in Fig. \ref{sec:wbh:fig3}, some of them lie on bond centers.

The upscaling code therefore offers a diverse range of options to take into account different kinds of Wannier center distributions. In cases such as graphene in Fig. \ref{sec:wbh:fig3}, fictitious atoms representing the bond centers can be introduced. Additionally it is also possible to specify matching positions manually or use the Wannier centers themselves. In any case, Wannier centers are matched to their closest position using a metric taking into account the periodicity of the lattice (see Appendix \ref{app:metric}). Subsequently, they can be viewed as orbitals describing the interactions between the (fictitious) atoms they are matched to. Matching schemes using criteria other than spacial proximity are not supported. \\

This matching process requires basic knowledge of the chemistry of the structure under investigation and furthermore depends on the preference of the user. It should be realized that for some operations such as the calculation of the bandstructure of a given unit cell, the positions of the Wannier centers do not matter, only the periodicity of the structure. The situation is different if the charge distribution should be accounted for, as in device simulations where Poisson's equation must be solved self-consistently with Schr{\"o}dinger's equation. If point charges are used, their exact location might influence the shape of the resulting electrostatic potential. The charge distribution in terms of Wannier centers can be followed more closely to address this issue, but at the expense of a large computational burden. \\

\subsection{Generating Interaction Data Along Atomic Bonds}
First of all, it is important to realize that Wannier functions are considered \emph{native} to the \emph{home cell} at $\mathbf{R} = 0$ no matter where the 
corresponding Wannier centers are located. Usually the Wannier centers are tightly clustered inside or around the home cell, but in some cases they can be more spread out such that multiple centers are found in other cells. In fact, there is no reference distribution and Wannier functions can be replaced by any of their images, but since the relative positioning must be reflected in the matrix elements of operators in Wannier representation, appropriate adaptations are necessary. Assuming a redistribution of Wannier functions $w_n$ native to $\mathbf{R} = 0$, to images $\tilde{w}_n$ of the same Wannier functions situated at $\mathbf{R} = \delta\mathbf{R}_n$, the following must hold for the Hamiltonian matrix elements:

\begin{equation}\label{sec:wbh:eq1}
 \braket{w_{n\mathbf{0}} | \mathcal{H} | w_{m\mathbf{R}}} = \braket{\tilde{w}_{n-\delta\mathbf{R}_n} | \mathcal{H} | \tilde{w}_{m\mathbf{R}-\delta\mathbf{R}_m}}.
\end{equation}

Both representations are equally valid and the Hamiltonian operators $\mathcal{H}(\mathbf{R})$ in Wannier representation are composed of the same matrix elements, except that they are permuted relative to each other as per Eq. \eqref{sec:wbh:eq1}. It is straightforward to show that the resulting Hamiltonian operators in k-space given by

\begin{equation}\label{sec:wbh:eq3}
 \mathcal{H}(\mathbf{k}) = \sum_{\mathbf{R}} \mathcal{H}(\mathbf{R}) e^{i\mathbf{k}\mathbf{R}}
\end{equation}

are equivalent up to a unitary transform $V_{mn}(\mathbf{k}) = \delta_{mn} e^{i\mathbf{k}\delta\mathbf{R}_m}$ such that $\mathcal{H}(\mathbf{k}) = V(\mathbf{k}) \cdot \tilde{\mathcal{H}}(\mathbf{k}) \cdot V(\mathbf{k})^\dagger$. Therefore, in terms of bandstructure reproduction the two representations are equivalent. \\

For the task of finding the correct interaction matrices between two (fictitious) atomic positions $\mathbf{p}_i$ and $\mathbf{p}_j$ with a set of Wannier functions $I_i$ and $I_j$ matched to them respectively, the relative positioning of the corresponding Wannier centers must be accounted for. Additionally, the closest images for the Wannier functions matched to a position $\mathbf{p}_i$ may not always be found at the same $\mathbf{R}$ vector. This situation may arise due to the choice of the positions $\mathbf{p}_i$ themselves, or due to the initial distribution of Wannier functions native to the home cell. Even for  the case where Wannier centers converge on to the atomic positions used in DFT, a scattering of Wannier centers among multiple images of the atomic positions can sometimes be observed. This appears to be a frequent occurrence in 2-D structures especially and can be difficult to avoid. The example of the treatment of two Wannierizations of the same MoS\tsc{2} monolayer, where the initial distribution of Wannier functions is different, is provided in Appendix \ref{app:mos2_relwannier}. \\

A bond between two positions $\tilde{\mathbf{p}}_i$ and $\tilde{\mathbf{p}}_j$ inside a unit cell is defined as the vector that points from one position to the other, i.e. $\tilde{\mathbf{b}}_{ij} = \tilde{\mathbf{p}}_j - \tilde{\mathbf{p}}_i$. All other bonds in the lattice can be recovered by allowing the placement of positions in image unit cells:

\begin{equation}\label{sec:wbh:eq9}
 \mathbf{b}_{ij}(\mathbf{R}) = (\tilde{\mathbf{p}}_j + \mathbf{R}_j) - (\tilde{\mathbf{p}}_i + \mathbf{R}_i) = \tilde{\mathbf{b}}_{ij} + \mathbf{R},
\end{equation}

where $\tilde{\mathbf{b}}_{ij}$ is called the \emph{principal bond}. This decomposition of a bond in a principal part and a relative placement among image unit cells preserves the basic property of Hamiltonians in a Wannier representation. The interactions along a bond $\mathbf{b}_{ij}(\mathbf{R})$ are given by a small Hamiltonian matrix $h^{(i,j)}(\mathbf{R})$, analogous to the Hamiltonian matrices $\mathcal{H}(\mathbf{R})$ describing interactions among unit cells. To find the matrices $h^{(i,j)}(\mathbf{R})$, we will first have a look at the principal bonds only. The initial situation consists of:

\begin{itemize}
 \item A list of (fictitious) positions $\mathbf{p}_i \in [0,1)^N, i \in \{1,...,N_a\}$ inside the unit cell.
 \item An index vector $I_i$ for each atomic position $\mathbf{p}_i$ holding the matched Wannier indices.
 \item A list of Wannier centers $\mathbf{w}_m^{(i)} \in \mathds{R}^N, m \in I_i$ matched to \textbf{p}\tsc{i}.
 \item A set of Hamiltonian matrices $\mathcal{H}(\mathbf{R})$ in a Wannier representation.
\end{itemize}

The challenge in finding the correct $h^{(i,j)}$ for each principal bond $\tilde{\mathbf{b}}_{ij}$, is the treatment for the case where Wannier centers are spread out among images of the positions defining the bond. To detect the placement of Wannier centers with respect to the positions inside the unit cell, a list of 'Wannier bonds' $\{\tilde{\mathbf{b}}_m^{(i)}\}, \ m \in I_i$, i.e. vectors pointing from an atomic position $\tilde{\mathbf{p}}_i$ to the closest image $\tilde{\mathbf{w}}_m^{(i)}$ of a matched Wannier center with index m, is generated. \\

The shift in \textbf{R} vectors of a Wannier center relative to the closest image matched to a position $\tilde{\mathbf{p}}_i$ can be derived from the corresponding 'Wannier bond':

\begin{equation}\label{sec:wbh:eq10}
 \mathbf{R}_m^{(i)} = \tilde{\mathbf{w}}_m^{(i)} - \mathbf{w}_m^{(i)} = \tilde{\mathbf{p}}_i + \tilde{\mathbf{b}}_m^{(i)} - \mathbf{w}_m^{(i)}.
\end{equation}

Thus, the relative position between two Wannier centers $\mathbf{w}_m^{(i)}$ and $\mathbf{w}_n^{(j)}$ matched to the atomic positions $\tilde{\mathbf{p}}_i$ and $\tilde{\mathbf{p}}_j$ respectively is

\begin{equation}\label{sec:wbh:eq11}
 \delta\mathbf{R}_{mn}^{(i,j)} = \mathbf{R}_n^{(j)} - \mathbf{R}_m^{(i)} = (\tilde{\mathbf{p}}_j + \tilde{\mathbf{b}}_n^{(j)} - \mathbf{w}_n^{(j)}) - (\tilde{\mathbf{p}}_i + \tilde{\mathbf{b}}_m^{(i)} - \mathbf{w}_m^{(i)}).
\end{equation}

\begin{figure}
 \centering
 \includegraphics[width=8cm]{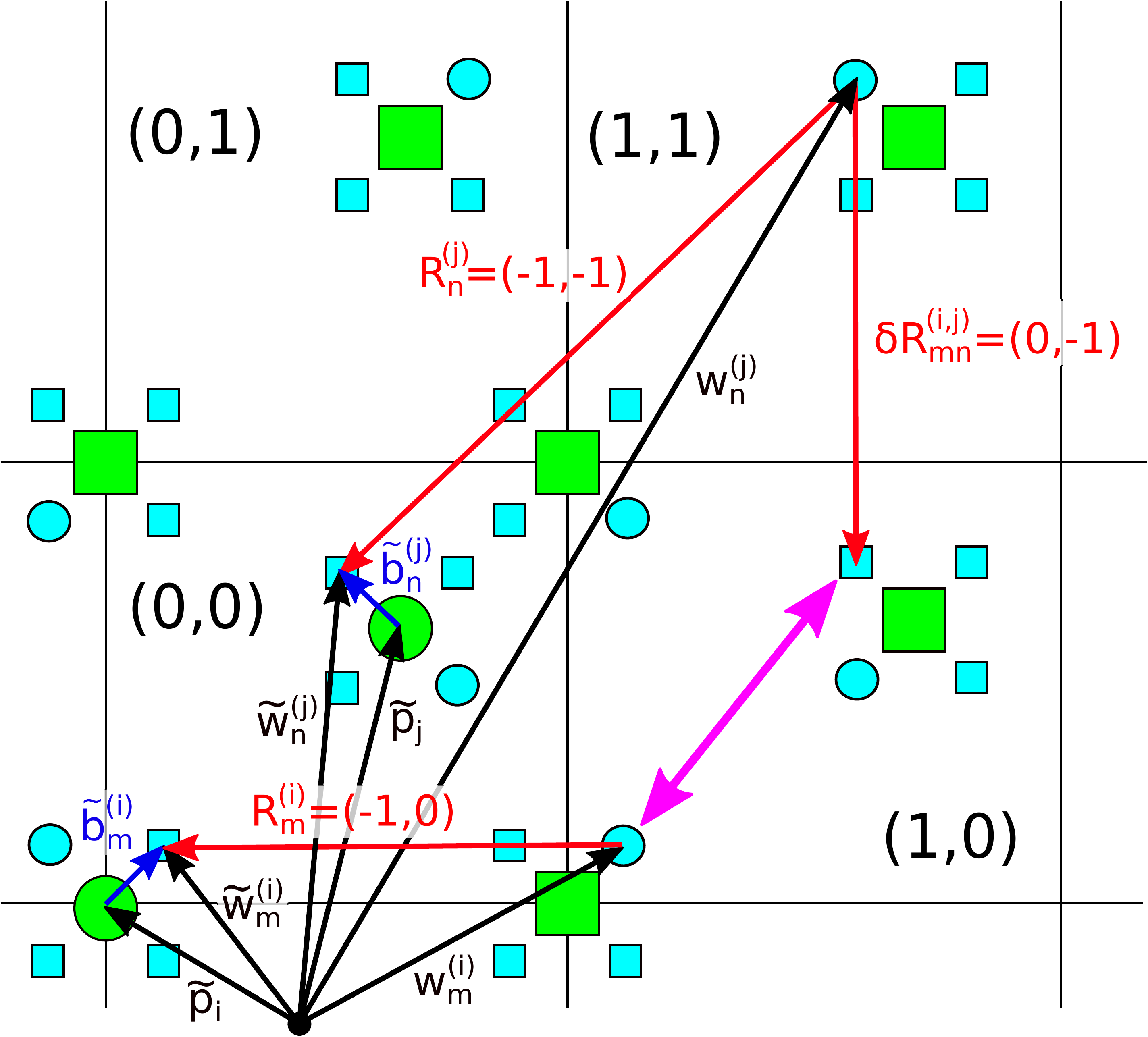}
 \caption{Schematic illustrating the detection of the relative position between two Wannier centers $\mathbf{w}_m^{(i)}$ and $\mathbf{w}_n^{(j)}$ matched to the atomic positions $\tilde{\mathbf{p}}_i$ and $\tilde{\mathbf{p}}_j$, respectively. The green circles (squares) are atomic positions (images) and the blue circles (squares) are Wannier centers (images). Even though most of the Wannier centers converged onto images of the atomic positions, they all belong to the home cell at $\mathbf{R} = (0,0)$. The matrix element needed for the connection between the images of these Wannier centers $\tilde{\mathbf{w}}_m^{(i)}$ and $\tilde{\mathbf{w}}_m^{(i)}$ closest to the atomic positions, is therefore found in $\mathcal{H}(\mathbf{R} \neq 0)$. The shift $\mathbf{R}_m^{(i)}$($\mathbf{R}_n^{(j)}$) for the Wannier center $\mathbf{w}_m^{(i)}$($\mathbf{w}_n^{(j)}$) to coincide with the image $\tilde{\mathbf{w}}_m^{(i)}$($\tilde{\mathbf{w}}_n^{(j)}$) is given by Eq. \eqref{sec:wbh:eq10} in terms of the 'Wannier bonds' $\tilde{\mathbf{b}}_m^{(i)}$($\tilde{\mathbf{b}}_n^{(j)}$). After shifting the Wannier center $\mathbf{w}_n^{(j)}$ by the relative placement $\delta \mathbf{R}_{mn}^{(i,j)} = \mathbf{R}_n^{(j)} - \mathbf{R}_m^{(i)}$, it finds itself in the same unit cell at $\mathbf{R} = (1,0)$ as $\mathbf{w}_m^{(i)}$. The correct matrix element is therefore located in $\mathcal{H}(\delta \mathbf{R}_{mn}^{(i,j)})$ as indicated by the large pink arrow.}\label{sec:wbh:fig7}
\end{figure}

Therefore, the matrix element $h_{mn}^{(i,j)}$ for the interaction along the principal bond $\tilde{\mathbf{b}}_{ij}$ is found in the matrix $\mathcal{H}(\delta\mathbf{R}_{mn}^{(i,j)})$ as:

\begin{equation}\label{sec:wbh:eq12}
 h_{mn}^{(i,j)} = \mathcal{H}_{I_i(m),I_j(n)} \big (\delta\mathbf{R}_{mn}^{(i,j)} \big ),
\end{equation}

where $\mathcal{H}_{I_i(m),I_j(n)} \big (\delta\mathbf{R}_{mn}^{(i,j)} \big )$ corresponds to the data produced by Wannier90 as introduced in Eq. \eqref{sec:mlwf:eq4}. A graphical representation of this process is presented in Fig. \ref{sec:wbh:fig7}.

The extension to general bonds $\mathbf{b}_{ij}(\mathbf{R})$ is done by shifting the position $\mathbf{p}_j = \tilde{\mathbf{p}}_j + \mathbf{R}$, which in turn shifts the attached Wannier center $\mathbf{w}_n^{(j)}$ in Eq. \eqref{sec:wbh:eq11} by the same $\textbf{R}$. This then translates into Eq. \eqref{sec:wbh:eq12} as

\begin{equation}\label{sec:wbh:eq13}
 h_{mn}^{(i,j)}(\mathbf{R}) = \mathcal{H}_{I_i(m),I_j(n)} \big (\delta\mathbf{R}_{mn}^{(i,j)} + \mathbf{R} \big ).
\end{equation}

The full data set of interactions along bonds $\{ \mathbf{b}_{ij}(\mathbf{R}) \}$ with $1 \leq i,j \leq N_a$ and \textbf{R} such that all significant elements of $\mathcal{H}(\mathbf{R})$ are included, can be constructed by placing $\mathbf{p}_i$ in the home cell and $\mathbf{p}_j$ in image cells as indicated by \textbf{R} for all pairwise combinations of i and j. For each such pairing, Eq. \eqref{sec:wbh:eq13} must be employed for all pairwise combinations of the matched Wannier functions as specified by $I_i$ and $I_j$ respectively to find all the elements of the interaction matrix $h(\mathbf{b}_{ij}(\mathbf{R}))$ associated with the bond in question. \\

Following the scheme outlined above, we note that $N_a \cdot N_a$ principal bonds exist, one for each pair of atomic positions inside the unit cell, including vanishing bonds where both positions are the same. The latter represent the self-interactions of the atoms with themselves. Additionally, each principal bond has a list of \textbf{R} vectors attached to it, forming the set of general bonds $\{ \mathbf{b}_{ij}(\mathbf{R}) \}$. The total number of bonds is then $\leq N_a \cdot N_a \cdot N_R$, because some of the Hamiltonian data describing long-range interactions can be neglected. \\

Since the main task of constructing Hamiltonian data in terms of interactions along bonds is querying for interactions with bond vectors, an efficient arrangement is required. The main building block is the unit cell containing the (fictitious) atomic positions generated during the matching process. Because the interactions between these positions are expressed in terms of mutually exclusive sets of Wannier functions, all atomic types are considered unique even if the chemical origin was equivalent, such as the two sulfur atoms in MoS\tsc{2}. As a direct consequence, this unit cell is always primitive. The interaction data in terms of bonds is arranged as follows:

\begin{itemize}
 \item The top level consists of a sorted list of index pairs $\{(i,j)\}$ with $1 \leq i,j \leq N_a$, indicating which of the positions $\{ \tilde{\mathbf{p}}_i \}$ form the principal bond vectors.
 \item Each element of $\{(i,j)\}$ has a sorted list of pairs $\{ \textbf{R}, h^{(i,j)}(\mathbf{R}) \}$ attached to it. The \textbf{R} indicates which image cell the position $\mathbf{p}_j = \tilde{\mathbf{p}}_j + \mathbf{R}$ is located in. The interactions along the bond $\mathbf{b}_{ij}(\mathbf{R})$ are then given by $h^{(i,j)}(\mathbf{R})$.
\end{itemize}

With the interaction data completely sorted, the querying for bonds given by a starting index $\mu$, an ending index $\nu$ and a bond vector \textbf{b}, is now a two stage process:

\begin{itemize}
 \item Search for $(\mu,\nu)$ in the list of $\{(i,j)\}$. If a matching entry is found, compute the corresponding principal bond $\tilde{\mathbf{b}}_{\mu\nu}$ and subtract it from the bond vector \textbf{b}.
 \item If $\mathbf{b} - \tilde{\mathbf{b}}_{\mu\nu} \in \mathds{Z}^N$ up to a numeric tolerance, search among the list of \textbf{R} vectors. If a matching entry is found, return the corresponding interaction matrix $h^{(\mu,\nu)}(\mathbf{R})$.
\end{itemize}

Note that $\mathbf{b} - \tilde{\mathbf{b}}_{\mu\nu}$ can be assigned to an $\mathbf{R} \in \mathds{Z}^N$ up to a tolerance of 1/2. In the extreme case, this allows for a spatial mismatch of an entire unit cell. In terms of accurate physical modeling, such an extreme warping of bonds is unlikely to produce sensible results. Especially for the approximate matching algorithms discussed in Section \ref{sec:approx}, the tolerance must be chosen such that acceptable regions of mismatch for different bonds do not overlap each other. Finally, since the querying algorithm above consists of sequential searches in two sorted ranges, the running time is $\mathcal{O}(log(N_a^2)) + \mathcal{O}(log(N_R)) = \mathcal{O}(2 \cdot log(N_a \cdot N_R))$. 

\section{Generating Hamiltonian Matrices}
\label{chap:hctor}

\subsection[Exact Upscaling]{Exact Upscaling Technique and Bandstructure Calculations}
\label{sec:hctor}

The preliminary work done in Section \ref{sec:wbh} eliminated various pitfalls of the raw Wannier Hamiltonians produced by Wannier90. For this purpose, the original data set was transformed into a different representation in terms of interactions along bonds. The generation of Hamiltonian matrices representing structures whose physics are encapsulated within the initial Wannier Hamiltonian is now a straightforward process.

Given two sets of atomic positions $\{\mathbf{p}^{(i)}\}$ and $\{\mathbf{p}^{(j)}\}$, a scan through all pairwise combinations must be done, whilst querying for the corresponding bond and then copying the interaction blocks into a large matrix container $\mathcal{H}(\{\mathbf{p}^{(i)}\},\{\mathbf{p}^{(j)}\})$, as summarized in the table below.

\begin{center}
 \begin{tabular}{ c | c c c c c | c }
 & $\mathbf{p}_1^{(i)}$ & $\mathbf{p}_2^{(i)}$ & ... & $\mathbf{p}_{N_i-1}^{(i)}$ & $\mathbf{p}_{N_i}^{(i)}$ & \\
 \hline
 $\mathbf{p}_1^{(j)}$ & $h(\mathbf{b}_{1,1}^{(i,j)})$ & $h(\mathbf{b}_{1,2}^{(i,j)})$ & ... & $h(\mathbf{b}_{1,N_i-1}^{(i,j)})$ & $h(\mathbf{b}_{1,N_i}^{(i,j)})$ & \\
 $\mathbf{p}_2^{(j)}$ & $h(\mathbf{b}_{2,1}^{(i,j)})$ & $h(\mathbf{b}_{2,2}^{(i,j)})$ & ... & $h(\mathbf{b}_{2,N_i-1}^{(i,j)})$ & $h(\mathbf{b}_{2,N_i}^{(i,j)})$ & \\
 $\vdots$ & $\vdots$ & $\vdots$ & $\ddots$ & $\vdots$ & $\vdots$ & \\
 $\mathbf{p}_{N_j-1}^{(j)}$ & $h(\mathbf{b}_{N_j-1,1}^{(i,j)})$ & $h(\mathbf{b}_{N_j-1,2}^{(i,j)})$ & ... & $h(\mathbf{b}_{N_j-1,N_i-1}^{(i,j)})$ & $h(\mathbf{b}_{N_j-1,N_i}^{(i,j)})$ & \\
 $\mathbf{p}_{N_j}^{(j)}$ & $h(\mathbf{b}_{N_j,1}^{(i,j)})$ & $h(\mathbf{b}_{N_j,2}^{(i,j)})$ & ... & $h(\mathbf{b}_{N_j,N_i-1}^{(i,j)})$ & $h(\mathbf{b}_{N_j,N_i}^{(i,j)})$ & \\
 \hline
 & & & & & &
 \end{tabular}
\end{center}

From Eq. \eqref{sec:wbh:eq13} we have that

\begin{equation}\label{sec:hctor:eq1}
 \mathcal{H}(\{\mathbf{p}^{(i)}\},\{\mathbf{p}^{(j)}\}) = \mathcal{H}(\{\mathbf{p}^{(j)}\},\{\mathbf{p}^{(i)}\})^\dagger,
\end{equation}

\begin{equation}\label{sec:hctor:eq2}
  \mathcal{H}(\{\mathbf{p}^{(i)}+\mathbf{R}\},\{\mathbf{p}^{(i)}\}) = \mathcal{H}(\{\mathbf{p}^{(i)}\},\{\mathbf{p}^{(i)}+\mathbf{R}\})^\dagger.
\end{equation}

Note that in principle, the two sets $\{\mathbf{p}^{(i)}\}$ and $\{\mathbf{p}^{(j)}\}$ do not have to be of the same size to generate interaction matrices between them. Since our objective is to transform an initial set of $\mathcal{H}(\mathbf{R})$, as produced by Wannier90, into a different set of $\tilde{\mathcal{H}}(\tilde{\mathbf{R}})$ representing the same lattice, but expressed in terms of a different unit cell, the special case $\{\mathbf{p}^{(i)}\} = \{\mathbf{p}^{(j)} + \mathbf{R}\}$ is the one relevant for us.

In the case where the new unit cell is a super cell, the Hamiltonian data must be upscaled, meaning that some of the original interactions must be repeated in the new representation. The situation in Fig. \ref{sec:upscaling:fig1} can now be readily understood, as it represents the ideal case where all Wannier centers converged onto a single atomic position inside the initial unit cell. \\

For large $\{\mathbf{p}^{(i)}\}$ where the bond length between atomic positions drastically exceeds the interaction range of the initial Wannier Hamiltonian, the produced device Hamiltonian matrices exhibit a high degree of sparsity whose pattern depends on the ordering of the positions in $\{\mathbf{p}^{(i)}\}$. The standard ordering relation for atomic positions used throughout this work is defined as

\begin{equation}\label{sec:hctor:eq3}
  \mathbf{p_i} < \mathbf{p_j} \Leftrightarrow [p_{i,1}<p_{j,1}] \vee [(p_{i,1}=p_{j,1}) \wedge (p_{i,2}<p_{j,2})] \vee ...
\end{equation}

\begin{figure}
 \centering
 \includegraphics[width=\textwidth]{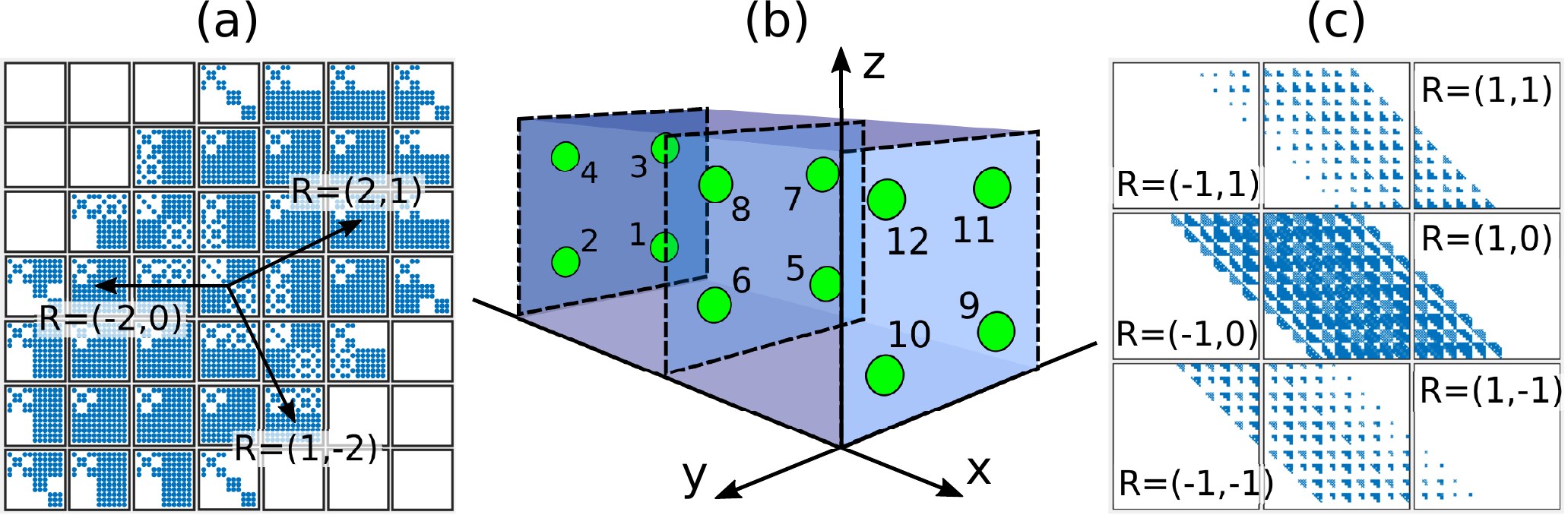}
 \caption{Example of the spatial ordering obtained with Eq. \eqref{sec:hctor:eq3} and the resulting Hamiltonian matrices. (a) Wannier Hamiltonian matrices $\mathcal{H}(\mathbf{R})$ on a grid of 37 \textbf{R} vectors (empty blocks are ommitted), represented using the primitive unit cell from Wannier90. (b) Atomic ordering according to Eq. \eqref{sec:hctor:eq3}. The green circles represent atomic positions with their ordering index displayed next to them. (c) Hamiltonian matrices $\mathcal{H}(\mathbf{R})$ exhibiting a block tri-diagonal sparsity pattern on a grid of 9 \textbf{R} vectors, represented using an orthorhombic super cell.}\label{sec:hctor:fig1}
\end{figure}

This ordering relation is also known as a \emph{lexicographical order}, which is used in sorting, for example, names in a phone book. It leads to atomic positions organized in slices along the first coordinate axis. For the orthorhombic unit cells typically used in quantum transport simulations, this is equivalent to cutting slices perpendicular to the x-axis, which is also the transport direction of electrons. In tandem with the limited range of Wannier functions, this ordering scheme results in a block tri-diagonal sparsity pattern of the Hamiltonian matrices, as exemplified in Fig. \ref{sec:hctor:fig1}.

\begin{figure}
 \centering
 \includegraphics[width=10cm]{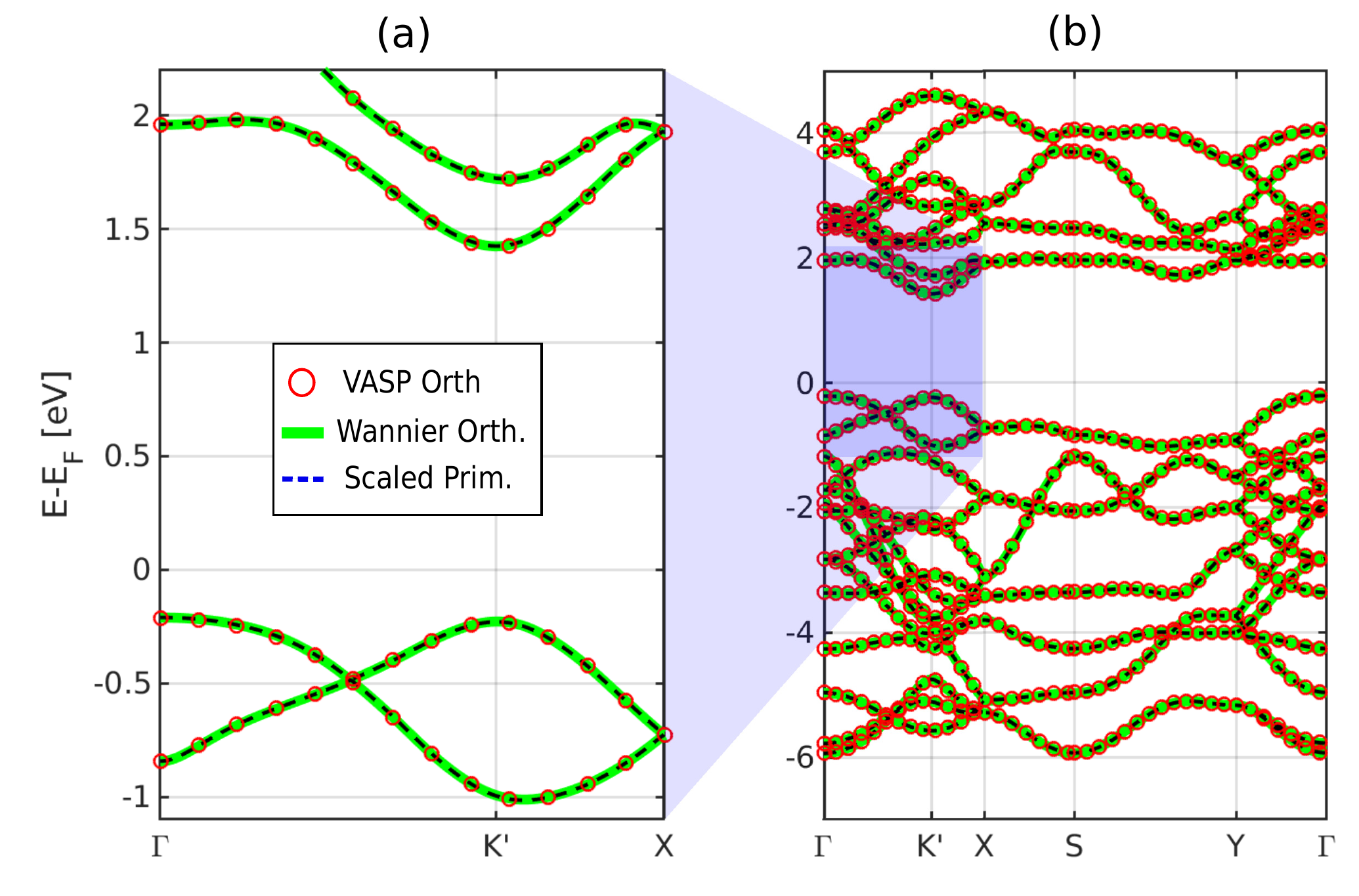}
 \caption{Comparison of the bandstuctures of the smallest orthorhombic cell of a MoS\tsc{2} monolayer. The dashed blue lines have been obtained with the proposed upscaling approach, based on bond-centered Wannier Hamiltonians as described in Sections \ref{sec:wbh}, using a primitive hexagonal cell as basis. The K'-point is where the K-point of the hexagonal symmetry is found in the orthorhombic symmetry. The red circles result from a direct simulation of an orthorhombic cell in DFT and the green lines from a wannierization of this data. The observed differences mainly come from the wannierization itself (3.22 meV on average between the red circles and the green curves), not from the upscaling method (0.42 meV between the green and the blue curves). (a) Zoom around the minimum of the conduction band and the maximum of the valence band. (b) Full bandstructure in the first Brillouin zone.}\label{sec:hctor:fig2}
\end{figure}

The generated Hamiltonian matrices and \textbf{R} vectors can then be used in bandstructure calculations using Eq. \eqref{sec:wbh:eq3}. To demonstrate the validity of our approach, a MoS\tsc{2} monolayer structure is considered. A DFT calculation, followed by a wannierization were performed on both the primitive hexagonal cell and an orthorhombic super cell. After employing the upscaling technique on the Hamiltonian represented in the primitive cell to match the orthorhombic cell, the bandstructures extracted from either case can be compared directly. The results are equivalent up to slight discrepancies coming from the transformation into Wannier functions and the Fourier interpolation of the bandstructure, as presented in Table \ref{sec:hctor:tbl1} and Fig. \ref{sec:hctor:fig2}.

\vspace{0.5cm}
\begin{table}[H]
\centering
\begin{tabular}{|c|c|c|c|c|c|}
\hline
          & $E_g (eV)$ & $m_e^*$ & $m_h^*$ & $\tilde{m}_e^*$ & $\tilde{m}_h^*$ \\
\hline
VASP      & 1.648      & 0.4639   & 0.5769   & -               & -             \\
\hline
wannier90 & 1.656      & 0.4644   & 0.5759   & 0.4619          & 0.5715        \\
\hline
scaled    & 1.656      & 0.4620   & 0.5765   & 0.4596          & 0.5720        \\
\hline
\end{tabular}
\caption{Band gaps and effective masses extracted at the K' point (in $\Gamma$-X direction) from the bandstructures in Fig. \ref{sec:hctor:fig2}. $m_e^*$ and $m_h^*$ were calculated using parabolic fits, $\tilde{m}_e^*$ and $\tilde{m}_h^*$ are derived from Hessian matrices (details given in Section \ref{sec:err}).}\label{sec:hctor:tbl1}
\end{table}

\newpage
\clearpage

\begin{figure}
 \centering
 \includegraphics[width=8cm]{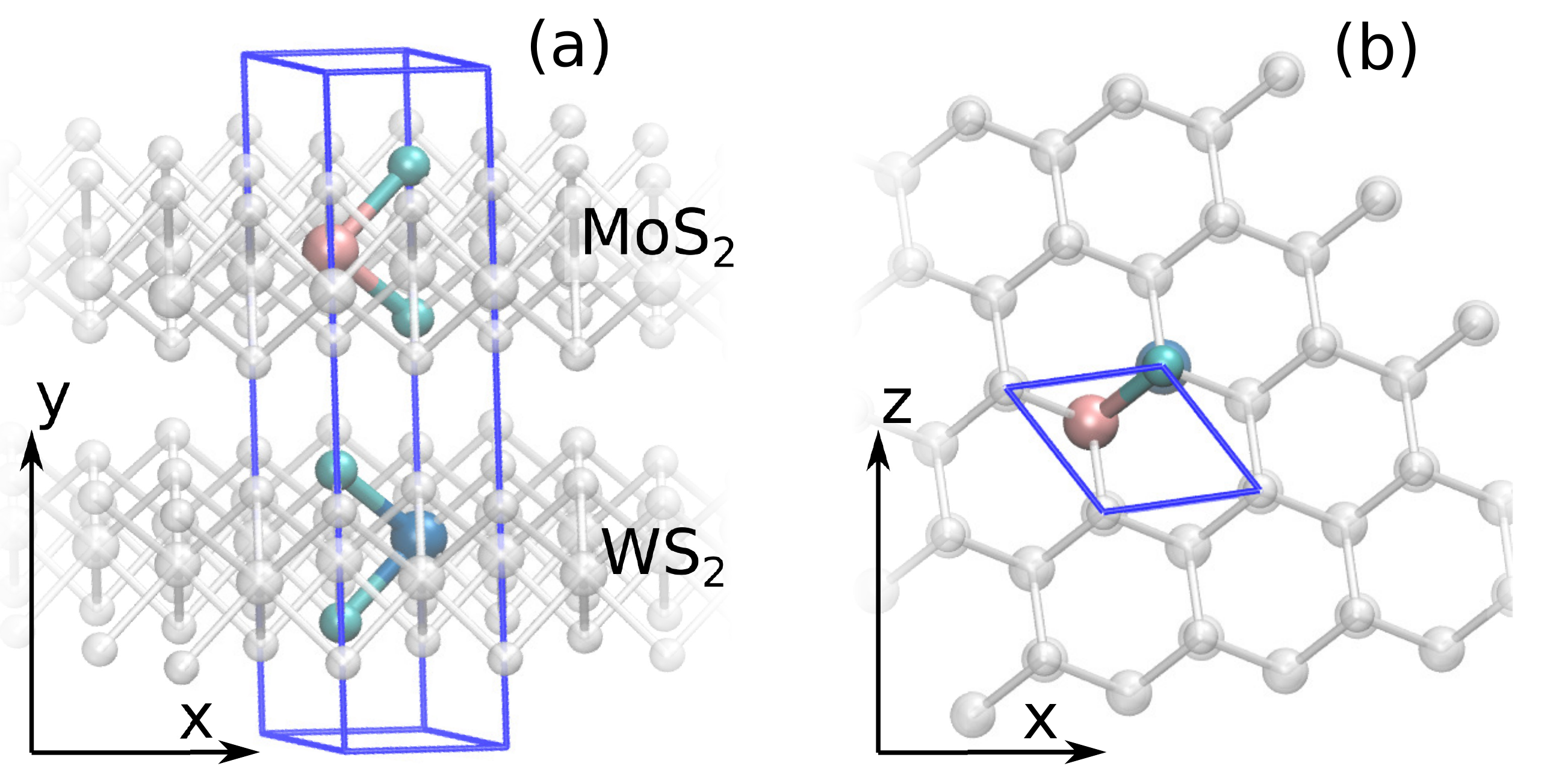}
 \caption{Van der Waals heterostructure made of a WS\tsc{2} (bottom) and a MoS\tsc{2} (top) monolayer. The primitive unit cell of this stack is made of 6 atoms only due to the close lattice constants. (a) Side view of the unit cell. (b) Top view of the same unit cell.}\label{sec:hctor:fig3}
\end{figure}

Our approach is not limited to structures made of exact reproductions of the initial DFT unit cell. Since the Hamiltonian data from Wannier90 is represented in terms of bonds, any domain where all or a sub-set of these bonds are present can be constructed. Since inter-atomic interactions depend on the surrounding environment, the physics may not be captured appropriately. One class of materials where removing or adding atomic layers is possible without significantly perturbing the local properties, are weakly interacting van der Waals heterostructures (vdWh) \cite{Geim2013}. A MoS\tsc{2}-WS\tsc{2} stack, as shown in Fig. \ref{sec:hctor:fig3}, ideally illustrates the concept of vdWh. Both materials have practically the same lattice constant and a very similar inter-layer distance in multi layered configurations.

First, we examine the coupling matrices between a single-layer of WS\tsc{2} placed below a monolayer of MoS\tsc{2}. To estimate the interaction strength inside each layer and between them, we should measure the so-called 'total energy' contained in each bond. As our interaction matrices are in general not quadratic, we cannot compute their eigenvalues and sum them. Instead, we can perform a singular value decomposition and add the results, as proposed in Table \ref{sec:hctor:tbl2}.

\begin{table}
 \centering
 \begin{tabular}{ c | c c c | c c c }
   & $W$ & $S_{W,l}$ & $S_{W,u}$ & $Mo$ & $S_{Mo,l}$ & $S_{Mo,u}$ \\
 \hline
 $W$ & 19.5859 & 3.4470 & 3.4494 & 0.0209 & 0.0684 & 0.0083 \\
 $S_{W,l}$ & 3.4470 & 19.0588 & 1.7342 & 0.0074 & 0.0333 & 0.0065 \\
 $S_{W,u}$ & 3.4494 & 1.7342 & 19.1133 & 0.0658 & \cellcolor{blue!25}0.4445 & 0.0315 \\
 \hline
 $Mo$ & 0.0209 & 0.0074 & 0.0658 & 21.3802 & 3.1517 & 3.1480 \\
 $S_{Mo,l}$ & 0.0684 & 0.0333 & \cellcolor{blue!25}0.4445 & 3.1517 & 18.6284 & 1.4716 \\
 $S_{Mo,u}$ & 0.0083 & 0.0065 & 0.0315 & 3.1480 & 1.4716 & 18.5698 \\
 \end{tabular}
 \caption{Sum of the singular values corresponding to atomic interactions inside the unit cell in Fig. \ref{sec:hctor:fig3}. The strongest MoS\tsc{2}-WS\tsc{2} interactions are highlighted in blue. The indices l(u) refer to the sulfur layer situated above (below) the transition metal layer.}\label{sec:hctor:tbl2}
\end{table}

\begin{figure}
 \centering
 \includegraphics[width=\textwidth]{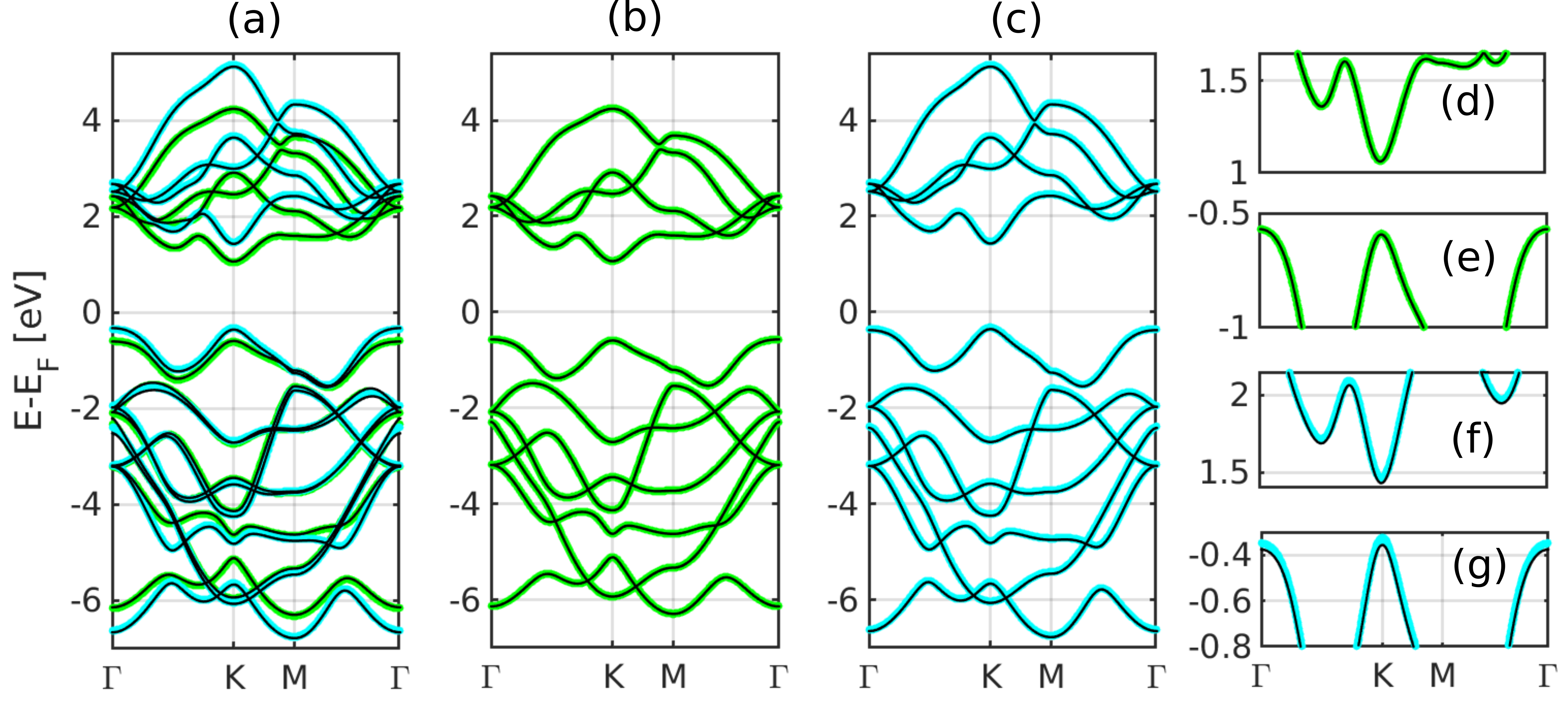}
 \caption{Bandstructures derived from the top layer of MoS\tsc{2} and the bottom layer of WS\tsc{2} from the heterostructure in Fig. \ref{sec:hctor:fig3} and comparison to the DFT simulations of pure MoS\tsc{2} and WS\tsc{2} monolayer. The thin black lines belong to the double layer simulation, the thick green (blue) lines to pure MoS\tsc{2} (WS\tsc{2}) monolayers. (a) Full bandstructure of the double layer heterostructure and of both monolayer simulations. (b) Comparison of the MoS\tsc{2} bandstructure extracted from the double layer simulation with that of the pure monolayer. (c) Same as (b), but for WS\tsc{2}. (d) Zoom into the conduction band from (b). (e) Zoom into the valence band from (b). (f) Zoom into the conduction band from (c). (g) Zoom into the valence band from (c).}\label{sec:hctor:fig4}
\end{figure}

We are first mainly interested in the off-diagonal elements as the diagonal ones are subject to an arbitrary shift in energy. From Table \ref{sec:hctor:tbl2} it is apparent that the intra-layer interactions are stronger in both the MoS\tsc{2} and WS\tsc{2} layers than the inter-layer ones, by about 2 orders of magnitude, except for the S-S connection at the layer interface. As expected, stacked MoS\tsc{2} and WS\tsc{2} only weakly interact with each other. However, to determine whether the properties of isolated MoS\tsc{2} (WS\tsc{2}) can be retrieved from the Hamiltonian of the MoS\tsc{2}/WS\tsc{2} heterostructure by removing the entries corresponding to WS\tsc{2} (MoS\tsc{2}), the bandstructure resulting from this process must be computed. To this end two sets of Hamiltonian matrices were constructed, one for the upper layer of MoS\tsc{2} and one for the lower layer of WS\tsc{2}. Bandstructure calculations for both sets can then be compared to their exact counterparts, i.e. DFT simulations of pure MoS\tsc{2} or WS\tsc{2} monolayers. Results are presented in Fig. \ref{sec:hctor:fig4}.

From these comparisons, it can be deduced that the approximation of using only one layer from the bilayer stack works well to extract the properties of an isolated 2-D material, which is also confirmed by the electron and hole effective masses extracted at the K-point, which are almost identical with less than 1\% difference between the ones coming from the individual monolayers and those computed from the bilayer stack, as presented in Table \ref{sec:hctor:tbl3}.

\begin{table}
\centering
\begin{tabular}{|c|c|c|c|c|c|}
\hline
                 & $E_g (eV)$ & $m_{e,x}^*$ & $m_{e,y}^*$ & $m_{h,x}^*$ & $m_{h,y}^*$ \\
\hline
MoS\tsc{2} extr. & 1.6558     & 0.4600      & 0.4627      & 0.5722      & 0.5636      \\
\hline
MoS\tsc{2} pure  & 1.6557     & 0.4596      & 0.4624      & 0.5720      & 0.5632      \\
\hline
WS\tsc{2} extr.  & 1.7864     & 0.2985      & 0.3053      & 0.4130      & 0.3981      \\
\hline
WS\tsc{2} pure   & 1.7862     & 0.2983      & 0.3051      & 0.4129      & 0.3979      \\
\hline
\end{tabular}
\caption{Band gaps and effective masses extracted at the K point along the x and y direction from the bandstructures in Fig. \ref{sec:hctor:fig4}. The effective masses were derived from Hessian matrices at the K-point (details given in Section \ref{sec:err}). The comparison is between MoS\tsc{2} (WS\tsc{2}) monolayers extracted from the bilayer stack in Fig. \ref{sec:hctor:fig3} with the pure variants of a monolayer of MoS\tsc{2} (WS\tsc{2}).}\label{sec:hctor:tbl3}
\end{table}

\begin{figure}
 \centering
 \includegraphics[width=7cm]{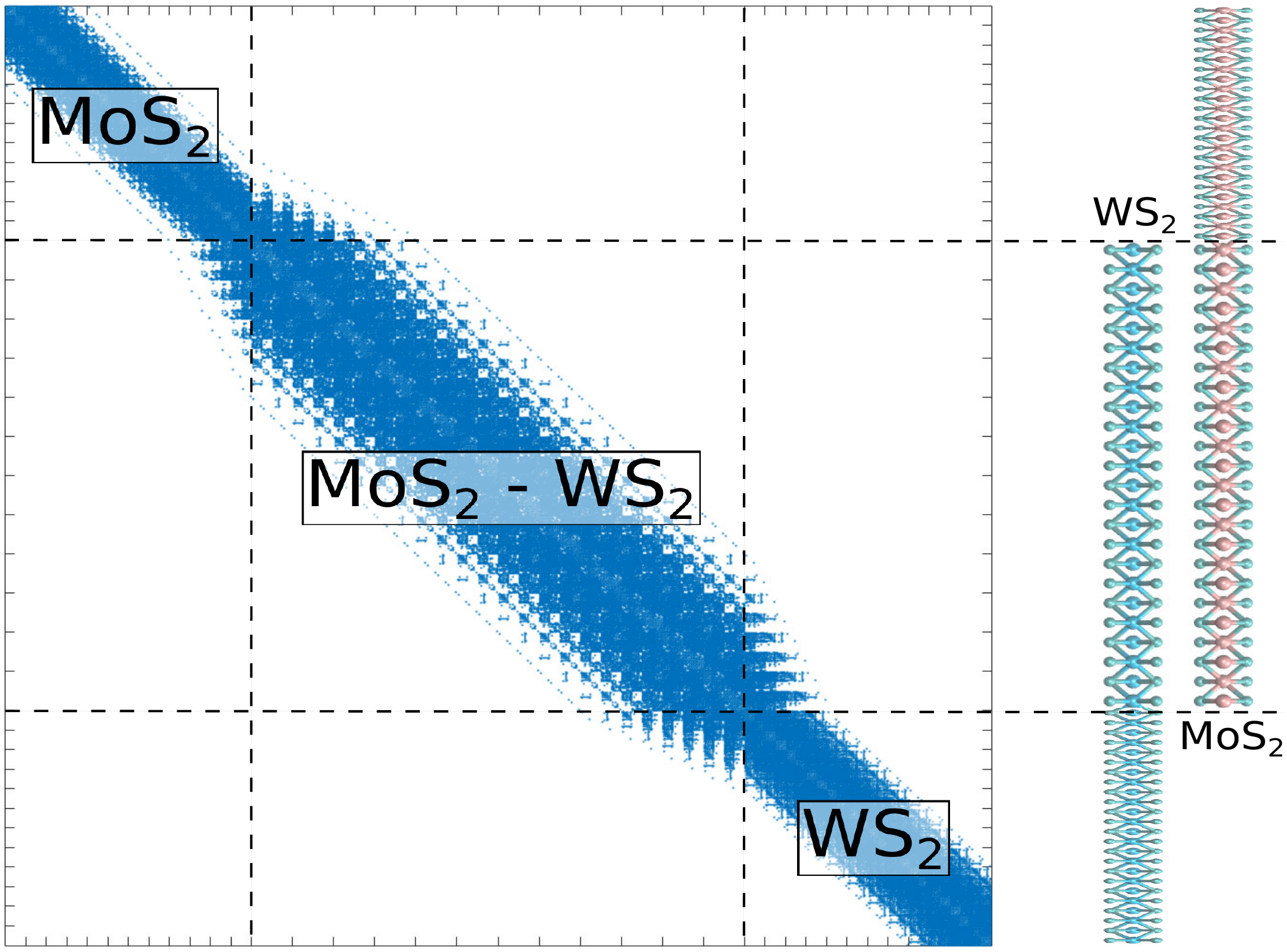}
 \caption{MoS\tsc{2}/WS\tsc{2} van der Waals heterostructure made of three parts: one  pure MoS\tsc{2} extension, one overlap region, and one pure WS\tsc{2} extension. The starting point is the unit cell from Fig. \ref{sec:hctor:fig3}, which is then upscaled to the smallest possible orthorhombic cell and then repeated 36 times. In the first 12 repetitions, the WS\tsc{2} entries were removed, whereas in the last 12 the MoS\tsc{2} entries were discarded. The Hamiltonian matrix (left) and the corresponding device structure (right) are plotted. The different sections clearly manifest themselves as sub-matrices of different sizes, with transition regions in between.}\label{sec:hctor:fig5}
\end{figure}

More complex device structures can thus be constructed on the basis of the bilayer stack, where in some sections both materials overlap, while in others only one compound is present. An example of such a heterojunction is given in Fig. \ref{sec:hctor:fig5}.

Such layered structures are particularly appealing to create p-n diodes at the ultimate thickness limit \cite{Lee2014} with two 2-D monolayers stacked on top of each other, as in Fig. \ref{sec:hctor:fig5}, one p-doped, the other one n-doped. Electrical doping through additional gate-like contacts is a commonly used approach for that. With overlap only in the central part, an ultra-thin depletion region can be obtained, as demonstrated experimentally in \cite{Deng2014}. In the specific MoS\tsc{2}/WS\tsc{2} example, removing the Hamiltonian entries corresponding to one layer and computing the bandstructure of the remaining components produces accurate results, but the success of this procedure might strongly depend on the materials in question. Often relaxed structures in a multi-layered arrangement do not exactly match their monolayer parent, thus leading to inaccurate bandstructures for the separated individual layers. Additionally, different materials usually do not share the same lattice constant, contrary to MoS\tsc{2} and WS\tsc{2}. In such cases, larger unit cells are required in the DFT calculations so that the lattice mismatch can be absorbed. The MoS\tsc{2}/WS\tsc{2} system is ideal and was chosen for demonstrational purposes. The separation of the Hamiltonian entries can still be applied to more complex material stacks and the algorithms presented here can tolerate some spatial warping. The limits of our approach will be explored and discussed in Section \ref{sec:approx}.
\section{Interface with OMEN}
\label{sec:interface}

\begin{figure}
 \centering
 \includegraphics[width=8cm]{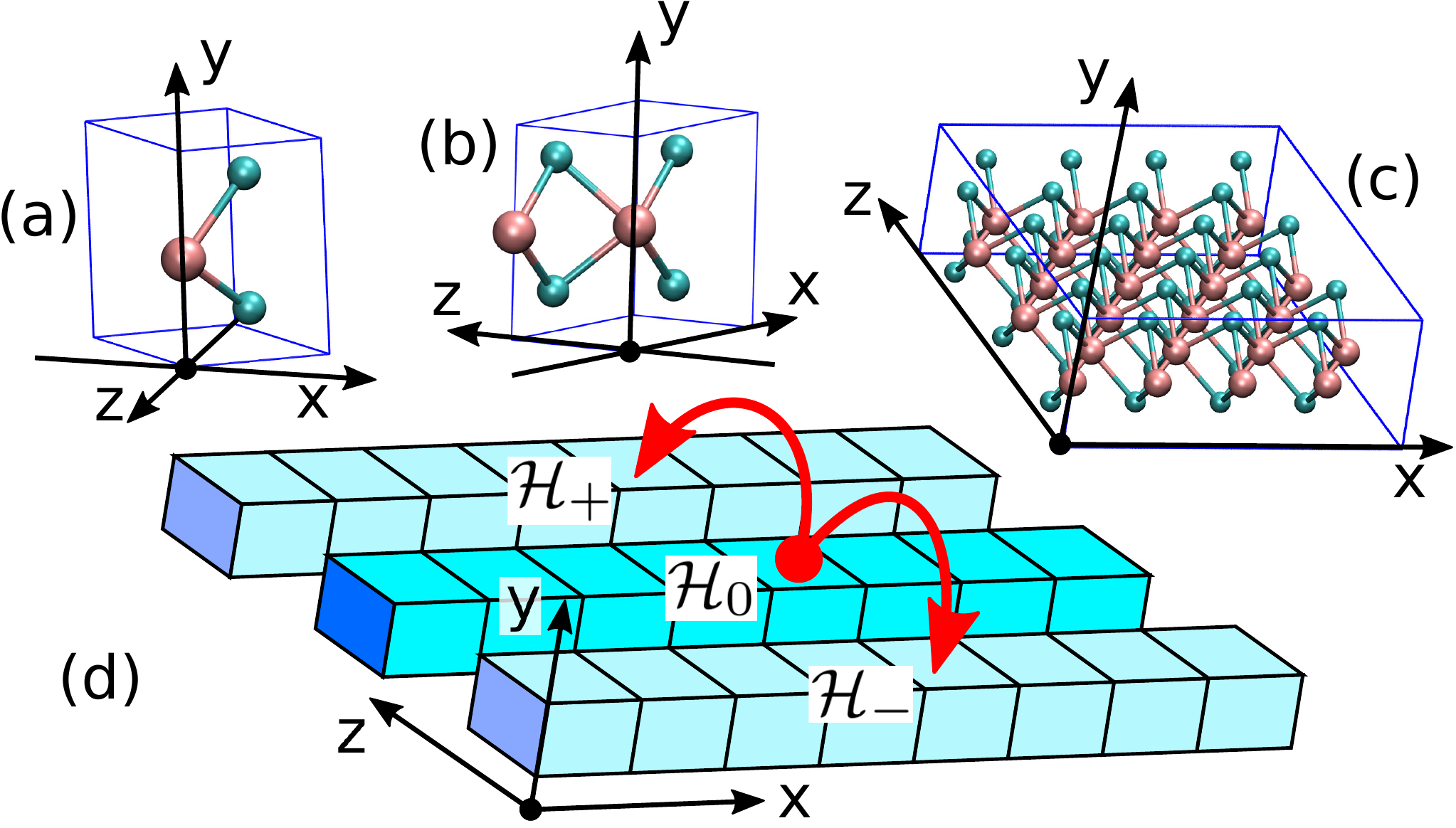}
 \caption{Schematic of the steps involved to interface Wannier90 with OMEN, starting from a conversion of Wannier90 outputs into a Hamiltonian in terms of interactions along bonds, as introduced in Section \ref{sec:wbh}. (a) Repeatable primitive cell (here hexagonal MoS\tsc{2}) from which the desired simulation domain can be constructed. (b) Corresponding orthorhombic cell. (c) Extended orthorhombic cell such that bonds with significant interactions do not extend beyond next-neighbor cells.
 (d) Schematic of the simulation domain. The unit cell in (c) is repeated both along the transport axis (x) and along the direction assumed periodic (z). Three Hamiltonian components are created, $\mathcal{H}_0$, which contains the interactions with neighbor cells along x and $\mathcal{H}_{+}$($\mathcal{H}_{-}$), which includes the interactions with the neighbor cells along $+$z($-$z) (see Eq. \eqref{sec:interface:eq1}).}\label{sec:interface:fig1}
\end{figure}

The concepts introduced in Sections \ref{sec:wbh} and \ref{sec:hctor} can now be used to manipulate outputs from Wannier90 and convert them to inputs for OMEN. Here, step (c) from Fig. \ref{sec:toolchain:fig1} is explained in detail. An outline is given in Fig. \ref{sec:interface:fig1}. \\

\begin{figure}
 \centering
 \includegraphics[width=10cm]{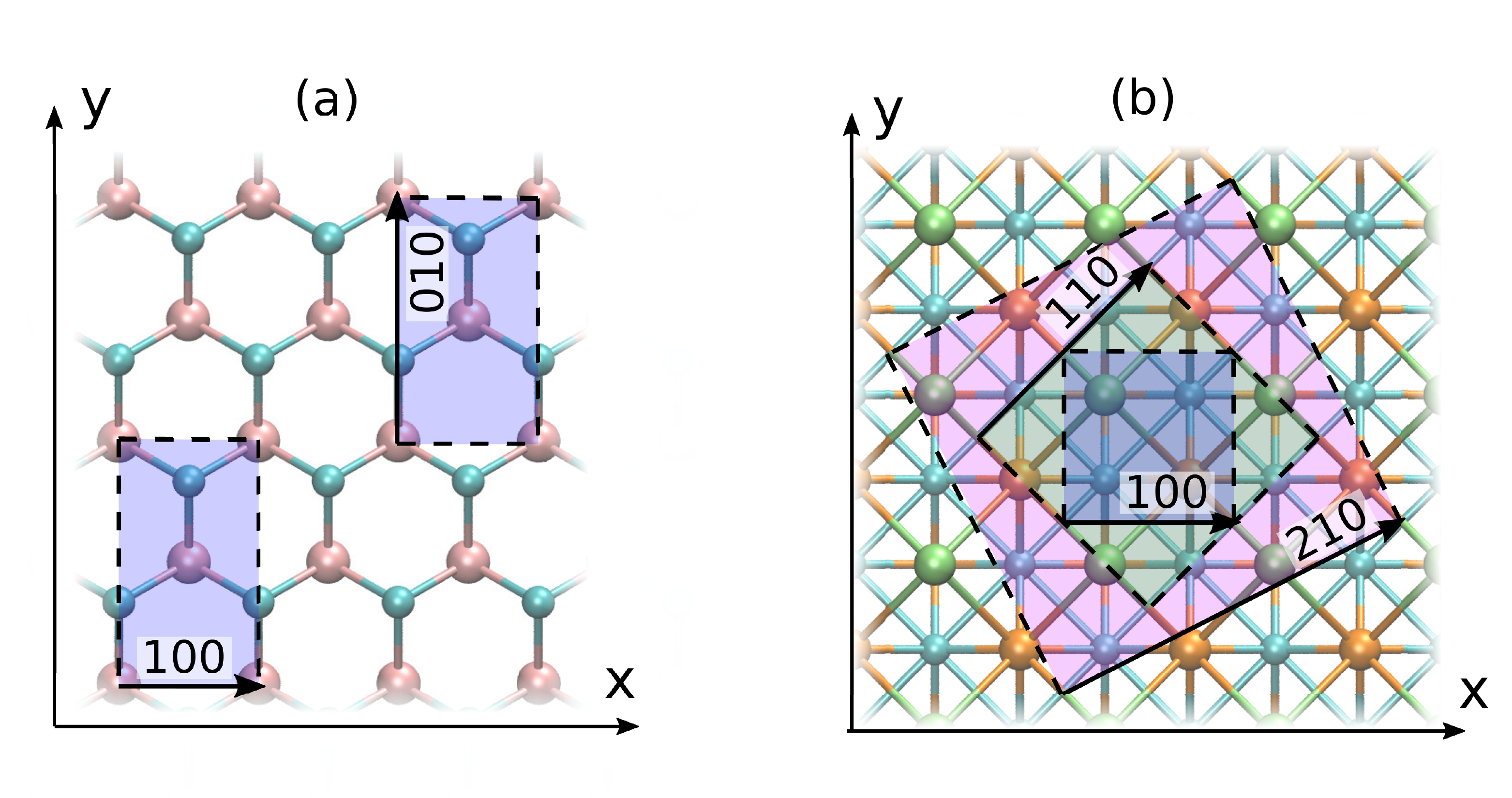}
 \caption{Examples of lattices with different symmetries and possible transport directions. (a) MoS\tsc{2} monolayer structure with hexagonal symmetry. The smallest orthorhombic cell is depicted as a blue frame, where the transport direction can be specified along (100) in the lower left corner or along (010) in the upper right corner. (b) BiIO monolayer structure with cubic symmetry, where an infinity of orthorhombic super cells exist. In the center, the primitive cell with transport along (100) is depicted. Due to the cubic symmetry, transport along (010) is equivalent. Transport along (110), (210) or in general (mn0) with $m,n \in \mathds{N}$ is also possible.}\label{sec:interface:fig2}
\end{figure}

The starting point is a Hamiltonian in terms of interactions along bonds, which is created on the basis of Wannier90 outputs. Since the underlying unit cell is always primitive, an algorithm was developed allowing for automatic detection of a super cell according to a template basis (see Appendix \ref{app:expansions}). By convention OMEN defines the x-axis as the transport direction, the y-axis as the restricted axis (direction of confinement), and the z-axis as the periodic axis. Since the first step is to find the smallest possible orthorhombic super cell, a template basis adhering to this convention must be specified. Depending on the lattice symmetry, there might be multiple directions along which transport can be simulated. Examples are presented in Fig. \ref{sec:interface:fig2}.

For the MoS\tsc{2} monolayer structure in Fig. \ref{sec:interface:fig2}, the template bases used for each case are:

\[
 \mathbf{T}_{100} =
 \begin{bmatrix}
  1 & 0 & 0\\
  0 & 0 & 1\\
  0 & 1 & 0\\
 \end{bmatrix},
 \hspace{1cm}
 \mathbf{T}_{010} =
 \begin{bmatrix}
  0 & 0 & 1\\
  1 & 0 & 0\\
  0 & 1 & 0\\
 \end{bmatrix}.
\]

The first column is the transport direction, the second one the restricted direction, and the third one specifies the direction assumed periodic. \\

The unit cell used as the basic building block for devices in OMEN must not only be of orthorhombic symmetry, but also the interactions may not exceed next neighbor cells along the periodic (z) direction. This condition is necessary such that the device Hamiltonian matrix $\mathcal{H}(k_z)$ can be written as

\begin{equation}\label{sec:interface:eq1}
 \mathcal{H}(k_z) = \mathcal{H}_0 + \mathcal{H}_{+} e^{ik_z\mathbf{\Delta}} + \mathcal{H}_{-} e^{-ik_z\mathbf{\Delta}}.
\end{equation}

where $\mathcal{H}_{-}$, $\mathcal{H}_{+}$, and $\mathcal{H}_{0}$ were introduced in Fig. \ref{sec:interface:fig1}, $\Delta$ is the width of the orthorhombic cell along the z-axis, and the wave vector $-\frac{\pi}{\Delta} \leq k_z \leq \frac{\pi}{\Delta}$ models the periodicity of the system according to Bloch's theorem. Expansion coefficients for the orthorhombic cell corresponding to (c) in Fig. \ref{sec:interface:fig1} should be extracted directly from the Hamiltonian in terms of interactions along bonds. As the longest-range interactions can usually be neglected without significantly affecting the transport simulations, a compromise between the matrix bandwidth and the accuracy can be made. There are further possibilities to decrease the size of the Hamiltonian matrices for transport calculations. The first and most important one is the size of the Wannier basis set. Even though the construction of the basis does not belong to the Winterface functionalities, it is still relevant to briefly discuss the wannierization process itself. \\

\subsubsection{Wannierization Process}
\label{sec:wannierization}

\begin{figure}
 \centering
 \includegraphics[width=12cm]{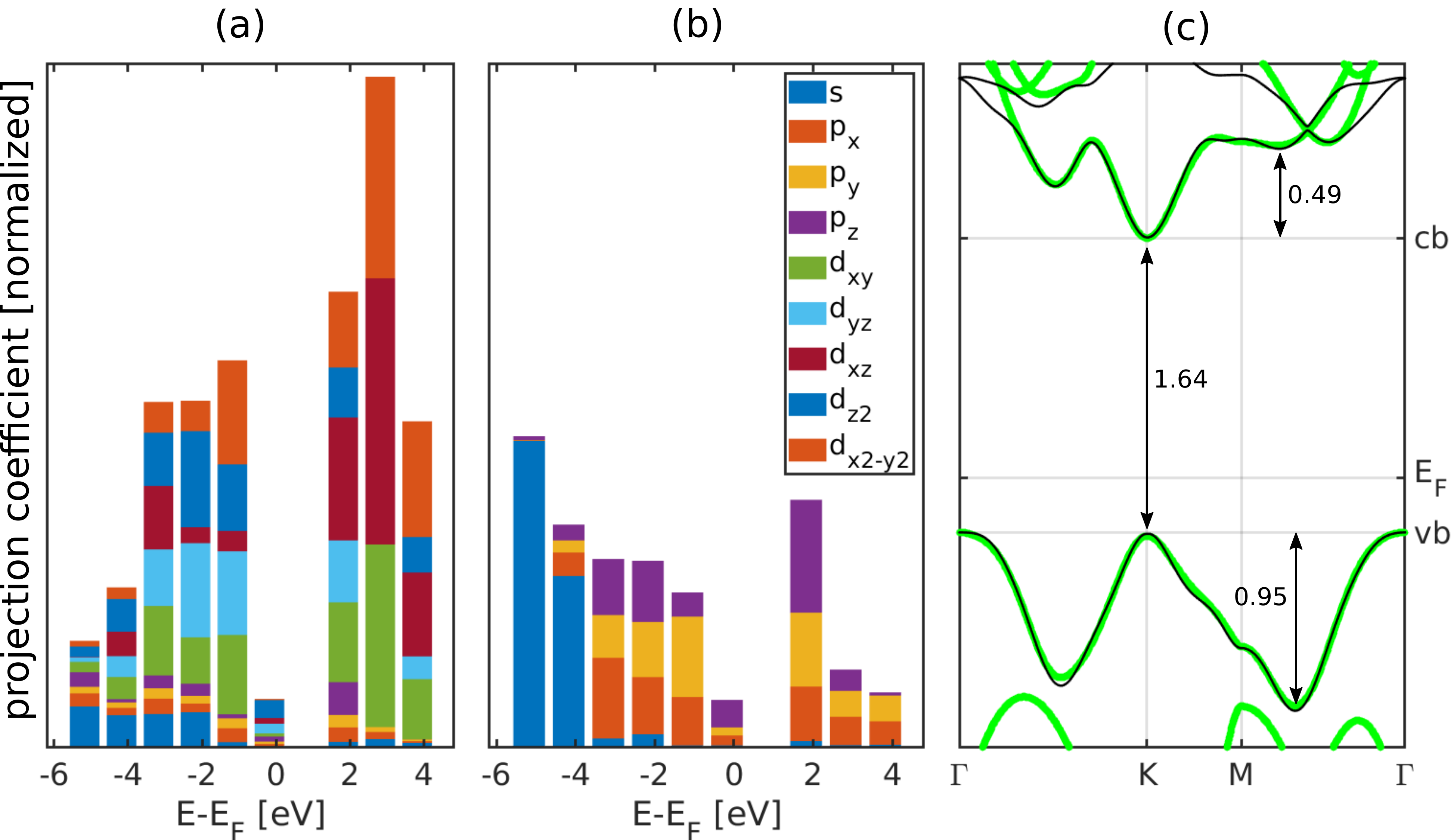}
 \caption{(a) Decomposition of the DFT plane-waves into 1s, 3p and 5d orbitals projected to the molybdenum atom. The region around the band gap appears to be predominantly of d-character. (b) Same as (a), but for the sulfur atoms, whose behavior around the band gap is dominated by p-like orbitals. (c) Comparison of the bandstructures resulting from two different initial guesses on the basis of (a) and (b). The green bands were derived using an initial guess containing 5 d-orbitals on molybdenum and 3 p-orbitals on each sulfur atom, resulting in a basis set of 11 Wannier functions. The black bands were derived using only d\tsc{x2-y2},d\tsc{x2} and d\tsc{xz} orbitals on molybdenum. The accuracy of the smaller basis set (black bands) is lower than the one with more Wannier functions, especially at energies larger 0.5eV into the conduction band.}\label{sec:interface:fig3}
\end{figure}

For quantum transport, accurate bandstructure modeling is required only around the gap separating the conduction from the valence band, considering a window of approximately 1eV on each side of it. A suitable initial guess for Wannier90 can often be found by analyzing the character of the site-projected wave functions. In VASP this information is stored in the PROCAR file \cite{procar}. A decomposition of each band of monolayer MoS\tsc{2} is presented in Fig. \ref{sec:interface:fig3}.

Wannier functions are constructed by mixing plane-wave eigenfunctions of the Hamiltonian operator, as defined in Eq. \eqref{sec:mlwf:eq2}. In general, each band receives contributions from all Wannier functions. Therefore, a reduction of the basis set post wannierization is in general not possible. \\

\subsubsection{Error Estimation}
\label{sec:err}

The basic principle when generating Hamiltonian matrices of minimal size for a given device structure consists of sorting out interactions along bonds whose absence does not significantly affect the bandstructure in the critical region around the band gap. This can be done by both setting a cutoff tolerance for the interaction strength, below which the bond is discarded, or by directly setting spacial limits and discarding bonds extending beyond them. Analytically, an upper bound to the perturbation of eigenvalues is given by Weyl's Theorem. Let $\nu_i \geq \ldots \geq \nu_n$ be the eigenvalues of a Hamiltonian operator $\mathcal{H}$, and $\mu_i \geq \ldots \geq \mu_n$ the eigenvalues of a perturbed operator $\mathcal{H} + P$, then

\begin{equation}\label{sec:interface:eq3}
 |\nu_i - \mu_i| \leq \Arrowvert P \Arrowvert_2 \hspace{1cm} \forall i \in \{1,...,n\}.
\end{equation}

This provides us with both an upper bound for the expected perturbation and the insight that the effect of perturbations is cumulative. For a detailed view, bandstructures must be computed based on a perturbed and an unperturbed Hamiltonian matrix, after which the influence of the perturbation can be estimated by directly comparing the results. To test perturbative effects and the functionality of the upscaling technique introduced in Sec. \ref{sec:wbh} and \ref{chap:hctor}, a second algorithm has been developed to compute the bandstructures of a super cell with Hamiltonian data belonging to a primitive cell. Such an algorithm relies on the zone folding concept. If two completely different methods, upscaling and zone folding, produce the same results, then the likelihood that both are working correctly is high. \\

In crystal theory, the basis \textbf{G} spanning the reciprocal unit cell is given by

\begin{equation}\label{sec:interface:eq4}
 \mathbf{G} = 2 \pi \mathbf{B}^{-T},
\end{equation}

\begin{figure}
 \centering
 \includegraphics[width=9cm]{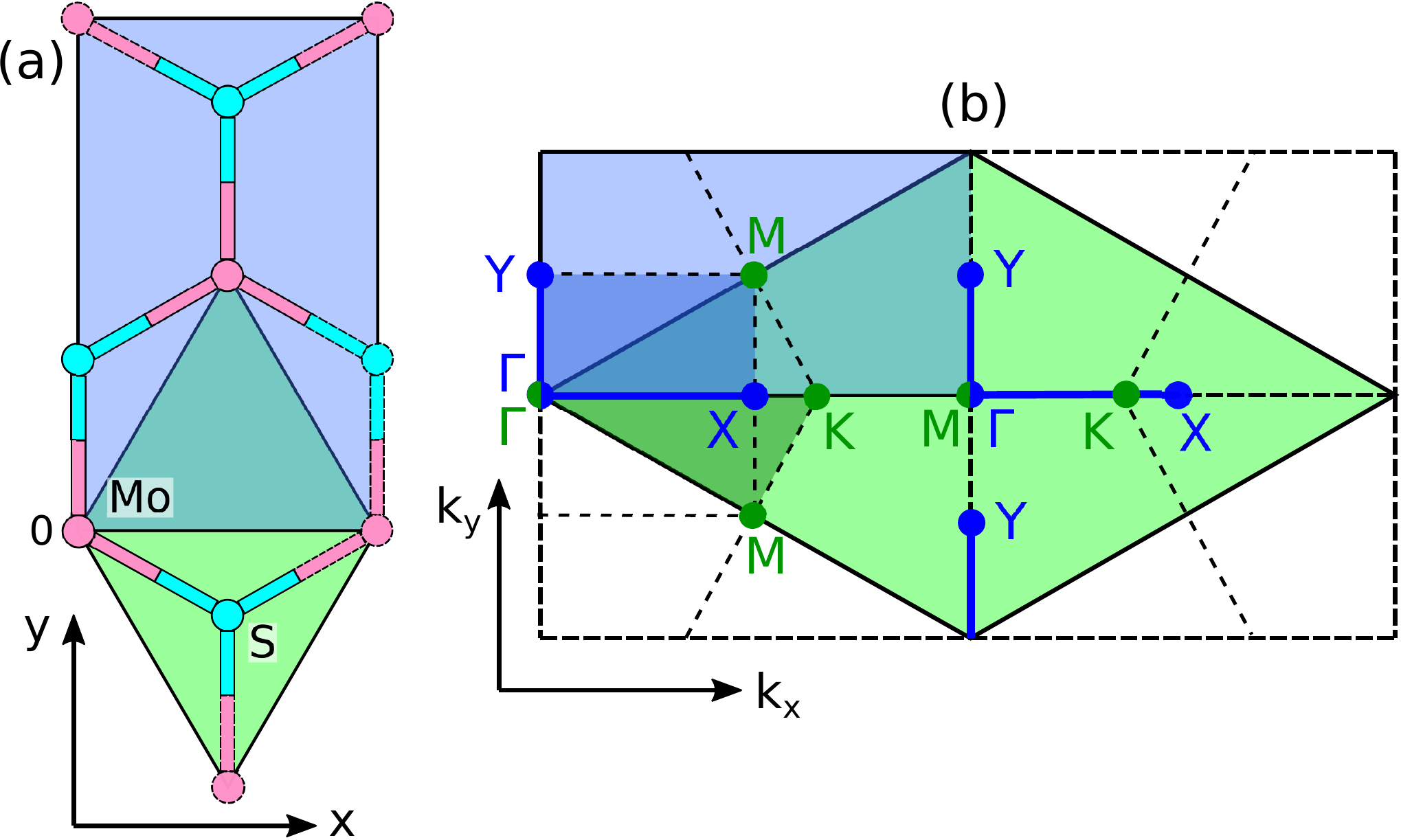}
 \caption{Schematic of a primitive hexagonal unit cell compared to an orthorhombic super cell in both real and reciprocal space. (a) Primitive hexagonal cell of MoS\tsc{2} in green and orthorhombic super cell in blue. (b) Same as (a), but in reciprocal space. Symmetry points in blue (green) belong to the blue (green) cell. The Wigner-Seitz cell for each case, i.e. the first Brillouin zone, is indicated by thin dashed lines and the irreducible parts by the stronger shading. The hexagonal green unit cell is completely covered by 4 orthorhombic blue cells, where in each case only half overlaps. The blue k-point path, $Y-\Gamma-X$, is defined in the irreducible part of the blue cell, with periodic images given in the center. Each k-point of the blue path is found twice in the green unit cell, but in different sections of the Brillouin zone, due to folding. For example, the $\Gamma$-point of the blue cell appears as both the $\Gamma$-point of the green cell, but also as the M-point in the middle, i.e. the M-point is folded into the $\Gamma$-point.}\label{sec:interface:fig4}
\end{figure}

The ratio between the volume of the reciprocal cell coming from the primitive ($|\mathbf{G}_p|$) and the super cell ($|\mathbf{G}_s|$) is equal to the ratio of the super cell ($|\mathbf{B}_s|$) and of the primitive cell ($|\mathbf{B}_p|$) volume, i.e. $|\mathbf{G}_p|/|\mathbf{G}_s| = |\mathbf{B}_s|/|\mathbf{B}_p|$. Each k-point on a path defined in the reciprocal cell of a supercell is therefore found multiple times in the reciprocal cell of the primitive unit cell, as demonstrated in Fig. \ref{sec:interface:fig4}

The algorithm to compute a bandstructure with the zone folding method includes the following steps:

\begin{itemize}
 \item Define a path of k-points in the reciprocal cell of the super cell.
 \item Map the k-points in the reciprocal cell of the super cell to those in the reciprocal cell of the primitive cell using the same algorithms as to find the atomic positions belonging to a super cell.
 \item Compute the bandstructure at these k-points with the Hamiltonian directly imported from Wannier90 and corresponding to the primitive cell.
 \item Group the energies of the k-points that are folded into the same location in the reciprocal unit cell of the super cell.
\end{itemize}

In the example of Fig. \ref{sec:interface:fig4}, the number of energies that must be grouped is equal to 2, as expected from the volume ratio of $|\mathbf{G}_p|/|\mathbf{G}_s|$. For the purpose of computing the bandstructure of super cells, this approach is much faster than the upscaling method, because it solves two eigenvalue problems of dimension N for each k-point, instead of solving one eigenvalue problem of dimension 2N per k-point. The bandstructure computed with this algorithm can be considered as an exact reference for the upscaling technique. The influence of perturbations to the Hamiltonian data can thus be determined by calculating bandstructures for the same cell using first the raw Wannier90 outputs with the folding algorithm laid out above and secondly with the upscaling method after sorting out certain interactions along bonds beforehand. \\

\begin{figure}
 \centering
 \includegraphics[width=11.5cm]{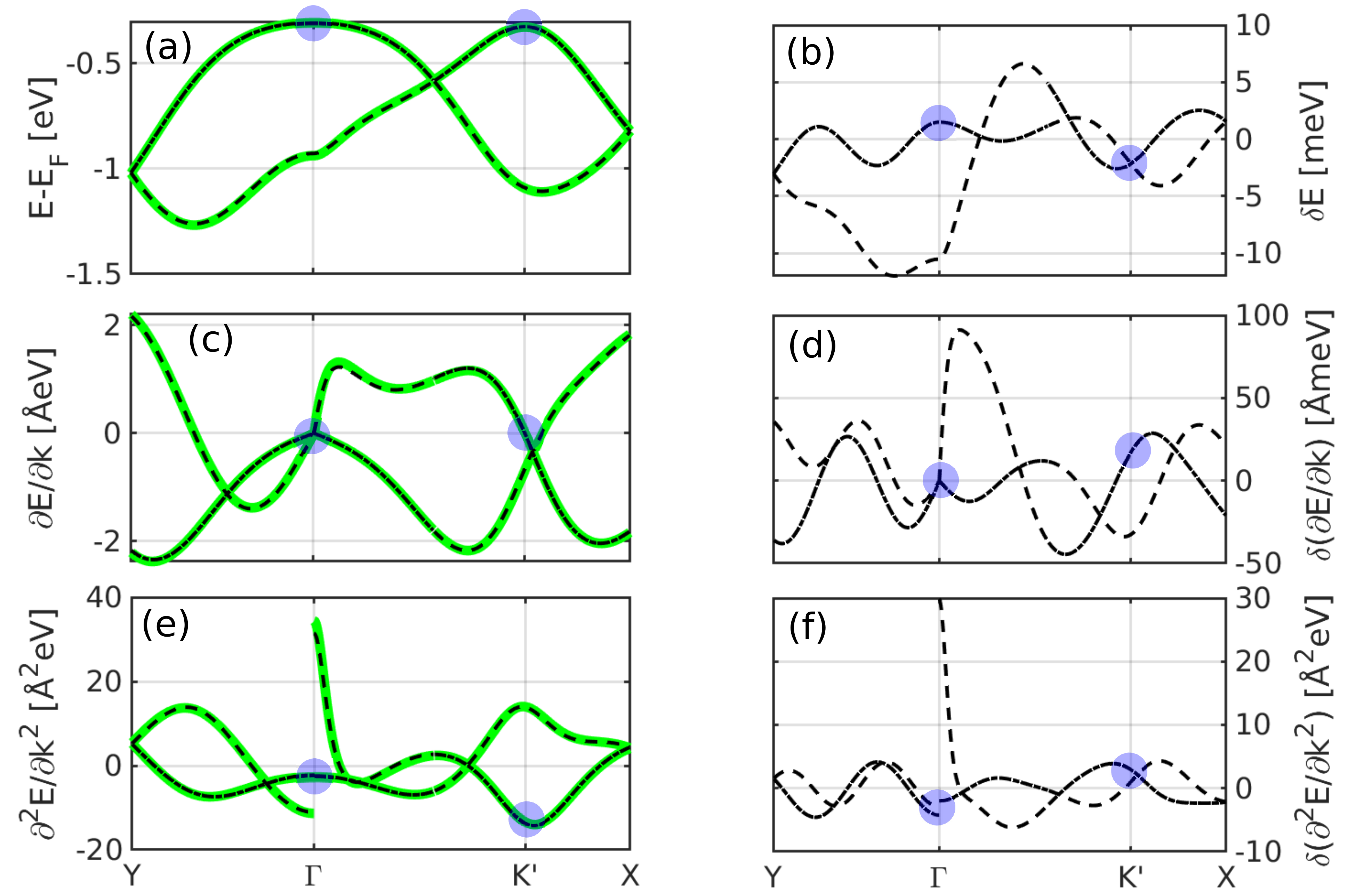}
 \caption{Comparison of the MoS\tsc{2} monolayer bandstructure around the top of the valence band along the $Y-\Gamma-X$ path of the super cell Brillouin zone (see Fig. \ref{sec:interface:fig4}). The K' point is where the K-point of the primitive cell is found in the super cell. The bands calculated using the folding algorithm (thick green lines) are considered the reference, whereas the bands calculated using the upscaling technique (dashed black lines) discard some of the long ranging interactions. The critical points at the top of the valence band at the $\Gamma$-point and the K'-point are highlighted. (a) Top two valence bands. (b) Error between the green and the dashed lines in (a). (c) Same as (a), but for the first derivative of the bandstructure. (d) Error in $dE/dk$. (e) Same as (a), but for the second derivative of the bandstructure. (f) Error in $d^2E/dk^2$.}\label{sec:interface:fig5}
\end{figure}

For quantum transport, errors in the first and second derivatives of the bandstructure are important as well, the latter being used to calculate effective mass tensors. These quantities can be computed together with the bandstructure with minimal overhead as shown in Fig. \ref{sec:interface:fig5}.

It can be seen that discarding the long range interactions produced by Wannier90 leads to relatively accurate results. For the bandstructure computed using the upscaling technique, the interactions were filtered such that those exceeding an expansion of $\mathbf{C} = diag([3,2,1])$ are discarded (the full expansion where are all bonds are included is $\mathbf{C} = diag([5,4,1])$). The user of Winterface can determine whether this accuracy is sufficient or not for his application. With the approximation presented in Fig. \ref{sec:interface:fig5}, the orthorhombic cell used in the quantum transport simulation is $20/6 \approx 3.33$ times smaller than if all interactions were included, which is advantageous from a computational viewpoint.
\section[Approximate Upscaling]{Approximate Upscaling Technique and Local Bandstructures}
\label{sec:approx}

In this Section an advanced simulation example will be presented. It requires additional work as compared to previous cases, as its device structures must be created manually. The interfacing scheme introduced in Section \ref{sec:interface} is not directly applicable here. Nevertheless, valid OMEN inputs can be produced. The purpose of this examples is to explore the limits of the proposed upscaling approach. Since a poorly converged wannierization is not compatible with the concepts presented here, it is assumed that the Wannier functions are maximally localized and the imaginary parts of the interactions insignificant. \\

The upscaling procedure explained in Section \ref{chap:hctor} relies on exact bonds in the sense that both the starting index i and the target index j, as well as the bond vector $\mathbf{b}_{ij}$ are provided. The target index is not strictly required, as a starting point and a bond vector are sufficient to extract an interaction matrix for the bond in question. The only criterion that should not be violated under any circumstances is that the interaction matrix along this bond has the correct dimensions. Otherwise, the Hamiltonian matrices cannot be properly generated. From this point of view, all positions with the same number of Wannier functions matched to them are potential valid targets when searching for interactions along bonds. This enables the creation of Hamiltonian matrices that do not correspond to exact super cells of the initial primitive cell simulated in DFT. In this way, it is possible to construct complex structures for quantum transport and to investigate their local properties such as the bandstructure of a well-specified region. \\

\begin{figure}
 \centering
 \includegraphics[width=10cm]{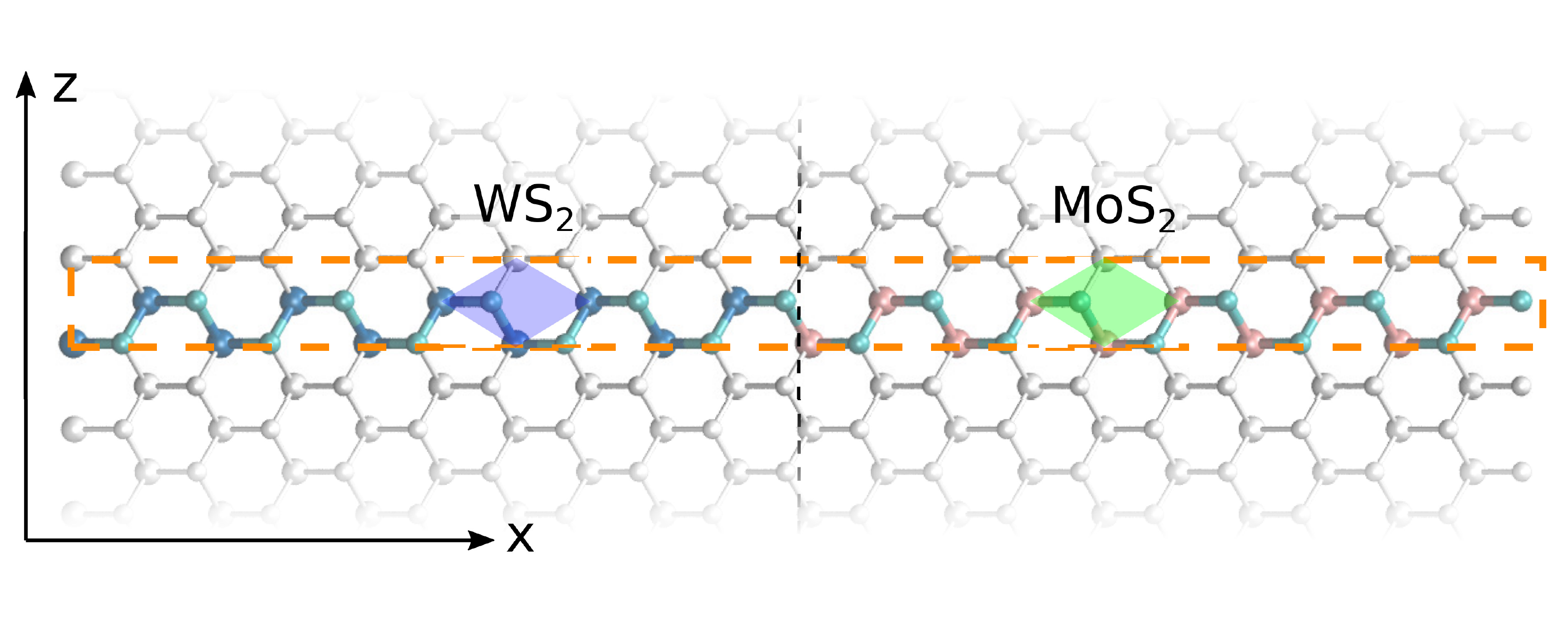}
 \caption{Top view of a MoS\tsc{2}-WS\tsc{2} heterostructure, where both materials are placed next to each other. Due to the applied periodicity along the x-axis in DFT, the interface region appears in the middle of the structure, as well as on both sides of the unit cell delimited by the dashed orange rectangle. The small blue rectangle corresponds to a primitive unit cell of WS\tsc{2}, the green one to a primitive unit cell of MoS\tsc{2}. Each of these two unit cells is only marginally affected by the presence of the other material, provided that the distance from the interface region is large enough. If wannierized, their Hamiltonian entries can be expected to match those of an isolated layer, up to a rigid energy shift.}\label{sec:approx:fig1}
\end{figure}

As an illustration, we again consider a heterostructure composed of MoS\tsc{2} and WS\tsc{2}, this time not placed on top of each other, but next to each other, as shown in Fig. \ref{sec:approx:fig1}

\begin{figure}
 \centering
 \includegraphics[width=9cm]{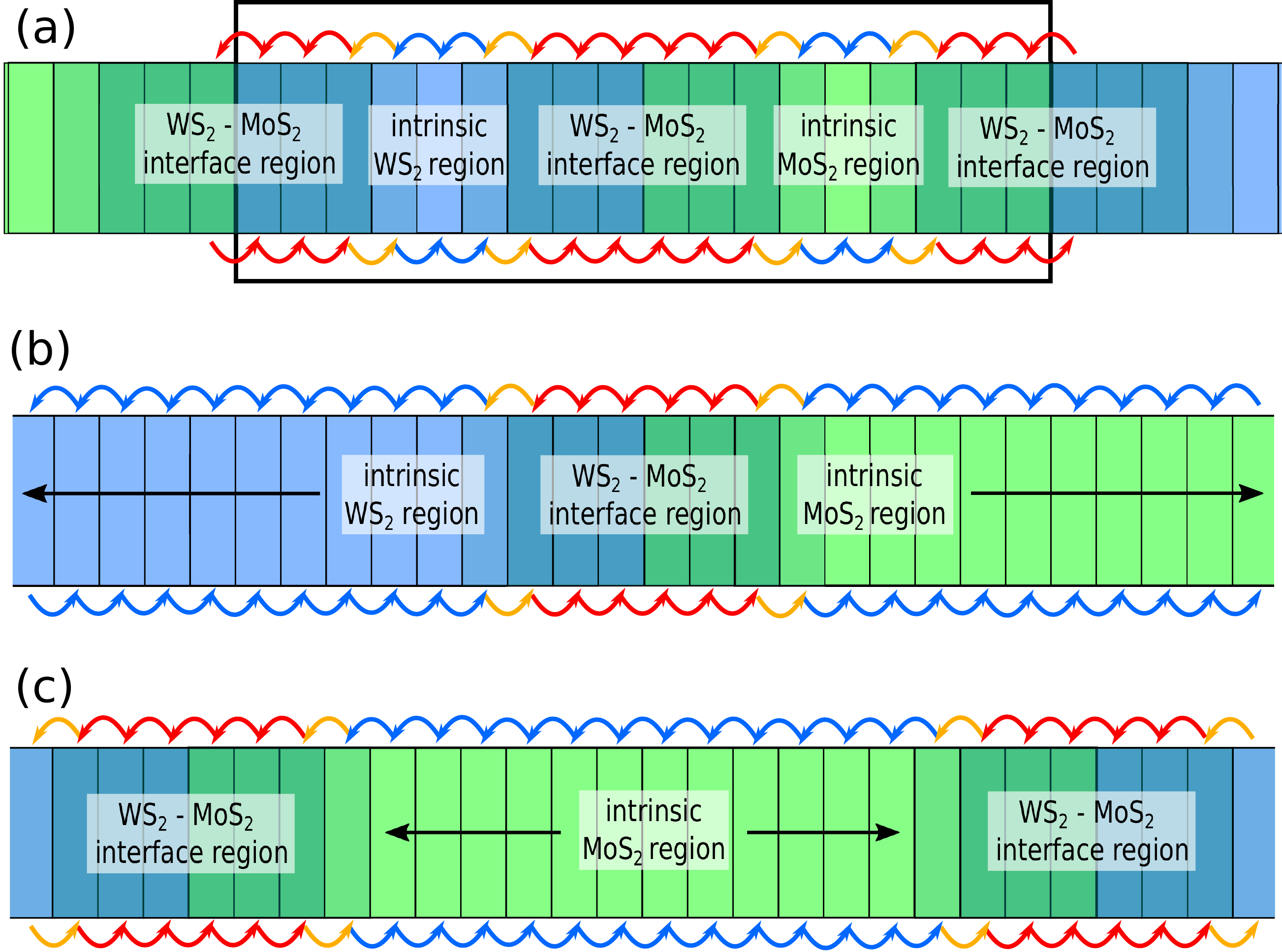}
 \caption{Examples of device structures that can be constructed after wannierizing the unit cell in Fig. \ref{sec:approx:fig1}. (a) WS\tsc{2}-MoS\tsc{2} super-lattice. The black frame represents the unit cells in \ref{sec:approx:fig1}. The red arrow refer to interactions close to the WS\tsc{2}-MoS\tsc{2} interface. The blue arrows represent interactions between two cells of pure WS\tsc{2} or MoS\tsc{2}. Finally, the orange arrows connect the pure material with the interface region. (b) Structure generated by repeating the subcells  in the intrinsic regions on both sides, leaving one interface region in the middle. (c) Structure generated by repeating subcells in the intrinsic region of MoS\tsc{2} with an interface region to WS\tsc{2} on both sides.}\label{sec:approx:fig2}
\end{figure}

From the original unit cell in Fig. \ref{sec:approx:fig1}, through repetition of certain subsections, larger structures can be generated, with different configurations. For example, the intrinsic regions on both sides of the interface can be made longer. Or one material can be sandwiched between two extensions of the other. To realize such structures, the method and approximations presented in Fig. \ref{sec:approx:fig2} must be followed.

The challenge with atomic arrangements such as the ones in Fig. \ref{sec:approx:fig2} resides in the association of the correct Wannier data with each bond. To ensure a proper Hamiltonian construction process, the matrix filling algorithm described in Section \ref{chap:hctor} must be modified when searching for an interaction matrix with indices and vector $\{i,j,\mathbf{b}_{ij}\}$:

\begin{itemize}
 \item If no exact interaction matrix can be found, determine all indices j\tsc{n} where the number of matched Wannier functions is equal to that at index j. For each such j\tsc{n}, identify the interaction matrix corresponding to $\{i,j_n,\mathbf{b}_{ij}\}$.
 \item Repeat the previous steps for the reverse bond $\{j,i,\mathbf{b}_{ji}\}$.
 \item The final interaction matrix $h(i,j,\mathbf{b}_{ij})$ is equal to the average $[ h(i,j_n,\mathbf{b}_{ij}) + h(j,i_n,\mathbf{b}_{ji})^\dagger ]/2$, thus ensuring self-adjointedness of the resulting Hamiltonian.
\end{itemize}

A few points should be further considered:

\begin{itemize}
 \item This scheme is not guaranteed to work as it can produce meaningless matrices if the coupling blocks are not chosen carefully. It is up to the user to make sure that the two small unit cells in Fig. \ref{sec:approx:fig1} extend far enough from the interface region so that the intrinsic properties of the targeted material is correctly captured in at least one subsection.
 \item The number of queries per bond is increased dramatically, which slows down the whole algorithm. As the time spent in Winterface is typically only a fraction of the overall time to simulate quantum transport, this is a minor issue.
 \item We must allow for some spatial tolerance when matching the supplied bond vector to those in the interaction data. For the example presented in Fig. \ref{sec:approx:fig2} this is minimal. For a different structure with distorted bonds at the interface, the situation might be more complicated.
 \item Because of these restrictions, it is recommended to manually check the produced Hamiltonian matrices before plugging them into a quantum transport simulator.
\end{itemize}

\begin{figure}
 \centering
 \includegraphics[width=\textwidth]{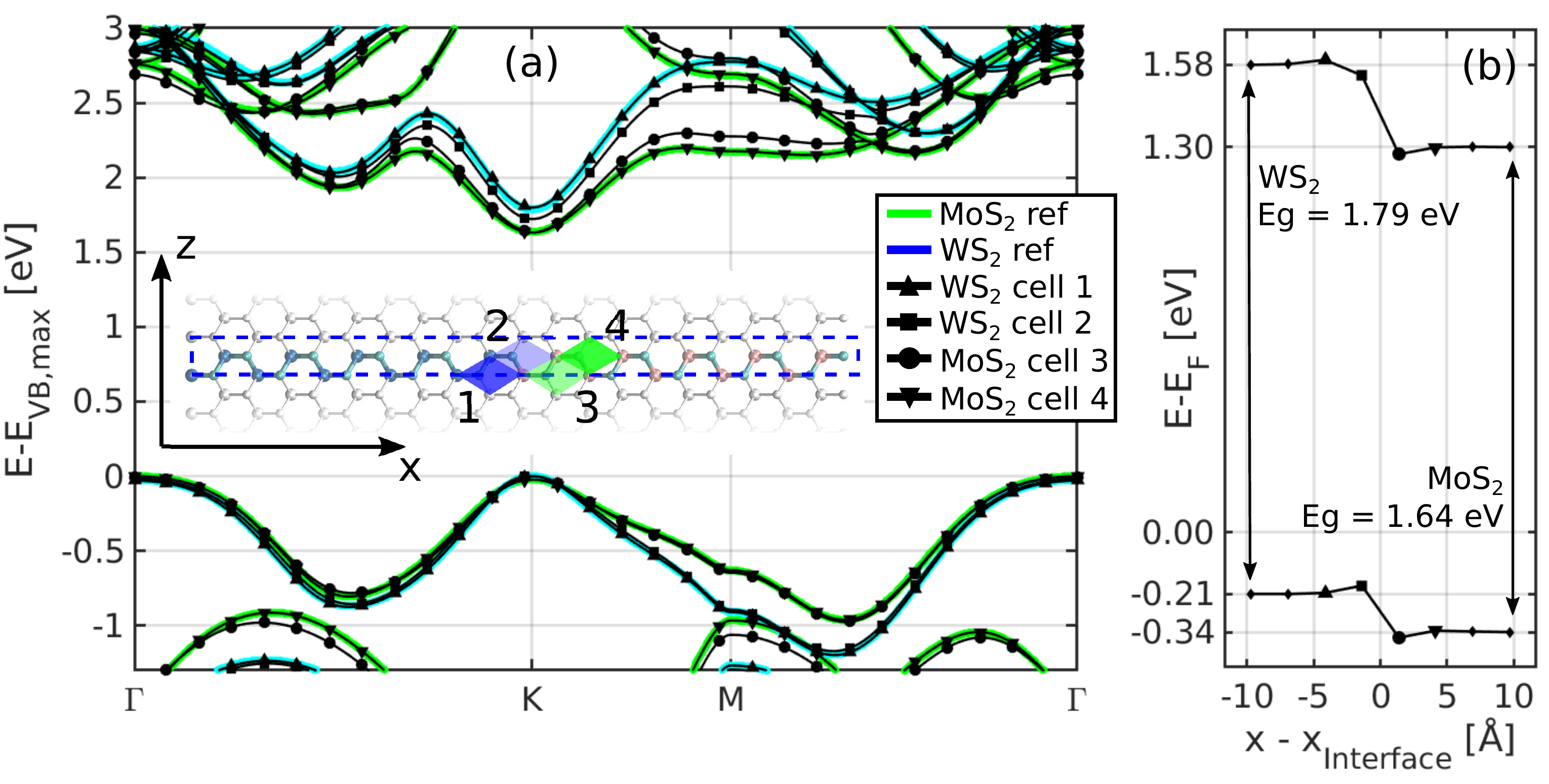}
 \caption{(a) Comparison of local bandstructures coming from unit cells extracted at different positions in Fig. \ref{sec:approx:fig1}. The results of pure MoS\tsc{2} and WS\tsc{2} are given as references, green for MoS\tsc{2} and blue for WS\tsc{2}. The top of the valence band is set to $E=0$ in all cases. The black lines refer to the bandstructures of the unit cells indicated by the shaded regions in the central unit cell. (b) Conduction (upper line) and valence (lower line) band edge as a function of the distance from the WS\tsc{2}-MoS\tsc{2} interface.}\label{sec:approx:fig3}
\end{figure}

Due to the localized nature of Wannier functions, each part of the structure only interacts with its immediate surroundings. This fact can be exploited with the modified scheme described above to generate a set of approximate Hamiltonian matrices for all subsections of the full structure. Since the proposed scheme ensures the spacial inversion symmetry of the Hamiltonian, the resulting matrices can be used in bandstructure calculations. This can be very useful when characterizing the local properties of a large structure such as determining whether the small unit cells in Fig. \ref{sec:approx:fig1} exhibit the same properties as pure MoS\tsc{2} or WS\tsc{2}. A comparison of local bandstructures computed for different small unit cells around the interface region in Fig. \ref{sec:approx:fig1} is shown in Fig \ref{sec:approx:fig3}. The bandstructures of pure MoS\tsc{2} and WS\tsc{2} are also given as references. It can be seen that the band gaps are accurately reproduced far away from the interface region, whereas a relatively smooth transition occurs at the interface. \\

\section{Results}
\label{chap:results}

To demonstrate that the Hamiltonian matrices generated using the ideas of Sections \ref{sec:wbh} and \ref{chap:hctor} are viable for quantum transport calculations, the transmission function $T(E,k_z)$ as introduced in Eq. \eqref{sec:negf:transm} was computed for various examples. With this quantity. the ballistic current defined in Eq. \eqref{sec:negf:current} flowing through them can be computed.

For devices where all unit cells are equivalent, the transmission function simply counts the number of available propagating modes at each energy $E$ and momentum $k_z$. For inhomogeneous devices, the picture is more complex. \\

\begin{figure}
 \centering
 \includegraphics[width=11cm]{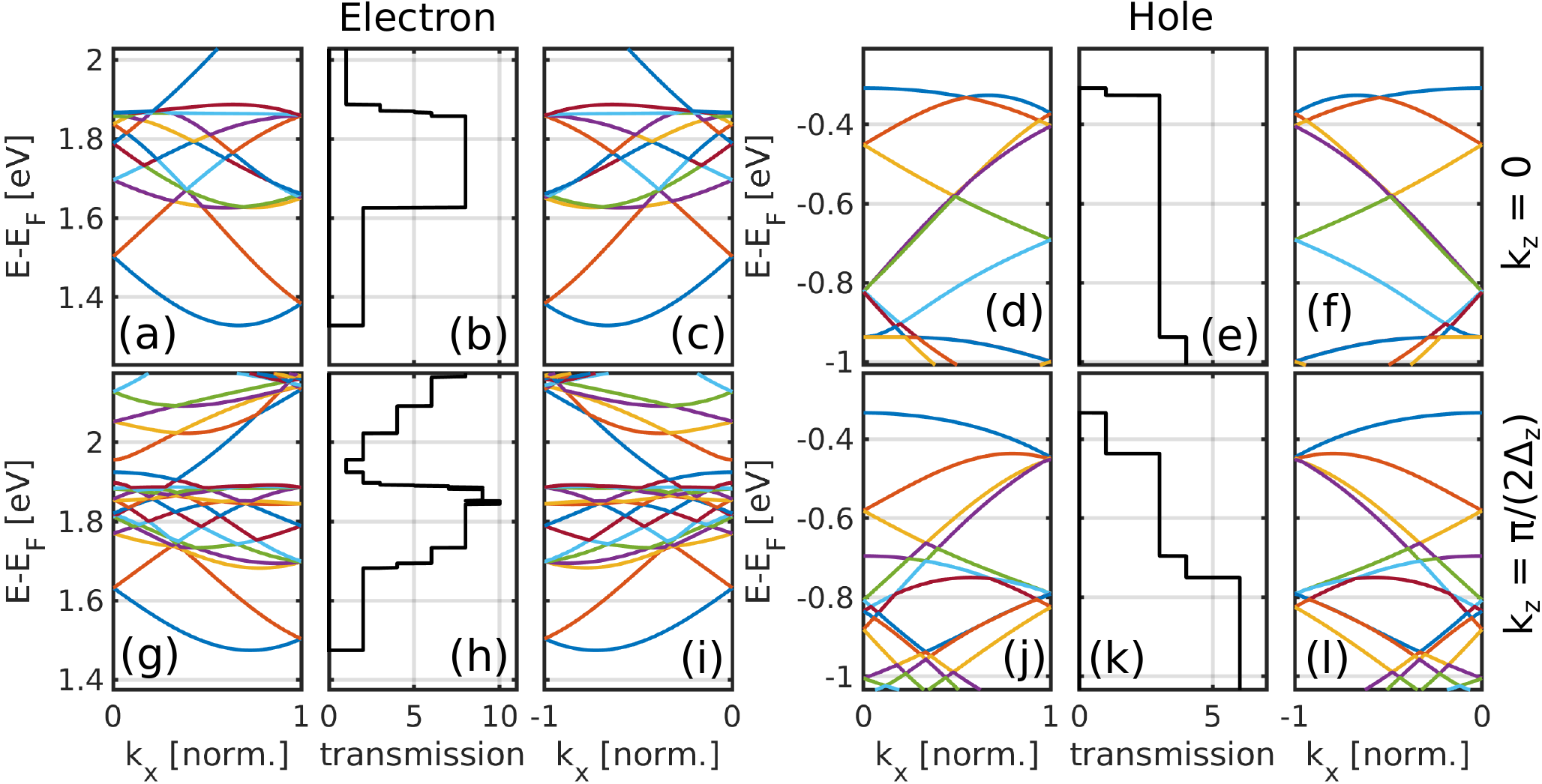}
 \caption{Transmission function of a MoS\tsc{2} monolayer with transport along the $[100]$-direction for $k_z = 0$ (top row) and $k_z = \pi/(2\Delta_z)$ (bottom row), for both electrons (left half) and holes (right half). (a) Conduction bands in the left contact. (b) Electron transmission function. (c) Conduction bands in the right contact. (d) Valence bands in the left contact. (e) Hole transmission function. (f) Valence bands in the right contact. (g-l) Same as (a-f) but for $k_z = \pi/(2\Delta_z)$. Since all unit cells composing the structure are identical and no bias is applied, $T(E,k_z)$ counts the number of modes propagating from one contact to the other at an energy $E$ and momentum $k_z$. The fact that this property is satisfied indicates that the upscaling procedure works as intended.}\label{sec:results:fig1}
\end{figure}

\begin{figure}
 \centering
 \includegraphics[width=11cm]{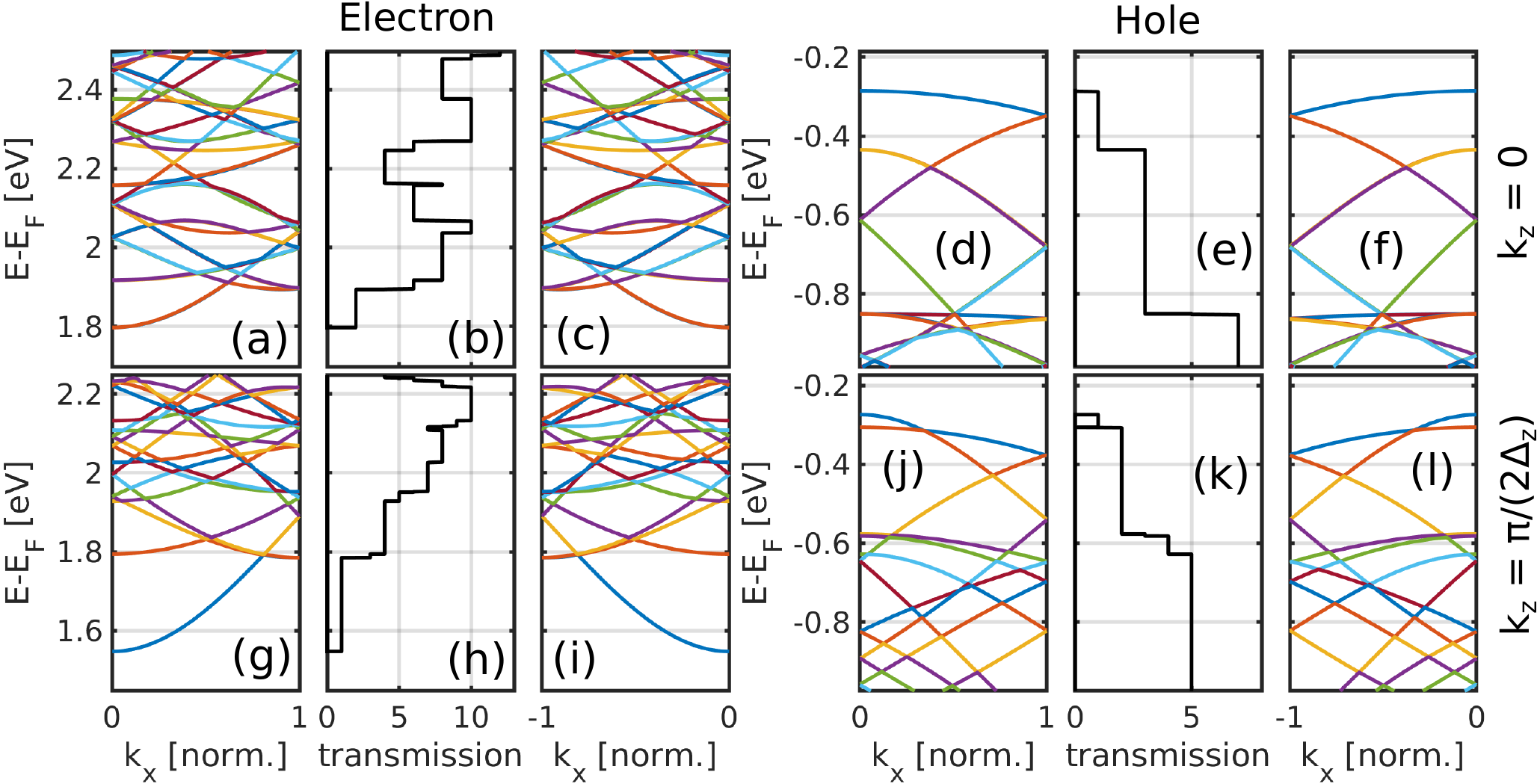}
 \caption{Same as Fig. \ref{sec:results:fig1}, but for a WS\tsc{2} monolayer with transport along the $[001]$-direction.}\label{sec:results:fig2}
\end{figure}

\begin{figure}
 \centering
 \includegraphics[width=11cm]{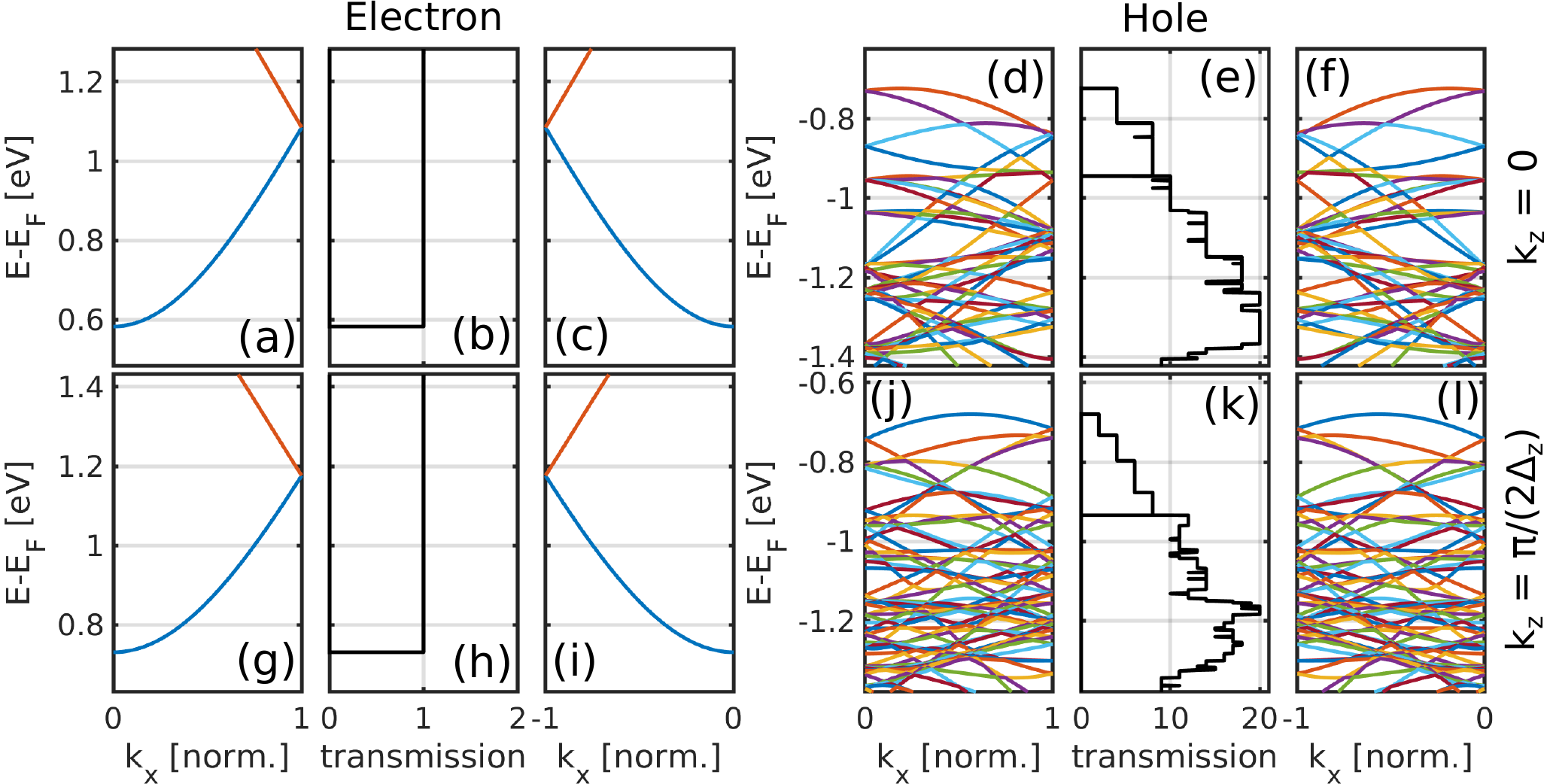}
 \caption{Same as Fig. \ref{sec:results:fig1} and Fig. \ref{sec:results:fig2}, but for BiIO with transport along the $[101]$ crystal axis.}\label{sec:results:fig3}
\end{figure}

\begin{figure}
 \centering
 \includegraphics[width=11cm]{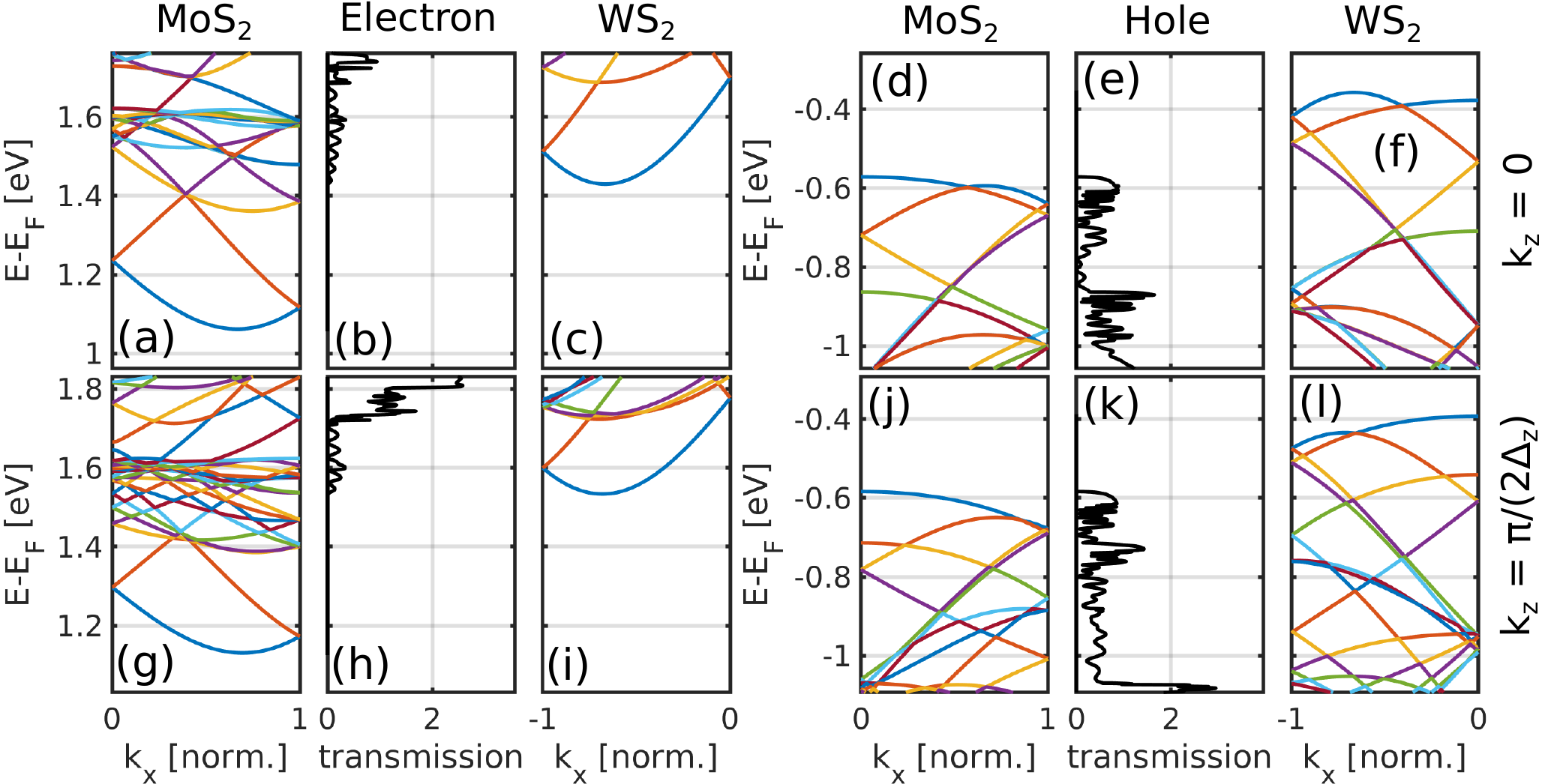}
 \caption{Same as Figs. \ref{sec:results:fig1} to \ref{sec:results:fig3}, but for a MoS\tsc{2}(top)-WS\tsc{2}(bottom) van der Waals heterostructure with three seperate regions, one made of pure MoS\tsc{2} on the left, an overlap area in the middle, and a pure WS\tsc{2} monolayer on the right.}\label{sec:results:fig4}
\end{figure}

\begin{figure}
 \centering
 \includegraphics[width=11cm]{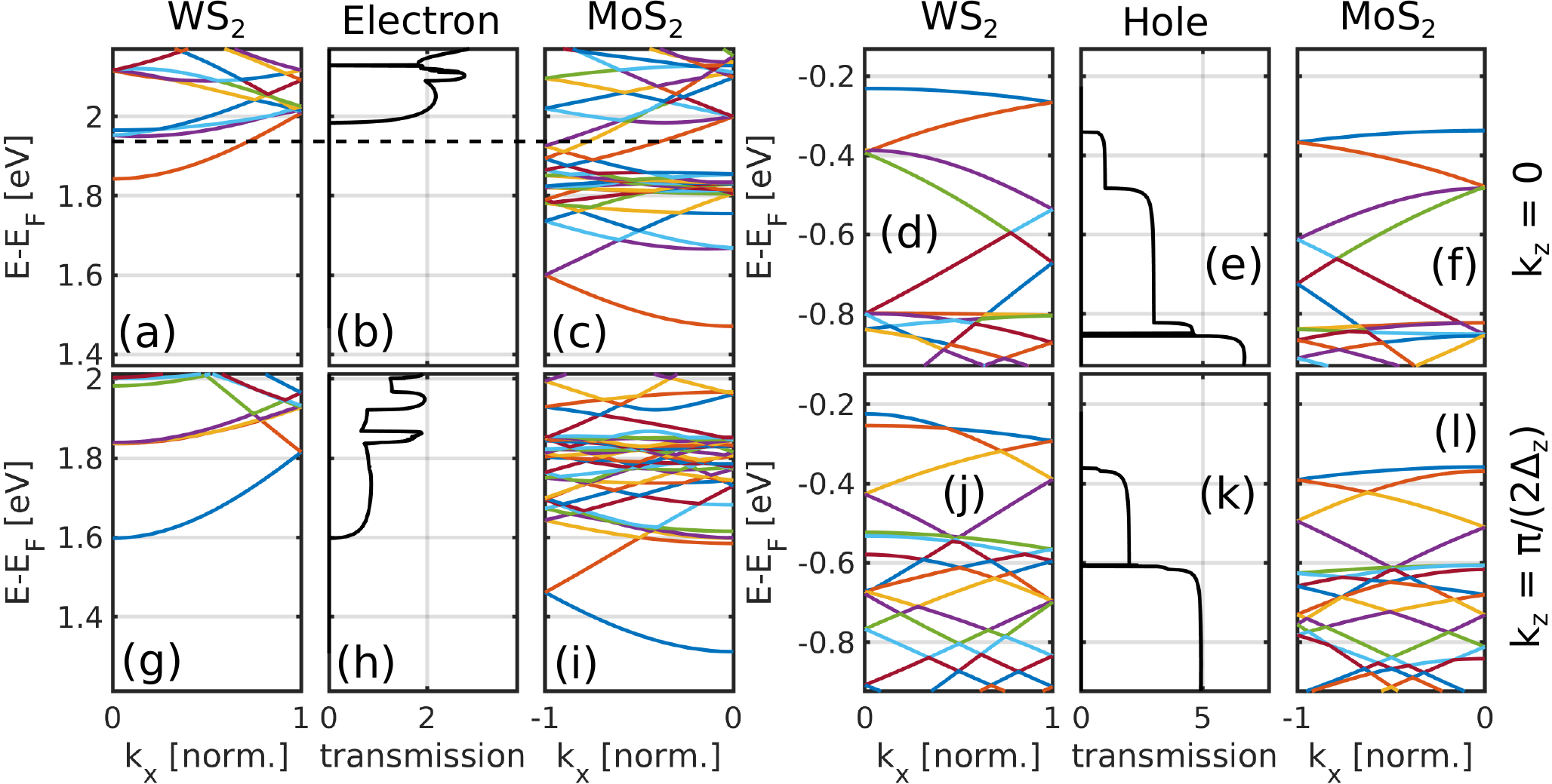}
 \caption{Same as Figs. \ref{sec:results:fig1} to \ref{sec:results:fig4}, but for a lateral MoS\tsc{2}-WS\tsc{2} heterostructure.}\label{sec:results:fig5}
\end{figure}

Modes are injected in to a given device from its contacts. If we consider the left contact, the modes of interest are those with $dE/dk > 0$ ($dE/dk < 0$) for electrons (holes), where $E(k)$ is the contact bandstructure. When plotting $E(k)$, only half of the Brillouin zone will be shown due to the symmetry with respect to $k = 0$. Note that the wave vectors are normalized with respect to their maximum value $k_{max} = \pi/\Delta_x$, where $\Delta_x$ is the length of a unit cell along the transport direction. In each example the transmission function will be given for $k_z = 0$ and $k_z = \pi/(2\Delta_z)$. The initial DFT simulations were all performed within the generalized gradient approximation of Perdew-Burke-Ernzerhof (PBE) \cite{pbe_paper}, except BiIO where van der Waals interactions were included (optB88-vdW) \cite{vdw1,vdw2}. The following examples were simulated:

\begin{itemize}
 \item A MoS\tsc{2} monolayer with transport along the $[100]$ direction in Fig. \ref{sec:results:fig1}. The device length is 40.8nm and it is made of 32 (identical) unit cells of 48 atoms each.
 \item A WS\tsc{2} monolayer with transport along the $[001]$ direction in Fig. \ref{sec:results:fig2}. The device length is 41.4nm, which is equal to 25 identical unit cells of 72 atoms each.
 \item A BiIO monolayer with transport along the $[101]$ direction in Fig. \ref{sec:results:fig3}. The device length is set to 40.9nm and is composed of 24 unit cells of 108 atoms each, all identical.
 \item A stack of MoS\tsc{2} and WS\tsc{2} monolayers, as presented in Fig. \ref{sec:hctor:fig5}, in Fig. \ref{sec:results:fig4}. The full device length is 40.8nm with an overlap region of 13.6nm. The whole structure is 32 unit cells in total, of 144 atoms in the overlap region and 72 on the left and right extensions.
 \item A monolayer of MoS\tsc{2} and WS\tsc{2} with a lateral interface region (see Fig. \ref{sec:approx:fig1}) in Fig. \ref{sec:results:fig5}. The device length is 39.8nm, which is equivalent to 18 unit cells of 96 atoms each. The first 9 unit cells are made of WS\tsc{2}, the 10th holds the interface region, whereas the last 8 are made of MoS\tsc{2}. Note that the MoS\tsc{2} and WS\tsc{2} layers are rotated such that transport is aligned with the $[001]$ direction when compared to pure monolayers of either flavor.
\end{itemize}

\newpage

\newpage

\clearpage

\newpage

Because of the conduction and valence band offsets between MoS\tsc{2} and WS\tsc{2}, the electron and hole transmission functions exhibit a more complex behavior than in the case of homogeneous materials. First, the transmission can only be different from 0 if the same band is available on both contacts. Secondly, if m (n) bands are injected from the left (right) contacts, because of quantum mechanical reflection, $T(E,k_z) \leq min(m,n)$. Besides these key features, it is difficult to interpret the transmission function results in Fig. \ref{sec:results:fig4}. It should however be noticed that the validity of the upscaling method can be verified in a different way: instead of calculating the transmission function from the left to the right contact, it can be evaluated between two adjacent cells along the transport direction. Due to current continuity, the transmission from cell $i$ to $i+1$ must be the same as between $j$ and $j+1$, where $i \neq j$. This property was verified for all results shown in this Section.

\newpage

In the lateral MoS\tsc{2}-WS\tsc{2} heterostructure in Fig. \ref{sec:results:fig5}, a perculiarity occurs at $E-E_f \approx 1.9 eV$ and $k_z = 0$. It can be seen in sub-plots (a) and (c) that bands are available on both the left and right contacts in this case and in spite of that, the transmission function is equal to 0. This can be attributed to the fact that the lowest energy band in (c) has an energy width smaller than the conduction band offset between MoS\tsc{2} and WS\tsc{2}. As a consequence, a state injected from the left contact at $E-E_F = 1.9 eV$ does not find any band with the same symmetry properties in the right contact. It is therefore reflected back to its origin.
\clearpage
\section{Conclusion and Outlook}
\label{chap:conclusion}

A general technique for upscaling Hamiltonians in MLWF representation of small unit cells up to the device level was established. The first step consists of a reformulation of raw MLWF Hamiltonian data into a representation in terms of interactions along bonds. This makes the second step, the actual upscaling, much more transparent and enables more complex, approximate, upscaling techniques, as well as investigations of local properties, e.g. bandstructures. The latter can be useful when generating complex geometries. \\

Even though the focus of this work is on 2-D structures, the algorithms presented in Sections \ref{sec:wbh} and \ref{chap:hctor} work equally well and without modification of the code for 1-D and 3-D domains. Additionally, once the first step of generating Hamiltonian data in terms of bonds is completed, generating Hamiltonian matrices for whole devices, as discussed in Section \ref{chap:hctor}, is independent of the chosen basis set. An extension of Winterface to other localized bases is therefore possible, such as tight-binding coefficients or Gaussian type orbitals (GTO), as implemented, for example, in the CP2K \cite{cp2k_package} package. \\

Winterface does not impose any restrictions on the geometry of the considered devices. Difficulties may arise at the DFT and wannierization stages, where it must be decided how the desired features are best included in a single unit cell. A few options to handle more complex geometries are: (i) inclusion of one or both contacts in a larger unit cell, (ii) a larger unit cell with localized defects, or (iii) inclusion of the oxide and/or the substrate. In each of these cases the limiting factor is the physical modeling of lattice mismatches between two materials that are put together, the ionic relaxation at interfaces or defects and in general the computational burden involved. Nevertheless, complex structures could be wannierized and then upscaled using the concepts of Section \ref{sec:approx}, e.g. Ti-TiO\tsc{2}-MoS\tsc{2} contact geometries, as demonstrated in Ref. \cite{Szabo2019}. To a certain extent, Winterface is able to assist in the process of generating such complex structures, as it allows to determine the coupling strength between two materials. Moreover, with the concept of localized bandstructures introduced in Section \ref{sec:approx}, a rough estimate of the local properties of each material can be made.

\section{Acknowledgment}
\label{sec:acknowledgment}

This work was supported by the European Union 7th Framework Programme DEEPEN (Grant agreement No. 604416) and by the MARVEL National Centre of Competence in Research of the Swiss National Science Foundation. We would also like to acknowledge CSCS for awarding us access to Piz Daint under Project s876.
\newpage

\begin{appendices}

\section{Unit Cell Example: MoS\tsc{2} monolayer}
\label{app:mos2_ex}

\begin{figure}
 \centering
 \includegraphics[width=\textwidth]{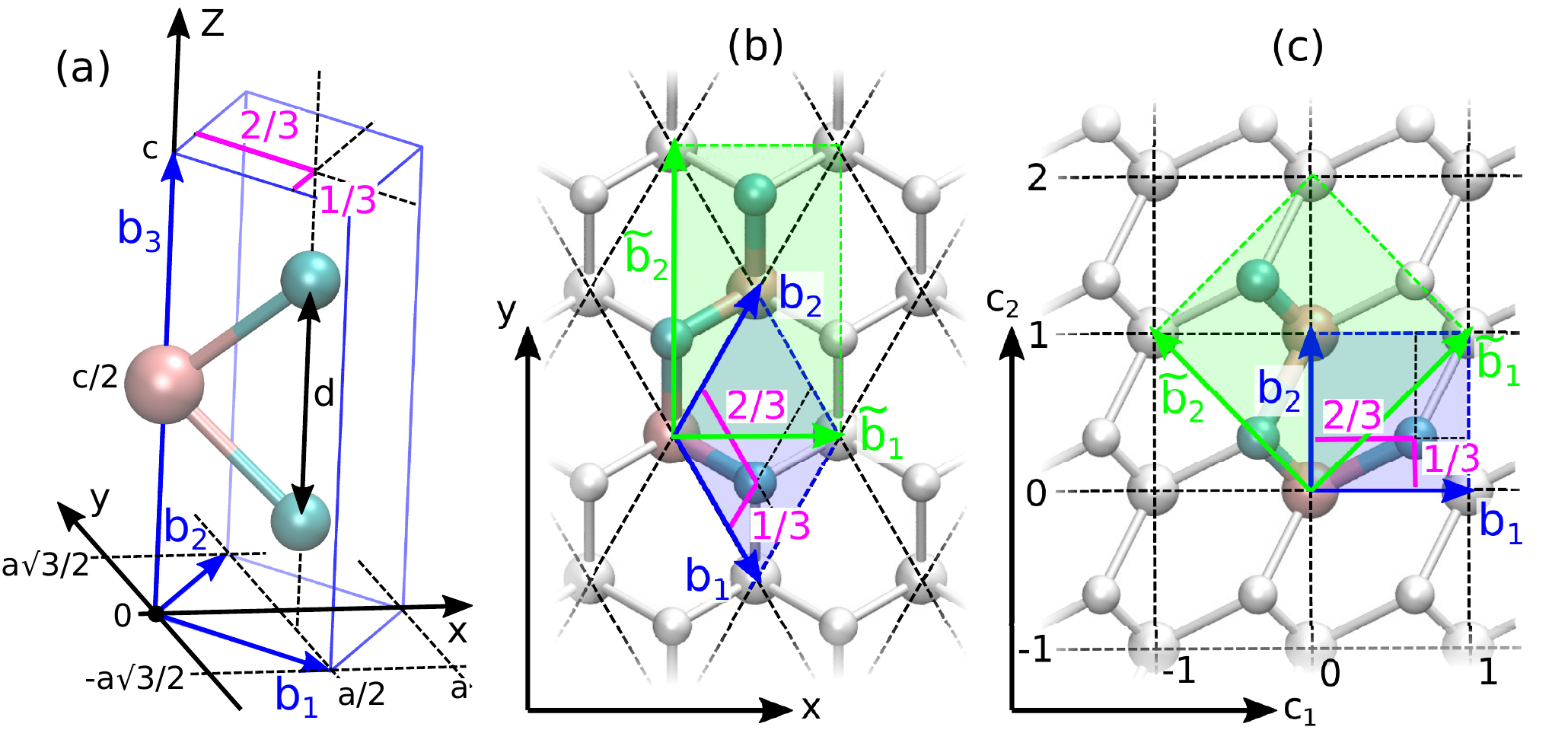}
 \caption{Example of a unit cell for a MoS\tsc{2} monolayer structure. (a) Isometric view of the primitive unit cell, where a is the lattice constant, c the interlayer distance (here displayed much shorter than in reality) and d the vertical distance between the two sulfur atoms (in cyan). The blue arrows are the three basis vectors of the unit cell and the pink fractions in the top plane of the parallelepiped correspond to the atomic positions in basis \textbf{B}. (b) Top view of the situation in (a). The green arrows are the basis vectors of an orthorhombic super cell. The colored atoms are either inside the blue or the green rectangle as they belong to either unit cell. The dashed lines delimit the image unit cells. (c) Same situation as in (b), but viewed from the primitive basis \textbf{B}. The axis c\tsc{1}(c\tsc{2}) represent the coordinates in the vectors \textbf{b}\tsc{1}(\textbf{b}\tsc{2}). From this perspective, the lattice exists on a grid of integers.}\label{sec:lattice:fig1}
\end{figure}

As an explicit example of a unit cell, a monolayer structure of MoS\tsc{2} is considered. The hexagonal unit cell for this structure can be defined using the basis \cite{bv}

\begin{equation}\label{sec:lattice:eq1}
\mathbf{B} = [\mathbf{b}_1,\mathbf{b}_2,\mathbf{b}_3] = 
 \begin{bmatrix}
  \frac{a}{2} & \frac{a}{2} & 0 \\
  \frac{-a\sqrt{3}}{2} & \frac{a\sqrt{3}}{2} & 0 \\
  0 & 0 & c \\
 \end{bmatrix},
\end{equation}

where a is the lattice constant and c the interlayer distance (for a monolayer $c \gg a$). The matrix \textbf{A} containing the atomic positions is then

\begin{equation}\label{sec:lattice:eq2}
 \mathbf{A} = [ \mathbf{p}_1^{(1)}, \mathbf{p}_2^{(2)}, \mathbf{p}_3^{(2)} ] = 
 \begin{bmatrix}
  0 & \frac{2}{3} & \frac{2}{3} \\
  0 & \frac{1}{3} & \frac{1}{3} \\
  \frac{1}{2} & \frac{1}{2}-\frac{d}{2c} & \frac{1}{2}+\frac{d}{2c} \\
 \end{bmatrix},
\end{equation}

where d is the vertical distance between the two sulfur atoms. The molybdenum atom resides at the origin in the xy-plane and in the middle of the unit cell in the z-direction. Finally, the choice of \textbf{id} $=$ $\{$'Mo','S'$\}$ completes the data set. A graphical representation is given in Fig. \ref{sec:lattice:fig1}.

\section{Metric in Periodic Space}
\label{app:metric}

When working with unit cells, it is important that the periodic space described in their terms qualifies as a metric space. The dot product of two positions \textbf{p} and \textbf{q} in basis \textbf{B} is

\begin{equation}\label{sec:lattice:eq7}
 \langle \mathbf{p} , \mathbf{q} \rangle_{B} := (\mathbf{B} \cdot \mathbf{p})^\top \cdot (\mathbf{B} \cdot \mathbf{q}) = \mathbf{p}^\top \cdot \mathbf{B}^\top\mathbf{B} \cdot \mathbf{q},
\end{equation}

implying the norm $\lVert \mathbf{p} \rVert_{B} = \sqrt{\langle \mathbf{p} , \mathbf{p} \rangle_{B}}$. A metric operating in periodic space must consider any position in the lattice. All equivalent images must therefore be taken into account to find the distance between two positions, defined as the length of the closest possible connection between them. Such a metric may then be defined as

\begin{equation}\label{sec:lattice:eq8}
 d(\mathbf{p},\mathbf{q})_{B} = \min_{\mathbf{R},\mathbf{R}^{'} \in \mathds{Z}^N} \Big \lVert (\mathbf{p} + \mathbf{R}) - (\mathbf{q} + \mathbf{R}^{'}) \Big \rVert_{B} = \min_{\mathbf{R} \in \mathds{Z}^N} \Big \lVert (\mathbf{p} - \mathbf{q}) + \mathbf{R} \Big \rVert_{B}.
\end{equation}

The range of \textbf{R} vectors to search can be restricted by forcing both positions into the same unit cell. In this case the shortest possible connection can be found either in the home cell or among the next-neighbor image cells. Since $\mathbf{p}-\mathbf{q} \in (-1,1)^N$ if $\mathbf{p},\mathbf{q} \in [0,1)^N$, the range of \textbf{R} vectors can be further restricted to the positive sector $\{0,1\}^N$ by taking the absolute value over $\mathbf{p}-\mathbf{q}$. The metric is thus defined as

\begin{equation}\label{sec:lattice:eq9}
 d(\mathbf{p},\mathbf{q})_{B} := \min_{\mathbf{R} \in \{0,1\}^N} \Big \lVert | \tilde{\mathbf{p}} - \tilde{\mathbf{q}} | - \mathbf{R} \Big \rVert_{B}, \quad \tilde{\mathbf{p}} = mod(\mathbf{p},1).
\end{equation}

As the product $\mathbf{B}^\top \mathbf{B}$ is positive definite, $d(\mathbf{p},\mathbf{q})_{B} \geq 0$ for all $\mathbf{p}$,$\mathbf{q}$. Furthermore, $d(\mathbf{p},\mathbf{q})_{B} = d(\mathbf{q},\mathbf{p})_{B}$ from Eq. \eqref{sec:lattice:eq9}. What is left to demonstrate is the triangle inequality:

\begin{equation}\label{sec:lattice:eq10}
\begin{split}
d(\mathbf{p},\mathbf{q}) + d(\mathbf{q},\mathbf{r}) & = \min_{\mathbf{R} \in \{0,1\}^N} \Big \lVert |\tilde{\mathbf{p}}-\tilde{\mathbf{q}}| - \mathbf{R} \Big \lVert_{B} + \min_{\mathbf{R} \in \{0,1\}^N} \Big \lVert |\tilde{\mathbf{q}}-\tilde{\mathbf{r}}| - \mathbf{R} \Big \lVert_{B}, \\
& = \Big \lVert |\tilde{\mathbf{p}}-\tilde{\mathbf{q}}| - \mathbf{R}_{min}^{pq} \Big \lVert_{B} + \Big \lVert |\tilde{\mathbf{q}}-\tilde{\mathbf{r}}| - \mathbf{R}_{min}^{qr} \Big \lVert_{B}, \\
& \geq \Big \lVert |\tilde{\mathbf{p}}-\tilde{\mathbf{q}}| + |\tilde{\mathbf{q}}-\tilde{\mathbf{r}}| - (\mathbf{R}_{min}^{pq} + \mathbf{R}_{min}^{qr}) \Big \lVert_{B}, \\
& \geq \Big \lVert |\tilde{\mathbf{p}}-\tilde{\mathbf{r}}| - (\mathbf{R}_{min}^{pq} + \mathbf{R}_{min}^{qr}) \Big \lVert_{B}, \\
& \geq \min_{\mathbf{R} \in \{0,1\}^N} \Big \lVert |\tilde{\mathbf{p}}-\tilde{\mathbf{r}}| - \mathbf{R} \Big \lVert_{B}, \\
& = d(\mathbf{p},\mathbf{r}). \\
\end{split}
\end{equation}

\section{Wannierizations exhibiting different initial distributions of Wannier centers for a monolayer of MoS\tsc{2}}
\label{app:mos2_relwannier}

\begin{figure}
 \centering
 \includegraphics[width=8cm]{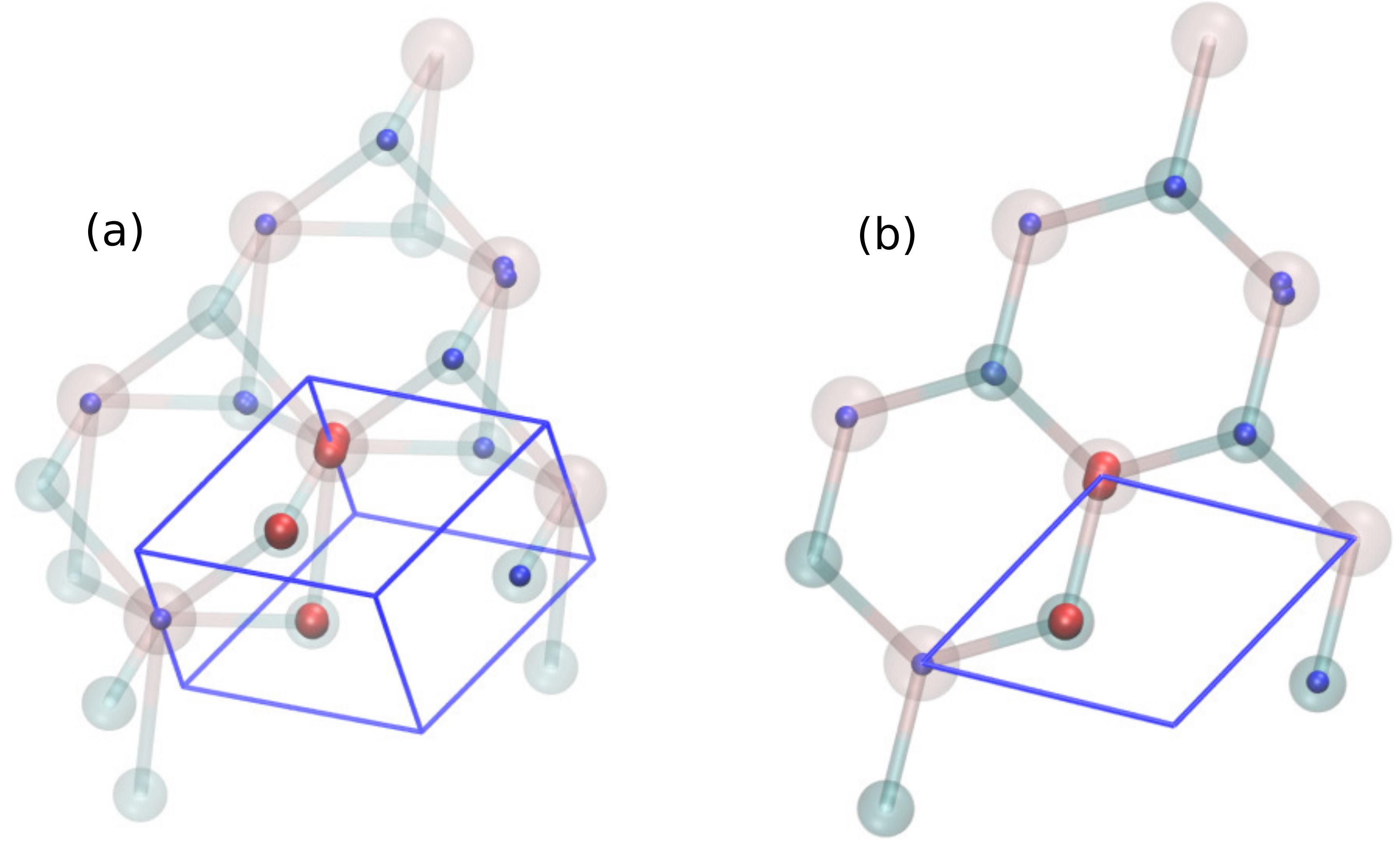}
 \caption{Small section of a MoS\tsc{2} monolayer structure. The primitive unit cell simulated in VASP holds one molybdenum (transparent red spheres) and two sulfur (transparent blue spheres) atoms. The small red marbles represent Wannier centers converged onto atomic positions within the parallelepiped delimited by the blue lines, while the blue marbles represent Wannier centers resulting from a different set of orbital projections. Both wannierizations are equally valid and produce 5 Wannier centers on molybdenum and 3 on sulfur. However, in the second case the centers are spread out among images in neighboring cells. (a) Isometric view. (b) View from the top.}\label{sec:wbh:fig6}
\end{figure}

To illustrate the treatment of two equally valid Wannierizations of the same structure, the MoS\tsc{2} monolayer structure introduced in Appendix \ref{app:mos2_ex} will serve as test bed. It was simulated in VASP within the generalized gradient approximation of Perdew-Burke-Ernzerhof (PBE) \cite{pbe_paper} using a lattice constant of 3.18\si{\angstrom}, a 400-eV plane-wave cutoff energy, a 21x21x1 Monkhorst-Pack k-point grid, and a 45\si{\angstrom} out-of-plane vacuum separation between the structure and its closest images. Due to the periodicity of the lattice, Wannier centers can be spread out among images of the same atomic positions, as can be observed Fig. \ref{sec:wbh:fig6}.

The wannierization corresponding to the red marbles in Fig. \ref{sec:wbh:fig6}, resulted from placing orbital projections directly onto atomic positions. The one corresponding to the blue marbles in Fig. \ref{sec:wbh:fig6} was arrived at by placing orbital projections on images of the atomic positions. For both cases, we are able to arrive at qualitatively equivalent results by taking the relative positioning of Wannier centers in to account. For instance, the following matrices describe the self-interactions of sulfur in each case:

\[
 H_{red,S}=
 \begin{bmatrix}
-5.4896 & 0.0000 & -0.0025\\
 0.0000 & -5.4920 & 0\\
-0.0025 & 0 & -5.6947\\
 \end{bmatrix}
\]

\[
 H_{blue,S}=
 \begin{bmatrix}
-5.5399+0.0000i &  0.0318-0.0076i & 0.0836-0.0032i\\
 0.0318+0.0076i & -5.5116+0.0000i & -0.0539-0.0104i\\
 0.0836+0.0032i & -0.0539+0.0104i & -5.6201+0.0000i\\
 \end{bmatrix}.
\]

Due to the different wannierizations, the matrices are not equivalent, but when performing a singular value decomposition to characterize the interactions, it becomes clear that they closely resemble each other for both cases:

\begin{center}
\begin{tabular}{ l | c | c }\label{sec:wbh:tab1}
 & $H_{red,S}$ & $H_{blue,S}$\\
 \hline
 $s_1$ & 5.6948 & 5.6951\\
 $s_2$ & 5.4920 & 5.4902\\
 $s_3$ & 5.4896 & 5.4865\\
\end{tabular}
\end{center}

The slight differences can be explained by the fact that the blue wannierization is of lower quality than the red one, as the Hamiltonian matrices exhibit non-vanishing imaginary parts. The total spreads for both cases are very similar however and as such, the blue wannierization can be considered as localized as the red one.

\section{Automatic Detection of Basis Expansions}
\label{app:expansions}

An algorithm was developed, allowing for automatic detection of expansion matrices \textbf{C} of a primitive basis \textbf{B}, given a template \textbf{T} for the desired basis $\tilde{\mathbf{B}}$, such that $\tilde{\mathbf{B}} = diag([\lambda_i,...,\lambda_N]) \cdot \mathbf{T}$ with $\lambda_i \in \mathds{R}_{>0}$ and $vol(\tilde{\mathbf{B}})$ is minimal. Hence, the task is to find the expansion \textbf{C} leading to the smallest possible super cell, such that the basis vectors of $\tilde{\mathbf{B}}$ are equivalent up to a positive constant to the template basis vectors in \textbf{T}. \\

In order for $\tilde{\mathbf{B}}$ to be a valid basis for the lattice initially expressed in \textbf{B}, the directions given by \textbf{T} must correspond to crystallographic directions, in which case \textbf{T} in basis \textbf{B} takes the form

\begin{equation}\label{sec:lattice:eq11}
 \mathbf{B}^{-1} \cdot \mathbf{T} = [ \lambda_1 \mathbf{q}_1, ... ,\lambda_N \mathbf{q}_N ],
\end{equation}

with $\lambda_i \in \mathbb{R}$ and $\mathbf{q}_i \in \mathbb{Q}^N$. The task of finding the smallest possible expansion is now equivalent to determining a factor $f_i$ for each column such that $f_i \cdot \lambda_i\mathbf{q}_i \in \mathbb{Z}^N$ and $\|f_i \cdot \lambda_i\mathbf{q}_i\|_2$ minimal. Since

\begin{equation}\label{sec:lattice:eq12}
 1 \geq \frac{|\lambda_i q_{ij} |}{\max_j |\lambda_i q_{ij} |} \in \mathbb{Q},
\end{equation}

it follows that

\begin{equation}
 \forall j \hspace{2mm} \exists \nu \in \mathbb{N} \quad \text{such that} \quad \nu \cdot \underbrace{\frac{\lambda_i q_{ij}}{\max_j|\lambda_i q_{ij}|}}_{p_{ij}} \in \mathbb{Z}.
\end{equation}

Determining the $\nu$ for each column $\mathbf{q}_i$, where $\| \mathbf{q}_i \|_2$ is minimal, corresponds to finding the least common multiple among the denominators of $p_{ij}$ for each i, over j. Since the denominators are not known, the implementation scans through all $\nu\geq1$ until either all $q_{ij}$ are integers (up to a small numerical tolerance), or an upper limit $\nu_{max}$ has been reached, i.e. the resulting basis would be so large as to render it useless. The $\nu_{max}$ might be exceeded only if the initial template \textbf{T} was poorly chosen, in which case Eq. \eqref{sec:lattice:eq11} is violated, or if the tolerance level was set too strict. A few points should be mentioned:

\begin{itemize}
 \item For this algorithm to be useful in real-world applications, the lattice must be aligned 'nicely' with the Cartesian axes. For example, to find the expansion from the primitive basis to the orthorhombic super cell in Fig. \ref{sec:lattice:fig1}, the template \textbf{T} can be set to the identity. However, if the basis vectors of the primitive basis were slightly rotated such that the Cartesian axes do not lie along crystallographic directions, the template would have to be rotated in the same way to arrive at the same result. In such a case, it is more practical to determine the correct expansion by other means, rather than attempting to detect the rotation.
 \item Small deviations from a 'nice' alignment can be compensated by increasing the tolerance level when detecting integers.
 \item For lattices that only slightly deviate from an exact symmetry as dictated by \textbf{T}, the tolerance level for detecting integers can be increased, upon which inexact expansions are automatically detected and compensated for by a deformation tensor.
\end{itemize}

\end{appendices}

\clearpage


\bibliography{mybibfile}


\end{document}